\documentclass[a4]{article}

\usepackage[utf8x]{inputenc}
\usepackage{ucs}
\usepackage{xcolor} 
\usepackage[round, sort, numbers, authoryear]{natbib}
\usepackage{amsmath}
\usepackage{amsfonts}
\usepackage{amssymb}
\usepackage{latexsym}
\usepackage{graphicx}
\usepackage[position=top]{subfig}
\usepackage{geometry}

\providecommand\bnabla{\boldsymbol{\nabla}}

\DeclareTextFontCommand\textsfbi{\usefont{OT1}{phv}{b}{it}}
\DeclareMathAlphabet\mathsfbi            {OT1}{phv}{b}{it}

\newcommand{\mylab}[3]{\raisebox{#2}[0mm][0mm]{%
\makebox[0mm][l]{\hspace*{#1}\textbf{#3}}}}
\def\spacce#1{\hskip #1pt}
\def\drawline#1#2{\raise 2.5pt\vbox{\hrule width #1pt height #2pt}}
\def\solid{\drawline{24}{.5}\nobreak}

\def\bdash{\hbox{\drawline{5.8}{.5}\spacce{2}}}

\def\dashed{\bdash\bdash\bdash\nobreak}

\def\chndot{\hbox%
{\drawline{4.6}{.5}\spacce{2}\drawline{1}{.5}\spacce{2}\drawline{4.6}{.5}\spacce{2}\drawline{1}{.5}\spacce{2}\drawline{4.6}{.5}}\nobreak }
\def\circle{$\circ$\nobreak }

\def\trian{\raise 1.25pt\hbox{$\scriptstyle\triangle$}\nobreak}

\def\dtrian{\raise 1.25pt\hbox%
{$\scriptscriptstyle\bigtriangledown$}\nobreak}

\def\squar{\raise 1.25pt\hbox{$\scriptstyle\Box$}\nobreak}

\def\diamon{\raise 1.25pt\hbox{$\scriptstyle\diamond$}\nobreak}

\newcommand{\soliddtrian}{$\blacktriangledown$\nobreak}

\def\solidcircle{$\bullet$\nobreak}

\def\linedtri1{\hbox{\bdash\hspace{-1.6mm}$\bigtriangleup$\hspace{-0.8mm}\bdash}\nobreak}
\def\soliddtrian1{$\blacktriangledown$\nobreak}
\def\solidrtrian2{$\blacktriangleright$\nobreak}
\def\solidltrian3{$\blacktriangleleft$\nobreak}
\def\Diamond{\diamond}
\def\dd{{\, \rm{d}}}
\def\dr{{\rm{d}}}

\def\bra{\langle}
\def\ket{\rangle}

\def\beq{\begin{equation}}
\def\eeq{\end{equation}}
\def\la{\label}

\def\uvec{\mbox{\boldmath $u$}}
\def\omvec{\mbox{\boldmath $\omega$}}
\def\Bvec{\mbox{\boldmath $B$}}
\def\Smat{{\mathsfbi{S}}}
\def\Surf{\Omega}
\def\r#1{(\ref{#1})}

\begin{document}
\author
{Guillem Borrell%
  \thanks{Email address for correspondence:
    guillem@torroja.dmt.upm.es},\ and Javier Jim\'enez \\
  School of Aeronautics, Universidad Polit\'ecnica de
  Madrid,\\
  Pza. Cardenal Cisneros, 3; 28040 Madrid. SPAIN }

\date{\today}

\title{Properties of the turbulent/non-turbulent interface in boundary layers.}

\maketitle

\begin{abstract}
The turbulent/non-turbulent interface is analysed in a direct
numerical simulation of a boundary layer in the range
$Re_\theta=2800-6600$, with emphasis on the behaviour of the
relatively large-scale fractal intermittent region. This requires the
introduction of a new definition of the distance between a point and a
general surface, which is compared with the more usual vertical
distance to the top of the layer. Interfaces are obtained by
thresholding the enstrophy field and the magnitude of the
rate-of-strain tensor, and it is concluded that, while the former are
physically relevant features, the latter are not. By varying the
threshold, a topological transition is identified as the interface
moves from the free stream into the turbulent core. A vorticity scale
is defined that collapses that transition for different Reynolds
numbers, roughly equivalent to the root-mean-squared vorticity at the
edge of the boundary layer. Conditionally averaged flow variables are
analysed as functions of the new distance, both within and outside the
interface. It is found that the interface contains a nonequilibrium
layer whose thickness scales well with the Taylor microscale,
enveloping a self-similar layer spanning a fixed fraction of the
boundary-layer thickness. Interestingly, the straining structure of
the flow is similar in both regions. Irrotational pockets within the
turbulent core are also studied. They form a self-similar set whose
size decreases with increasing depth, presumably due to break-up by
the turbulence, but the rate of viscous diffusion is independent of
the pocket size. The raw data used in the analysis are freely available from 
our web page  (\tt {\tt http://torroja.dmt.upm.es}).
\end{abstract}

\section{\label{Intro} Introduction}

It has been known since the early days of turbulence research that the
near-wall region of boundary layers follows the law of the wall, but
that the outer region is influenced by the interaction between
turbulence and the free stream \citep{NACA:1247}, whose most obvious
manifestation is the `wake' component of the mean velocity profile
\citep{FLM:367043,jimenez2010turbulent}. Early work by
\citet{NACA:W-94} revealed the presence of a sharp but irregular
boundary between turbulent and non-turbulent flow, and the
intermittent character of the flow near that boundary. It is also
known that, although the outer part of boundary layers has some
similarities to a wake \citep{FLM:367043}, intermittency does not
behave identically in different flows \citep{FLM:381094}. This is true
even if the extent of the intermittent region, quantified by
\cite{Townsend} in terms of the fraction of time during which a given
point is turbulent, is found to be similar in many flows.

Much of the research on the turbulent/non-turbulent (T/NT) interface
has dealt with the entrainment process by which the irrotational flow
acquires vorticity. An important early result was that the surface
area of the T/NT interface is much larger than its projected area in
the direction normal to the wall, and that it is intensely folded
\citep{Fiedler}. This observation was the origin of two conjectures
summarized by \cite{TownsendBook}. The first one is that the interface
itself has similar mass transfer per unit area in all turbulent flows,
and that the different entrainment rates (stronger in jets and wakes,
weaker in boundary layers and in plane mixing layers) are due to
different folding intensities. The second conjecture has to do with
the details of the entrainment mechanism. It is clear that
irrotational fluid can only acquire vorticity by viscous diffusion
\citep{NACA:1244}. But if the interface is folded enough, large
pockets of irrotational flow can be trapped by large coherent
structures and driven deep into the turbulent region before acquiring
vorticity. To add some nomenclature, small-scale entrainment is called
nibbling, while the process by which large blobs of irrotational fluid
are swallowed by the turbulent flow before becoming vortical is called
engulfment \citep{mathew:2065}. The ongoing discussion on the relative
importance of the two entrainment mechanisms hinges in part on the
understanding of the geometry of the T/NT interface.

Capturing this geometry is challenging in both experiments and
simulations, partly because of its complexity. The thickness of the
intermittent zone is comparable to the boundary layer thickness
$\delta_{99}$ \citep{NACA:1244}, while we will see that the strong
gradients present in the interface contain some of the shortest scales
in the flow. The interface inherits the fractal-like properties of the
underlying turbulent flow \citep{Sreenivasan09011989} and, since
turbulent flows typically contain eddies of all possible sizes between
the smallest and largest scales, all of them have to be considered
when the interface geometry is studied. As a result, important
questions about entrainment in turbulent flows had to wait for the
necessary data to be available.

Some experimental techniques are able to capture the interface with
considerable detail, and the methods described in
\citet{PrasadSreenivasan} are still used today. The properties of the
flow surrounding the interface could not be measured until the advent
of particle image velocimetry \citep{WesterweelEiF} and particle
tracking velocimetry \citep{FLM:1757608}. However, experiments are
typically restricted to two-dimensional sections of the flow, and the
three-dimensional description of the field requires direct numerical
simulations (DNS).

Just as experiments, simulations have limitations. The range of
available scales is the most obvious, and is crucial if the scaling
properties of a phenomenon are to be studied. Direct numerical
simulations at Reynolds numbers large enough to observe a reasonable
scale separation are a recent achievement. While there have been
boundary layer simulations at moderate Reynolds numbers for some time
\citep{jimenez2010turbulent,FLM:7881038,leesung2013}, a domain size
sufficiently large to obtain a deep T/NT interface requires
state-of-the-art DNSes
\citep{SilleroJimenez,pirozzoli2013probing}. The Reynolds numbers of
these newer simulations is comparable to that of most experiments for
which the interface has been analysed in any detail.

These larger and more accurate representations of the flow field, and
better analysis tools, have called into question the consensus of what
is the dominant mechanism of entrainment. \citet{DahmDimotakis},
\citet{Ferreetal}, \citet{Mungaletal} and \citet{Dimotakis} suggested
that engulfment is the dominant process, but later works like
\citet{mathew:2065}, and \citet{WesterweelPRL} emphasize again the
importance of nibbling. The dichotomy may have something to do with
the level of description desired, since it is clear that viscosity is
the ultimate mechanism for vorticity diffusion, but it is equally
clear that the complex geometry of the interface has to be taken into
account in determining the rate of diffusion.

To determine which scales are most relevant to entrainment requires
the study of the turbulent structures in the vicinity of the
interface, which implies the analysis of the properties of the flow in
a reference frame linked to the interface itself.  \citet{Fiedler}
presented results obtained from hot wires, but it was not until the
work of \citet{FLM:95049}, \citet{WesterweelEiF}, \citet{FLM:8400021}
and \cite{Ree:Holz:14} that conditional profiles relevant to the
scaling of the interface were shown. \citet{FLM:95049} mentioned that
the T/NT interface could contain at least two layers with possibly
different scaling properties: a turbulent region where the major
exchanges between the irrotational fluid and the fully turbulent core
occur, and a thinner viscous superlayer at its outer boundary, already
conjectured by \citet{NACA:1244}. A similar observation was made
recently by \cite{Ishihara2015analysis} for a boundary layer. A recent
review of the state of the art is \cite{SilvaAR14}.

The length scales of the interface provide information about the
configuration of the nearby eddies, and about how they are affected by
the irrotational flow. We define in this paper the T/NT interface as
the region in which the properties of the flow are neither fully
turbulent nor completely irrotational, and we are interested in
describing how this transition is structured. Two important questions
are what is the thickness of the transition layer, and whether it can
be further subdivided into distinct sublayers. The main candidates for
the scaling of the T/NT interface are the Kolmogorov viscous length
$\eta$ and the Taylor microscale $\lambda$. The thickness of the
vorticity interface of a DNS temporal jet was computed by
\citet{SilvaTaveira} and found to be of the order of the Taylor
microscale, and \citet{GampertRetau} were able to scale quite
accurately with $\lambda$ the average thickness of the interface of a
passive scalar in a jet over the range $Re_\lambda=61-141$.  This
would agree with the theory described in \citet{HuntDurbin} who, on
the assumption that the interface is subject to a relatively strong
local shear, noted that eddies would be blocked and squeezed instead
of escaping to the irrotational side. Such an interface would have
different dynamics from the rest of the flow and a characteristic
thickness of $O(\lambda)$.

The goal of this paper is to study the properties of the T/NT
interface in a turbulent boundary layer, with emphasis on the
relatively large-scale interactions across the fractal intermittent
layer, rather than on the thinner viscous superlayer.  We also analyse
the consequences of the threshold used for interface detection. New
methods are developed for the geometric characterisation of surfaces
of arbitrary complexity in three-dimensional space, and for the
conditional analysis of scalar fields with respect to those
surfaces. These methods are used to describe the properties of the
flow depending on its position relative to the T/NT interface, and to
determine the characteristic thickness of the interface layer. The
choice of the identification threshold is given special attention, as
well as the choice of the variable being thresholded.

The paper is organized as follows. The next section is a short description of the data used
in this research. The characteristics of the intermittent zone that are relevant to the
detection of a T/NT interface based on a vorticity isocontour are presented in
\S\ref{sec:intermittency}, followed in \S\ref{sec:Geometry} by a quantitative analysis of
the geometrical properties of the interface and its dependence on the threshold. Section
\ref{sec:conditional} presents the conditional analysis of the flow using the interface as a
reference frame. In particular, \S\ref{sec:Layer} and \S\ref{sec:Strain} describe the
structure of the vorticity and of other velocity gradients within the T/NT interface layer,
and \S\ref{sec:Thickness} discusses the determination of its thickness. Finally,
\S\ref{sec:shear} explores the behaviour of the velocity magnitude across the interface, and
\S\ref{sec:Conclusions} concludes.

\section{\label{sec:Description} Description of the data.}

\begin{figure}
\centering
\includegraphics[width=0.70\linewidth]{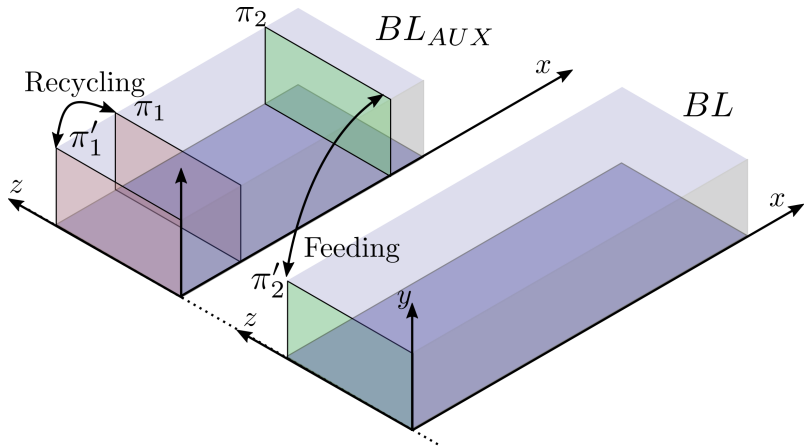}
\caption{Sketch of the simulation and the boundary conditions. The
  inflow boundary conditions for $BL$ are obtained from $BL_{AUX}$,
  copying the plane $\pi_2$ to to the first plane of $BL$ at
  $\pi_2^\prime$. The streamwise location of $\pi_2$ is chosen so that
  the flow has recovered from the recycling scheme ($\pi_1$ is
  recycled to $\pi_1^\prime$) used to start $BL_{AUX}$ from a
  turbulent inflow condition.}
\label{fig:sketch-2tbls-pof}
\end{figure}

The boundary layer is simulated in a rectangular box over a no-slip
smooth wall. The spanwise boundary conditions are periodic, and inflow
and outflow conditions are imposed in the streamwise direction. A
transpiration velocity in the boundary opposite to the wall keeps the
pressure gradient very close to zero. The simulation code and its
implementation are thoroughly explained in \cite{Simens}, and the
modifications to achieve higher Reynolds numbers are presented in
\cite{Borrell}. The axes in the streamwise, wall-normal and spanwise
directions are $x$, $y$, and $z$ respectively. The total velocity
vector is $\uvec$, with components along each axis $u$, $v$, and $w$,
respectively. Wall units are defined in terms of the friction velocity
$u_\tau$ and of the kinematic viscosity $\nu$, and are denoted by a
`+' superscript.  The brackets $\langle \cdot \rangle$ denote the
ensemble average at a given wall-normal location, and primes denote
root-mean-squared values. Both are functions of $x$ and $y$, and are
obtained from field snapshots sufficiently separated in time (about
0.2 flow turnovers) to discard spurious correlations between the
samples. The boundary layer thickness is $\delta_{99}$, defined as the
distance to the wall at which $\langle u \rangle$ is 99\% of the
free-stream velocity. The Kolmogorov length is $\eta=(\nu^3/\langle
\varepsilon \rangle)^{1/4}$, where
\begin{equation}
  \label{eq:epsilon}
  \langle \varepsilon \rangle = \nu \left[ 
    \left \langle |\nabla(\uvec-\bra\uvec\ket)|^2 \right \rangle
\right]
\end{equation}
is the turbulent kinetic energy pseudo-dissipation rate. A third relevant length 
is the Taylor microscale
\begin{equation}
  \label{eq:lambda}
  \lambda=\sqrt{\frac{15\nu u'^2}{\langle \varepsilon \rangle}},
\end{equation}
where $u'^2=[\bra|\uvec|^2\ket-|\bra\uvec\ket|^2]/3$ is the
one-component velocity fluctuation intensity computed under the
assumption of isotropy. Table \ref{table:parameters} and figure
\ref{fig:sketch-2tbls-pof} summarize the important parameters and
characteristics of the simulation, which was designed to achieve
convergence of all the scales of the flow in the domain labelled $BL$,
over a range of Reynolds numbers as wide as possible. Two simulations
are run simultaneously with a synchronized time step, but the purpose
of the auxiliary simulation $BL_{aux}$ is just to provide inflow
boundary conditions for $BL$. Only data from $BL$ are used in this
paper. A detailed discussion of the effects of the inflow and of the
evolution with $x$ of the flow properties towards their asymptotic
behaviour can be found in \cite{SilleroJimenez}.

\begin{table}
\centering
\begin{tabular}{lcccccc}
  \hline Case & $N_x$, $N_y$, $N_z$  & $\delta_{99}^+$ &
  $Re_\lambda$ & $\delta_{99}/\eta$ & $\delta_{99}/\lambda$ & $Tu_\tau/\delta_{99}$\\ 
  \hline  $BL_{aux}$ & $3585 \times 315 \times 2560$ & $630-1100$ 
  &  \  & \  & \   \\ 
  $BL$       & $15361 \times 535 \times 4096$ & $1000-2000$
  & $75-108$ & $242-440$ & $14.2-21.4$ & 11.5\\ 
  \hline 
\end{tabular}  
\caption{\label{table:parameters} Summary of important parameters of
  the simulation. $N_x$, $N_y$ and $N_z$ are the size of the
  computational grid. The Taylor-microscale Reynolds number
  $Re_\lambda$, the Kolmogorov length, $\eta$, and the Taylor
  microscale, $\lambda$, are estimated at $y=0.6\delta_{99}$. The
  running time $T$ is normalised with properties at the middle of the
  {\it BL} box. }
\end{table}

Because both $\lambda$ and $\eta$ depend on the distance to the wall, especially in the
intermittent region, the values used below to normalise lengths are taken at
$y=0.6\delta_{99}$. This is the point farthest from the wall which can be assumed to be
roughly free of intermittency corrections. We will see below that the dissipation decays
in the turbulent parts of the layer approximately as in non-intermittent internal
turbulent flows (see figure \ref{fig:intermittency}a), so that $\lambda\propto y^{1/2}$
and $\eta\propto y^{1/4}$ \citep{tennekes1972first}. As a consequence, the reference values used
below are proportional to `notional' values at the edge of the layer,
$\eta(0.6\delta_{99})/\eta(\delta_{99})\approx 0.6^{1/4}=0.88$ and
$\lambda(0.6\delta_{99})/\lambda(\delta_{99})\approx0.77$, and the scaling properties with
respect to the mid-layer values are the same as with respect to those at the
boundary-layer edge.

The simulation agrees excellently with previous experiments and direct
numerical simulations \citep{SilleroJimenez,sillero14}. The Taylor
microscale Reynolds number, $Re_\lambda=\lambda u^\prime / \nu \simeq
O(100)$, is comparable to those of most experiments and simulations
used in the analysis of the T/NT interface in free shear flows, and
higher than those in the boundary layers previously used for that
purpose. The microscale Reynolds numbers in table
\ref{table:parameters} are evaluated at $y=0.6\delta_{99}$, but they
vary little in the range $y/\delta_{99}=0.3-0.6$ and only start to
decrease where the flow intermittency becomes important. They can be
taken as representative of the `turbulent' $Re_\lambda$ near the T/NT
interface. The friction Reynolds number $\delta_{99}^+$ ranges over a
factor of two (see table \ref{table:parameters}), allowing it to be
used as a parameter in the analysis.  The resulting ratio of
$\delta_{99}/\eta$ ranges over a factor of 1.8, easily allowing the
distinction between outer $(\delta_{99})$ and viscous $(\eta)$
scaling. The corresponding range of $\lambda/\eta$ is only about 1.2,
but still sufficient to distinguish between scalings with the two
quantities. The averaged properties of the data set have been
accumulated over the complete history of the simulation after
discarding the initial transient. Some of the more detailed results
have been obtained from at least eight flow snapshots, sufficiently
separated to ensure statistical independence.

\section{\label{sec:intermittency} The interface detection criterion.}

The first step to study the T/NT interface is to define the criterion
that discriminates between turbulent and non-turbulent
flow. Unfortunately, different methods produce different interfaces,
and the criteria found in the literature are variable enough to be
difficult to compare consistently. Historically, the first interface
detections were based on a cut-off frequency for the one-point
streamwise-velocity signal, in the expectation that turbulent
fluctuations could be easily distinguished by their faster time
scales. As better descriptions of the flow became available, the
interface came to be defined by an indicator function with two
components: a scalar field related to the turbulent fluctuations, and
a threshold. \citet{PrasadSreenivasan} use the concentration of a
passive scalar injected in the turbulent side, and threshold it at the
least probable value of the concentration.  \Citet{FLM:8400021} in
jets, and \citet{FLM:95049} in wakes, use the vorticity magnitude as
indicator, and a particularly low vorticity value as the threshold. In
boundary layers, \citet{jimenez2010turbulent} also uses the vorticity
magnitude, and a threshold based on a sharp jump in its probability
density function (PDF) at $y=\delta_{99}$. \citet{Chauhan:14} use a
measure of the kinetic-energy fluctuations as their scalar, and choose
the highest threshold for which the PDF of the height of the interface
above the wall can be fitted by a gaussian.

Our criterion is based on the magnitude of the total vorticity,
$\omega = |\omvec| = |\nabla \wedge \uvec|$, which has several
desirable properties as a turbulence indicator and can easily be
obtained from DNS. In the first place, the incompressible identity
$\nabla^2 \uvec = - \nabla \wedge \omvec$ implies that the
characteristic turbulent dissipation of energy requires
vorticity. Secondly, while velocity gradients can be created by
pressure fluctuations in potential flow, there is no inviscid
mechanism to create vorticity fluctuations. As a consequence, even if
vorticity is not conserved, any vorticity in the boundary layer is
ultimately linked to the wall. Note that some of these desirable
properties of the vorticity magnitude do not extend to its individual
components. For example, there can be energy dissipation in the
absence of one vorticity component, and any component can easily
appear or disappear by simple rotation. More significantly, although
vorticity is very approximately isotropic away from the wall
\citep{jimenez2013near}, the solenoidality of the curl requires that
the vorticity vector should be roughly parallel to any surface across
which its magnitude drops or increases sharply. The geometry of the
vorticity magnitude and of any one of its component cannot be assumed
to be similar.

We thus define a point as turbulent if 
\begin{equation}
  \label{eq:definition_turbulent}
  \omega(x,y,z,t) > \omega_0,
\end{equation}
and the T/NT interface by $\omega = \omega_0$, and turn our attention to determining the
threshold $\omega_0$, either from the properties of the resulting interface or
from comparisons with previous investigations.

\begin{figure}
\centering
\subfloat[]{
  \includegraphics[width=0.48\textwidth, clip=true, 
  trim=0em 0em 0em 0em]{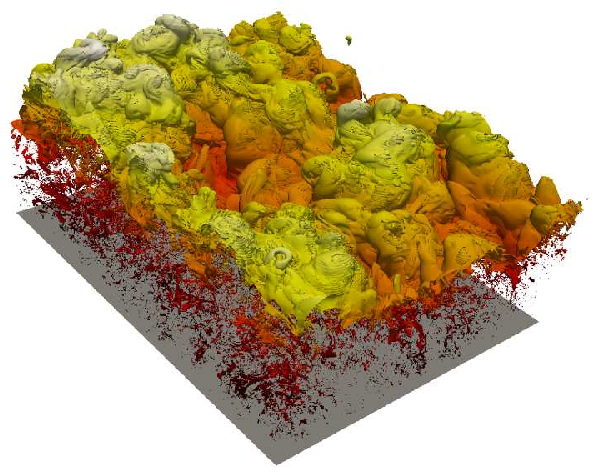}
}
\phantom{\hspace{-1em}}
\subfloat[]{
  \includegraphics[width=0.48\textwidth, clip=true, 
  trim=0em 0em 0em 0em]{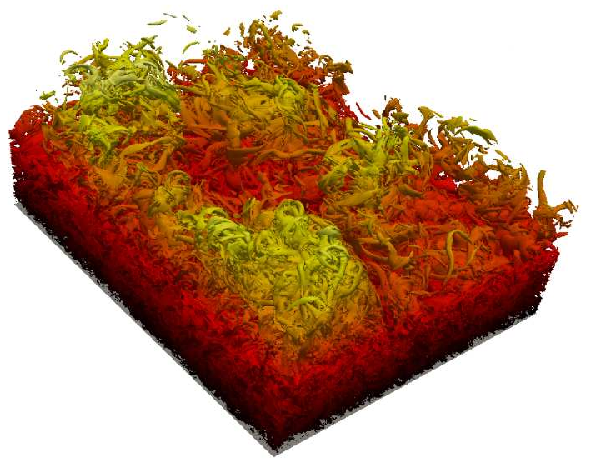}
}
\caption{\label{fig:vorticity_isocontour} Vorticity magnitude
  isosurfaces of the same region of the present data at $\delta_{99}^+
  \simeq 2000$, of size $3\delta_{99}\times 2\delta_{99}$ in the
  streamwise and spanwise directions, respectively.
(a) $\omega_0^+=5\times 10^{-4}\, (\omega_0^*=0.022)$. 
(b) $\omega_0^+=5\times 10^{-3}\, (\omega_0^*=0.22)$. 
For the definition of $\omega^*$, see \eqref{eq:vortouter}.  The flow
is from top-left to bottom right, but note that the spanwise and
streamwise directions are barely distinguishable.  }
\end{figure}

The simplest tool is three-dimensional visualization, preferably of a
relatively large part of the interface. Figures
\ref{fig:vorticity_isocontour}(a,b) show the interface in a domain
whose wall-parallel size is several boundary layer thicknesses, for
two thresholds separated by an order of magnitude. The two figures are
different.  Figure \ref{fig:vorticity_isocontour}(a) can be described
as a moderately complex envelope with scattered small regions of low
vorticity within the turbulent side, while figure
\ref{fig:vorticity_isocontour}(b) has a large number of handles and
contortions that span a significant fraction of the boundary layer
thickness.

\begin{figure}
\centerline{%
\includegraphics[width=0.49\textwidth,clip=true,trim=0em 1em 1em 1em]{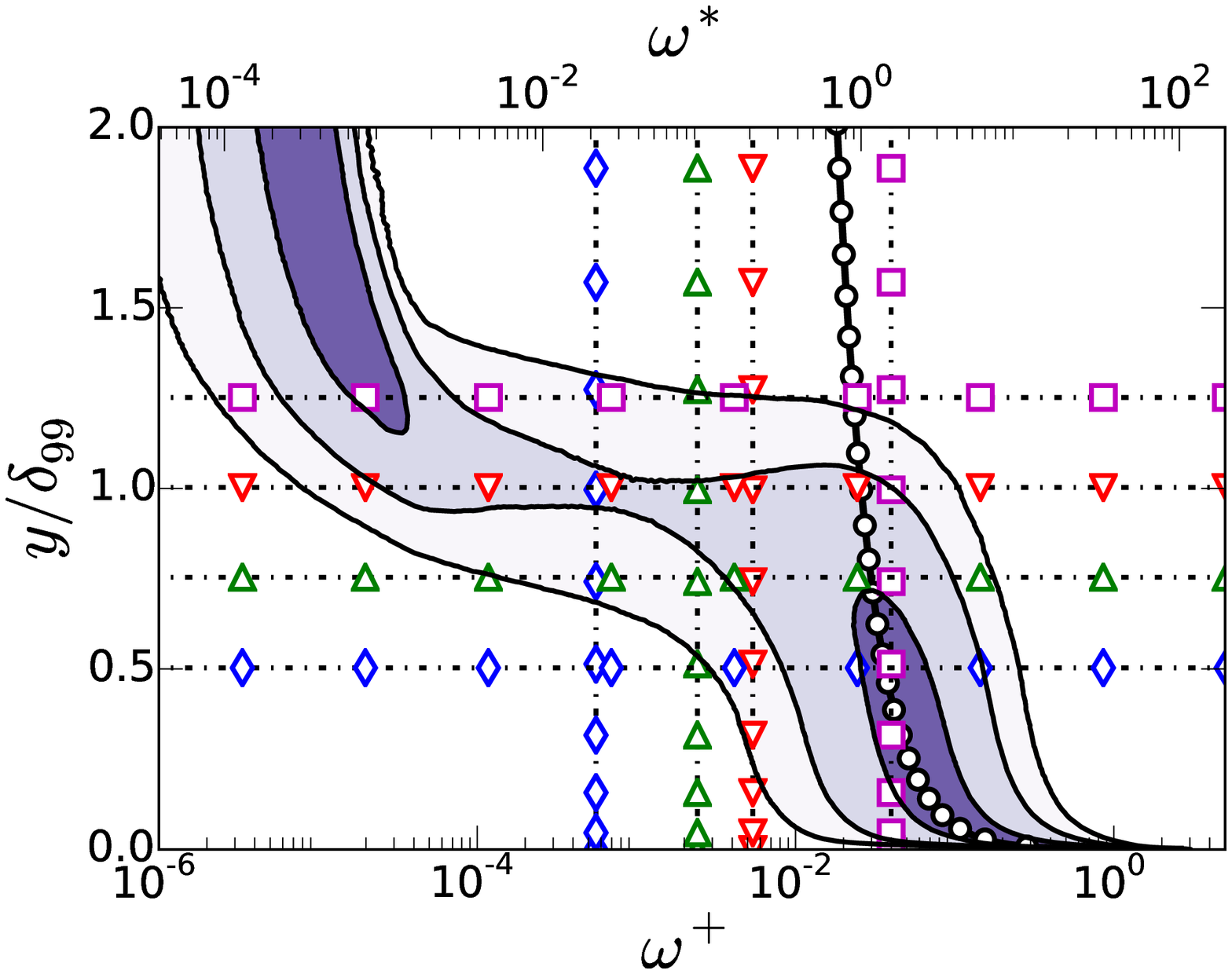}%
\mylab{-0.1\textwidth}{0.28\textwidth}{(a)}%
\hspace{2mm}%
\includegraphics[width=0.49\textwidth,clip=true,trim=0em 1em 1em 1em]{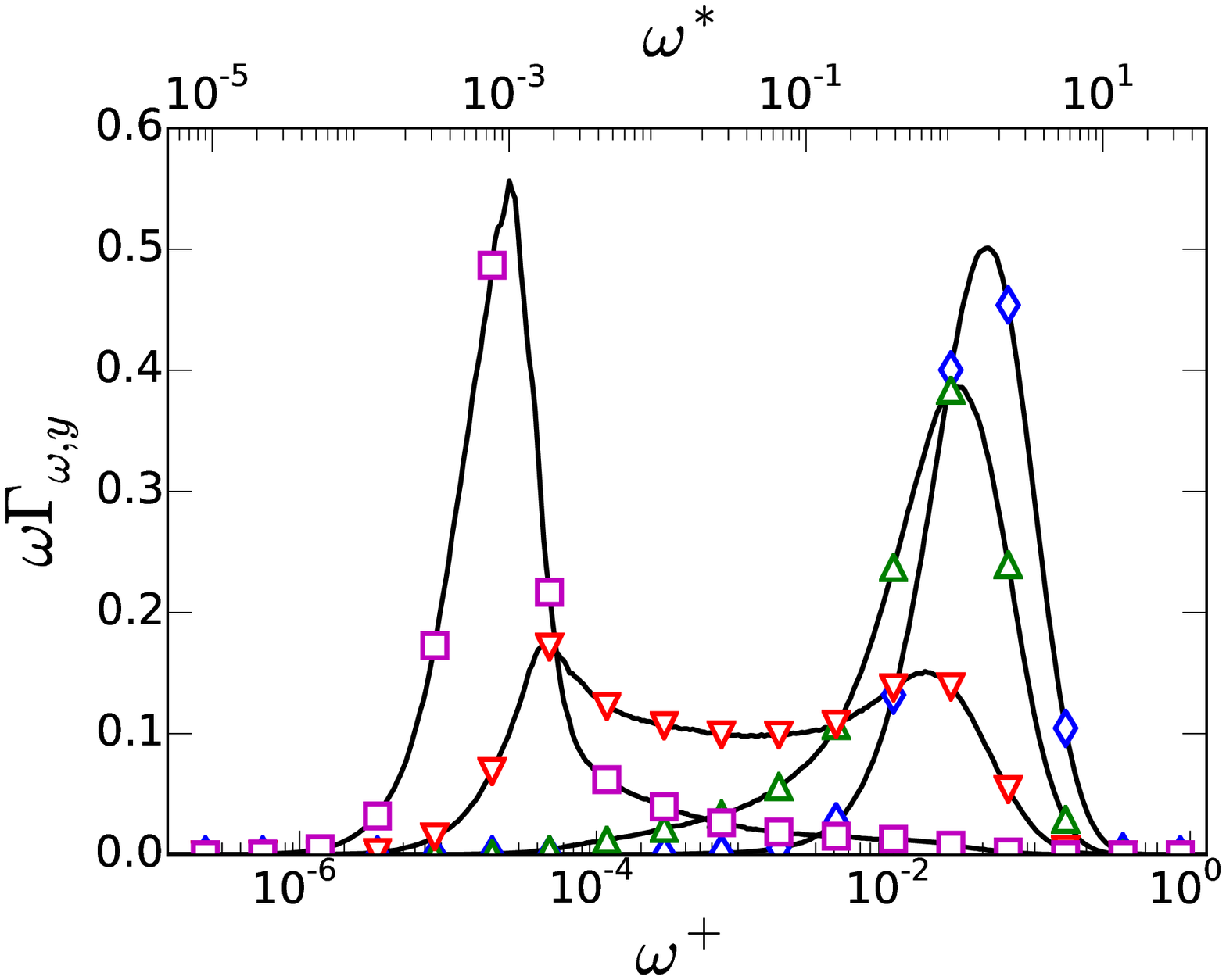}%
\mylab{-0.1\textwidth}{0.28\textwidth}{(b)}%
}
\vspace{1ex}%
\centerline{%
\includegraphics[width=0.49\textwidth, clip=true,
trim=0em 1em 1em 1em]{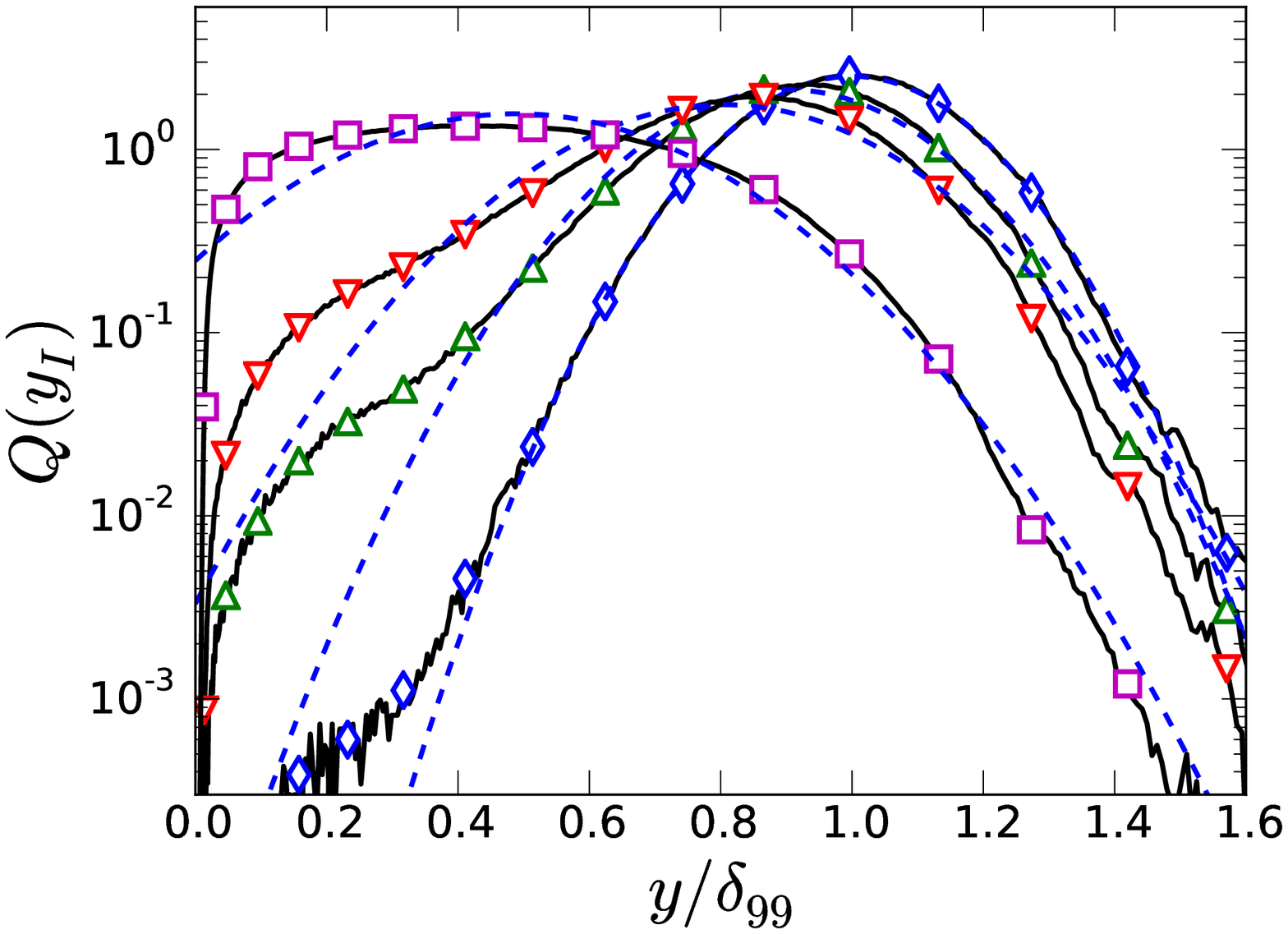}%
\mylab{-0.1\textwidth}{0.28\textwidth}{(c)}%
\hspace{2mm}%
\includegraphics[width=0.49\textwidth,clip=true,trim=0em 1em 1em 1em]{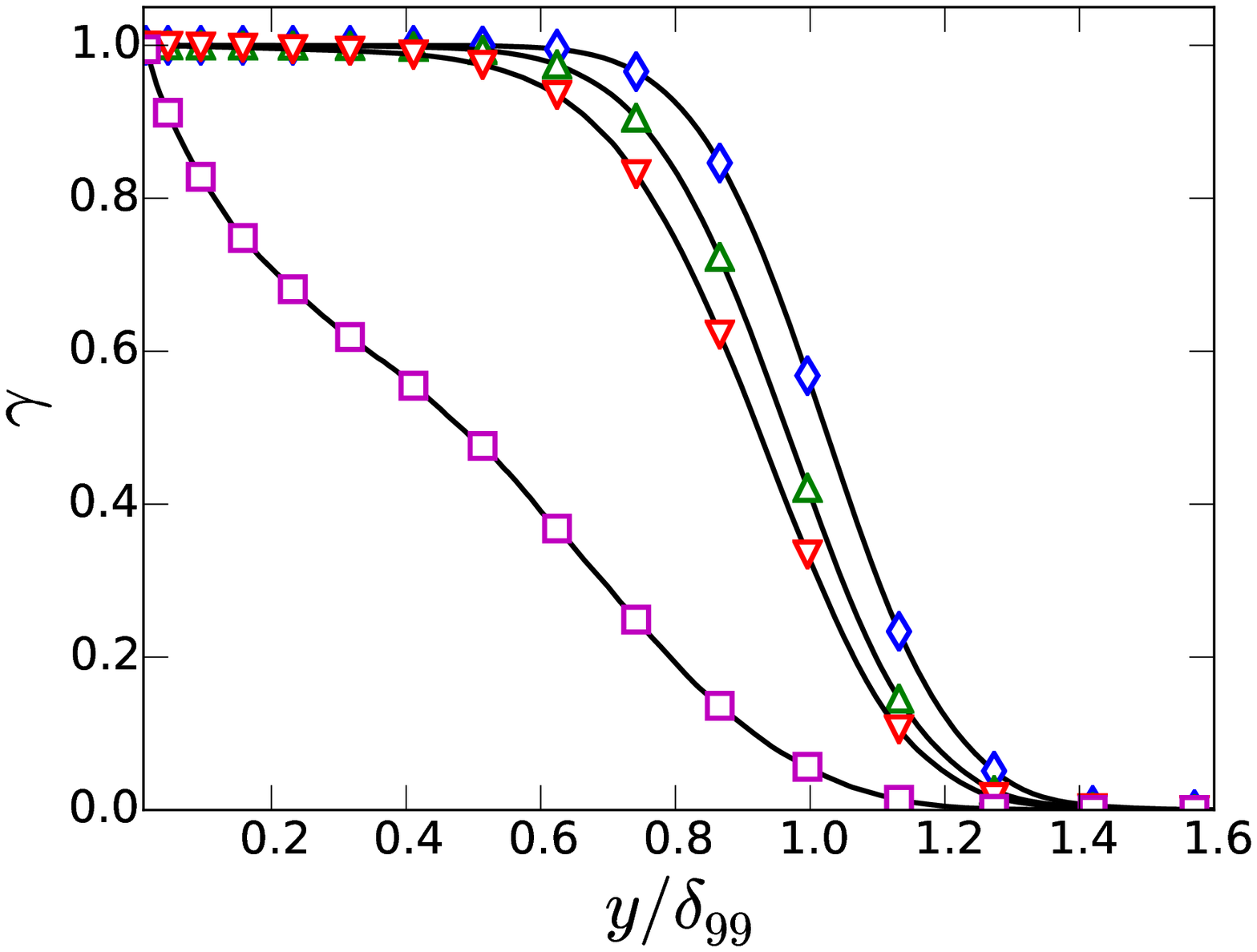}%
\mylab{-0.1\textwidth}{0.28\textwidth}{(d)}%
}
\caption{%
(a) Premultiplied joint PDF, $\omega \Gamma_{\omega,y}$, of
  the wall-normal distance and the vorticity magnitude. Contours
  contain 50\%, 90\%, and 99\% of points, respectively. Two vorticity
  scales are provided, wall units $\omega^+$, and the $\omega^*$
  defined in \eqref{eq:vortouter}. The line with open circles is
  $\omega^+=(y^+)^{-1/2}$. The horizontal and vertical lines
  correspond to the one-dimensional sections in (b, c), using the same
  markers.
(b) Sections of $\omega \Gamma_{\omega,y}$ at four different distances
to the wall: $\diamond$, $y/\delta_{99}=0.5$; $\triangle$, $0.75$;
  $\triangledown$, 1; $\square$, $1.25$.
(c) One-dimensional PDF, $Q(y_I)$, of the vertical position of the
interface when its average position is: $\Diamond$, $\langle y_I
\rangle/ \delta_{99} = 1$ $(\omega_0^*=0.022)$; $\triangle$, $0.9$
$(\omega_0^*=0.09)$; $\triangledown$, $0.8$ $(\omega_0^*=0.2)$;
$\square$, $\omega^*_0=2.0$. The vorticity threshold of the first and
third curves are those of the isosurfaces in figures
\ref{fig:vorticity_isocontour}(a,b), respectively. The dashed line
fitting each curve is the gaussian distribution with the same mean and
standard deviation.
(d) Intermittency factor for the four thresholds in (c).  In all
cases, $\delta_{99}^+ \simeq 1500$.  }
\label{fig:intermittency}
\end{figure}

Another useful tool is the joint PDF of the vorticity magnitude and of
the vertical distance to the wall, which is presented in figure
\ref{fig:intermittency}(a) as a premultiplied PDF, $\omega
\Gamma_{\omega,y}$, to account for the logarithmic scale of the
vorticity. It has two well-defined regions. The high-vorticity
near-wall points of the turbulent core of the boundary layer are in
the lower right-hand corner. Points far from the wall with very low
vorticity, representing the ideally irrotational non-turbulent free
stream, are in the top left corner. Their residual vorticity is due to
the finite accuracy of the inflow condition, but it is about four
orders of magnitude weaker than the turbulent values, and easily
distinguished from them. In the present data set, the details of the
joint PDF depend only weakly on the Reynolds number.

The quantity $\omega\Gamma_{\omega,y}$ was also obtained by
\citet{silva2014characteristics} for other external turbulent flows,
showing that the intermittent region is comparable in all the tested
cases. There is always a sharp jump in vorticity, and a relative wide
range of choices for the threshold.

On the turbulent side of the PDF, the mode of the vorticity
distribution follows closely the expected $y$-dependence of its
root-mean-squared value, $\omega'$, which can be estimated from the
approximate balance between the production and the pseudo-dissipation
of the turbulent kinetic energy,
\begin{equation}
  \nu \omega'^2 \simeq - \langle u v \rangle \frac{\partial \langle u
    \rangle}{\partial y} \simeq \frac{u_\tau^3}{\kappa y},
  \label{eq:omp}
\end{equation}
where $\kappa\simeq 0.4$ is the von K\'arm\'an constant. Equation
\eqref{eq:omp} holds above $y^+\simeq50$ \citep{jimenez2013near}, and
provides a characteristic magnitude for the vorticity fluctuations,
\begin{equation}
  \langle \omega^+ \rangle \simeq \left(\kappa y^+\right)^{-1/2}.
  \label{eq:epsilon_y}
\end{equation}
We will use this dependency, particularised at the edge of the boundary
layer, to define dimensionless \emph{star} units for the vorticity,
\begin{equation}
\omega^*= \omega\,(\delta_{99}^+)^{1/2} (\nu/u_\tau^2),
\label{eq:vortouter}
\end{equation}
which are linked to the interface. The usual scaling $\omega^+=
\omega\nu/u_\tau^2$ is linked to the wall. The ratio
$\omega^*/\omega^+$ varies by a factor of 1.4 in our range of Reynolds
numbers, and we will see below that star units collapse most
properties of the interface substantially better than wall units.

The definition of $\omega^*$ can be adapted to flows other than the
boundary layer by normalising the vorticity with the root-mean-squared
value of the enstrophy fluctuations of the turbulent fluid close to
the interface. We will occasionally do this for the purpose of
comparison.

There is a band connecting the turbulent and non-turbulent regions of
figure \ref{fig:intermittency}(a) that spans several orders of
magnitude of the vorticity and extends over $y/ \delta_{99}=0.3$
--1.5. Four sections of the premultiplied PDF at different $y$ are
presented in figure \ref{fig:intermittency}(b). The one at $y =
\delta_{99}$ is particularly interesting, because it shows the
separation between the two regions of the flow. Its two mild peaks
bracket a plateau three orders of magnitude wide, from the expected
turbulent value $\omega^*\simeq\kappa^{-1/2}=1.6$ on the right, to the
free-stream residual vorticity on the left. Any vorticity within this
plateau could in principle be used as a threshold for the interface
but, even with generous safety margins at both ends, this leaves a
full order of magnitude of possible choices. This would not be a
problem if thresholds within this range produced similar results, but
they do not. The two isosurfaces in figure
\ref{fig:vorticity_isocontour} are obtained with thresholds within the
plateau. They correspond to the first and third left-most vertical
lines in figure \ref{fig:intermittency}(a).

Other quantities frequently used to analyse the properties of the edge of
boundary layers can be obtained from $\Gamma_{\omega,y}$. The intermittency
parameter 
\begin{equation}
  \label{eq:intermittency_from_jpdf}
  \gamma(y; \omega_0) = \int_{\omega_0}^\infty \Gamma_{\omega,y}\dd\omega
  \Big/ \int_0^\infty \Gamma_{\omega,y}\dd\omega,
\end{equation}
is the probability that a point at a given distance from the wall is
turbulent according to \eqref{eq:definition_turbulent}. The sections
of $\Gamma_{\omega,y}$ at constant $\omega$ provide the marginal
probability distribution of the distance, $y_I$, to the wall of the 
interface defined as a vorticity isocontour,
\begin{equation}
\label{eq:heightpdf}
Q(y_I; \omega_0) = -\partial \gamma/\partial \omega_0 =
\Gamma_{\omega_0,y} \Big/\int_0^\infty \Gamma_{\omega_0,y}\dd y.
\end{equation}
Note that, because of several approximations analysed in detail below,
the effective definition of the interface does not usually exactly
coincide with this vorticity isosurface.  Four examples of $Q(y_I)$
and $\gamma(y)$ are presented in figures
\ref{fig:intermittency}(c,d). The thresholds of the first three are
chosen so that the average height of the T/NT interface is $\langle
y_I\ket/ \delta_{99} = 1$, 0.9, and 0.8, respectively, and are within
the plateau in figure \ref{fig:intermittency}(a). The first and third
ones are used in figure \ref{fig:vorticity_isocontour}. This confirms
that the threshold has an important effect on the geometry of the T/NT
interface, even for properties that are easily measurable. Note that,
although $Q(y_I)$ and $\gamma(y)$ are linked by the first equality in
\eqref{eq:heightpdf}, $\gamma$ is not very sensitive to the changes in
$Q$, and always tends to look approximately gaussian. The fourth line
in figures \ref{fig:intermittency}(c,d), marked with open squares, is
$\omega^*_0=1.6$ and corresponds to the right-most end of the plateau
in figure \ref{fig:intermittency}(a). It behaves differently from the
other three PDFs, and neither $Q(y_I)$ nor $\gamma(y)$ can be
approximated as gaussian. This threshold does not represent the
interface any more, and can best be understood as describing the
internal structure of the turbulent vorticity.

\begin{figure}
  \centering
  \includegraphics[width=0.6\textwidth, clip=true,
  trim=1em 1em 1em 1.5em]{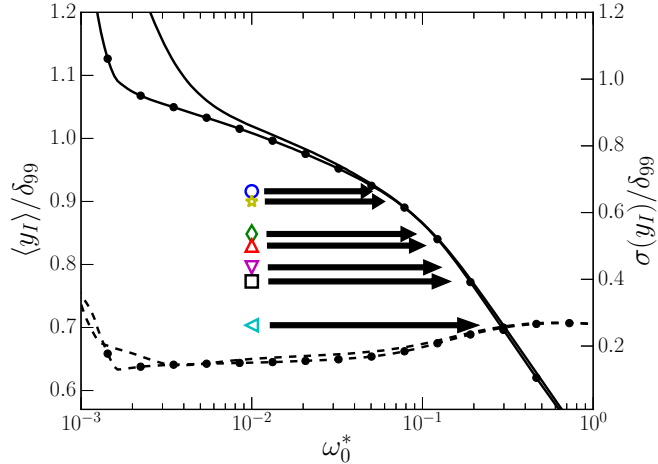}
\caption{%
The solid lines and left vertical axis are the average,
  $\bra y_I\ket$, of the height of the vorticity isosurface, and the
  dashed lines and right vertical axis are its standard deviation,
  both as functions of $\omega_0^*$. The values of $\langle y_I
  \rangle$ in table \ref{table:gamma} are marked by their
  corresponding symbols, with an arrow pointing to the matching
  $\omega_0^*$. Lines without symbols are $\delta_{99}^+=1100$; those
  with symbols are $\delta_{99}^+=1900$.  }
\label{fig:mean_std}
\end{figure}

The mean, $\langle y_I \rangle$, and standard deviation,
$\sigma(y_I)$, of the interface height are presented in figure
\ref{fig:mean_std} as functions of $\omega_0$. Three regimes can be
distinguished. The first one, below $\omega_0^*=2\times10^{-3}$,
reflects the vorticity fluctuations in the free stream, and therefore
is basically a numerical artefact.  In the second one, between
$\omega_0^*=2\times10^{-3}$ and $\omega_0^*=0.1$, the average position
of the interface is $\langle y_I \rangle \simeq \delta_{99}$, and
$Q(y_I)$ is well approximated by a normal distribution with symmetric
tails (figure \ref{fig:intermittency}c). Above $\omega_0^*=0.1$, the
left tail of $Q(y_I)$ begins to penetrate the turbulent core, $\langle
y_I \rangle$ drops faster with the threshold, and the standard
deviation increases slightly. Note that figure \ref{fig:mean_std}
includes the two extreme Reynolds numbers in our simulation, which
agree well except for thresholds low enough to represent the free
stream. This good collapse with the Reynolds number suggests that the
extent of the intermittent region is not expected to change
significantly with increasing $\delta_{99}^+$.

%
\begin{table}
\centering
\begin{tabular}{lccccc}
  \hline
  Case & Symbol &$\langle y_I \rangle/\delta_{99}$  & $\sigma(y_I)/\delta_{99}$ & $\omega_0^*$ & $\delta_{99}^+$  \\ 
  \hline
  \citet{jimenez2010turbulent} & $\circ$ & $0.92$  & $0.018$ & $0.068$ & 692  \\
  \citet{eisma2015interfaces} & $\star$ & $0.90$ & $0.018$ & $0.081$ & 2053 \\
  \citet{NACA:1244}    & $\Diamond$      & $0.83$  & $0.021$ & $0.127$ & N/A \\ 
  \citet{FLM:382907}   & $\triangle$  & $0.82$  & $0.022$ & $0.146$ & 1237 \\ 
  \citet{Murlis}     & $\triangledown$ & $0.8$   & $0.024$ & $0.182$ & 1450 \\ 
  \citet{NACA:1247}   & $\square$       & $0.78$  & $0.024$ & $0.208$ & N/A \\
  \citet{Chauhan:14}   & $\lhd$        & $0.71$  & $0.026$ & $0.311$ & 14500 \\
  \hline 
\end{tabular}  
\caption{\label{table:gamma} Properties of $Q(y_I)$ for the different
  values of $\langle y_I \rangle$ found in the literature. The
  standard deviation $\sigma(y_I)$ and the threshold $\omega^*_0$ are
  obtained from the present data set, and correspond to the threshold
  required to match $\langle y_I \rangle$ for each entry.  }
\end{table}

The values of $\langle y_I \rangle$ available in the literature are
compiled in table \ref{table:gamma} and marked with their
corresponding symbols in figure \ref{fig:mean_std}. They can be used
as guides in choosing our threshold. Note that there is a fairly large
spread between the choices of \citet{jimenez2010turbulent} and of
\citet{Chauhan:14} which, if translated to vorticity thresholds using
figure \ref{fig:mean_std}, would imply half an order of magnitude in
$\omega_0^*$. The thresholds in figures
\ref{fig:vorticity_isocontour}(a,b) corresponds to $\langle y_I
\rangle \approx \delta_{99}$ and $0.8\delta_{99}$, respectively.

Figure \ref{fig:intermittency}(b) suggests that $\omega_0^*=0.022$,
for which $\langle y_I \rangle=\delta_{99}$, should be a reasonable
threshold, since it is at this height that the vorticity PDF is widest
and bimodal. However, figure \ref{fig:mean_std} and table
\ref{table:gamma} show that this threshold is an order of magnitude
lower than most values used in previous works.

The definition in \citet{PrasadSreenivasan} can be adapted to cases
without a passive scalar, using the vorticity magnitude as a tracer
\citep{silva2014characteristics,FLM:9282512}. Applying this criterion
to the present data would imply $\omega_0^*=0.05$ and $\langle y_I
\rangle=0.95\delta_{99}$, which is comparable to figure
\ref{fig:vorticity_isocontour}(a), and again lower than the values
found in the literature. On the other hand,
\citet{silva2014characteristics} found that $\omega^*=0.01$,
corresponding to roughly $\omega/\omega^\prime=0.1$, is a reasonable
choice in jets and shear-free turbulence, and it can be shown that
there is very little difference in our case between this choice and
the value $\omega^*=0.022$ suggested
above. \citet{silva2014characteristics} also includes a proof that
this definition is related to the one used by
\citet{watanabe2015turbulent} to detect his \emph{irrotational
  boundary}.

If we try to apply the criterion of \citet{Chauhan:14} to the
vorticity field, we find that $Q(y_I)$ is approximately gaussian for
$\omega_0^* \in (2\times 10^{-3}$-- 0.1). This corresponds to $\langle
y_I \rangle/\delta_{99} \in (1.1\mbox{--}0.9) $.  Although the lowest
end of this range agrees with the mean interface height in
\citet{jimenez2010turbulent}, it is very far from the value $\langle
y_I \rangle/\delta_{99}=0.71$ obtained by \citet{Chauhan:14}. This
shows that the vorticity and velocity interfaces are different, and
that the criterion in \citet{Chauhan:14} should not be used for the
vorticity.

In summary, since neither the intermittency properties of the
interface nor previous studies provide guidance on a unique vorticity
threshold, we defer our decision until we study the evolution of the
interface over the rather wide range $\omega_0^*\in(0.001-10)$.

\section{\label{sec:Geometry} The geometry of the T/NT interface.}

In this section we study the geometry of the T/NT interface defined as
an isosurface separating vortical from irrotational fluid. As such, we
can use its dependence on $\omega_0$ to explore the geometry of the
two flow regimes as the isosurface moves from one to the other. This
will also help us to decide which threshold is best suited for each
particular purpose. For example, figure
\ref{fig:vorticity_isocontour}(a) appears to represent better the free
stream, while figure \ref{fig:vorticity_isocontour}(b) is more
representative of the interior of turbulence.

\subsection{\label{sec:drops}Bubbles and drops}

\begin{figure}
\centering
\subfloat[]{
  \includegraphics[width=0.32\textwidth]{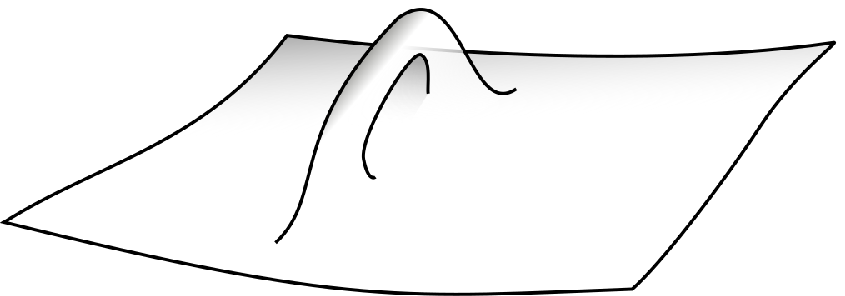}
}
\subfloat[]{
  \includegraphics[width=0.32\textwidth]{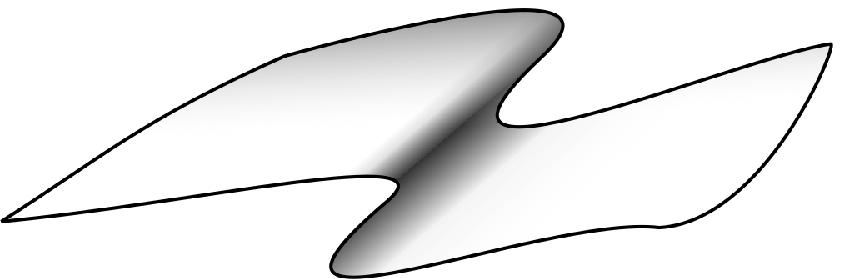}
}
\subfloat[]{
  \includegraphics[width=0.32\textwidth]{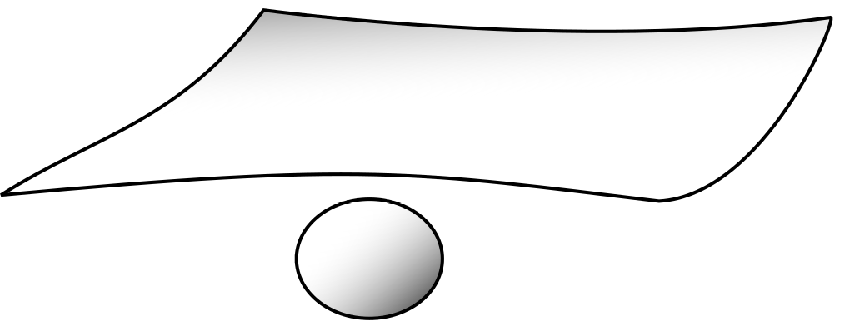}
}
\caption{Sketch of the three basic geometrical features in the vorticity
isosurface: (a) handles, (b) pockets and (c) bubbles.}
\label{fig:features}
\end{figure}

The first step is to define the interface separating the flow into
turbulent and non-turbulent regions. This is not as straightforward as
the previous section may suggest. Figure
\ref{fig:vorticity_isocontour}(b) shows that the vorticity isocontour
is not usually a singly connected surface. Depending on the threshold,
there may be a few or several thousands of disconnected components of
the isosurface, but one of them is typically much larger than the rest
and divides the computational box into two large disjoint regions. The
remaining isosurface components can be classified as envelopes of low
vorticity \emph{bubbles} within the turbulent region (figure
\ref{fig:features}c), or of high vorticity \emph{drops} in the free
stream. It will be shown in \S\ref{sec:conditional} that, although
there can be a large number of bubbles, they are usually too small to
contribute significantly to most quantities related to the T/NT
interface. There are generally very few drops. In consequence, the
rest of the paper defines the interface as the largest singly
connected component of the vorticity isosurface that separates
`smoothed' irrotational and vortical regions from which drops and
bubbles have been eliminated.

\begin{figure}
\centering
\subfloat[]{
  \includegraphics[width=0.32\textwidth]{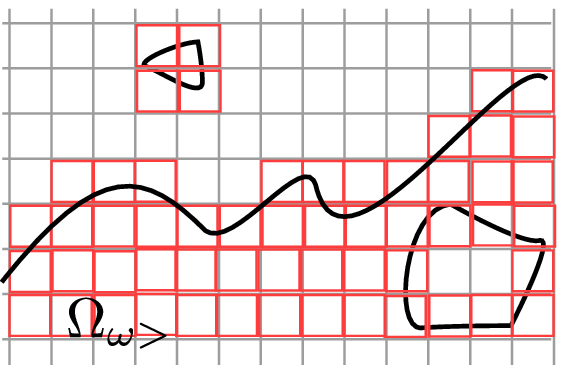}
}
\subfloat[]{
  \includegraphics[width=0.32\textwidth]{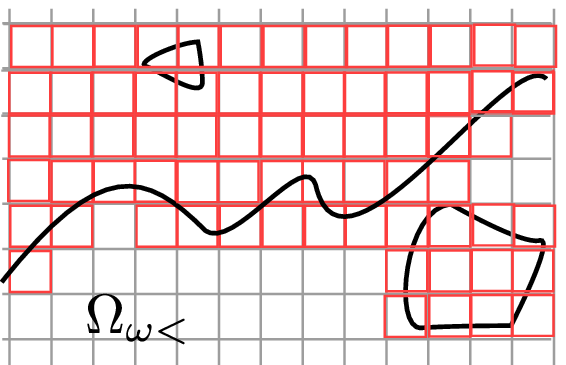}
}
\subfloat[]{
  \includegraphics[width=0.32\textwidth]{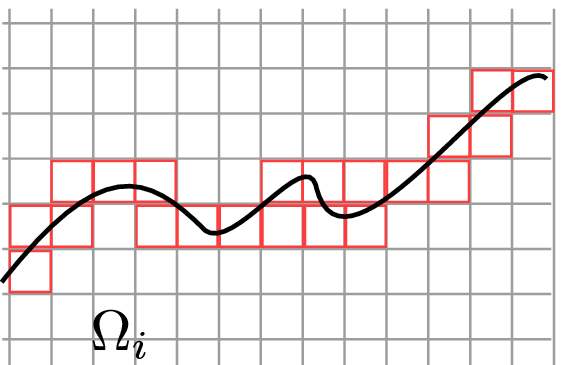}
}
\caption{(a) Set $\Surf_{\omega>}$ of voxels in which at least one
  vertex is $\omega > \omega_0$.  (b) Set $\Surf_{\omega<}$ of voxels
  where $\omega < \omega_0$.  (c) The set $\Surf_i$ of voxels that
  contain the interface, from \eqref{eq:interface_algorithm}.  }
\label{fig:algorithm}
\end{figure}

The algorithm to obtain this largest component is sketched in figure
\ref{fig:algorithm}. We first decompose the computational domain into
computational cells (voxels). The flow properties are defined at their
vertices. We next obtain the set $\Surf_{\omega>}$ of voxels for which
at least one vertex has a vorticity higher than the threshold (figure
\ref{fig:algorithm}a). This set contains the drops and the bulk of the
turbulent flow. Similarly, we obtain the set $\Surf_{\omega<}$ for
which at least one vertex has a vorticity lower than the threshold,
containing the bubbles and the bulk of non-turbulent flow (figure
\ref{fig:algorithm}b). Each of these sets has a connected component
many times larger than the rest (about seven orders of magnitude in
our case). In the case of $\Surf_{\omega>}$, this set represents the
bulk of the turbulent flow, $\Surf_t$, while in the case of
$\Surf_{\omega<}$ it represents the bulk of the free stream,
$\Surf_n$. The sets of voxels containing the drops, $\Surf_d$, and the
bubbles, $\Surf_b$, are obtained by subtracting these largest
components from their respective sets. Thus, $\Surf_d =
\Surf_{\omega>}-\Surf_t$ and $\Surf_b = \Surf_{\omega<}-\Surf_n$. The
final step is to define the set of voxels of the cleaned T/NT
interface as (figure \ref{fig:algorithm}c)
\begin{equation}
  \label{eq:interface_algorithm}
  \Surf_i = \underbrace{(\Surf_{t} \cup \Surf_{b})}_{\text{Turbulent side}} \cap
  \underbrace{(\Surf_{n} \cup \Surf_{d})}_{\text{Non-turbulent side}}.
\end{equation}
Note that each term of this equation is an effective definition of the smoothed
turbulent region (the bulk of the turbulent flow plus the bubbles), and of the
smoothed non-turbulent region (the bulk of the non-turbulent flow plus the drops).

Drops and bubbles should not be confused with other complications of
the interface, such as the handles and overhangs or `pockets'
represented in figures \ref{fig:features}(a,b). The former complicate
the topology of the flow and cannot be eliminated. The latter are
topologically neutral, but may be important from the experimental or
dynamical point of view. They hide part of the surface to some
observational procedures, and may be precursors of large-scale
engulfing. At this point, the interface is still a set of voxels that
has to be converted into a surface, but this representation is
sufficient for the analysis in the next two sections.

\subsection{\label{sec:fractal_dimension}Fractal dimension}

\citet{FLM:386975} was the first to suggest that the hierarchy of turbulent eddies can be
approximated by a fractal when the Reynolds number is large enough. This was first verified
by \citet{FLM:392671} for the bulk of the flow, and by \citet{Sreenivasan09011989} for the
T/NT interface. The latter also proposed a simple theory to relate both results. The fractal
dimension of the vorticity isosurface measures how contorted it is, and is a useful
statistical measure of its complexity. The most widely used definition is the box-counting
Kolmogorov capacity: if $N_b$ is the number of boxes of size $r$ required to cover a set
$\Surf$, such as the interface, the fractal dimension $D$ is defined by $N_b \propto r^D$.

In practice, the computation of fractal dimensions is complicated because
turbulence is only self-similar in a limited range of scales. Vorticity
is smooth at scales of the order of the Kolmogorov microscale, and the
largest eddies responsible for the energy input are not self-similar. In cases
in which an extended power law is not immediately obvious, a reasonable
redefinition of the box-counting dimension is
\begin{equation}
\dim_b = -\lim_{r \rightarrow \varsigma} \frac{\log N_b}{\log r},
\label{eq:fdimension}
\end{equation}
where $\varsigma$ stands for the smallest possible box size at which
the data set remains self-similar or, in the present case, for the
size of the computational voxels. This requirement is difficult to
define, and it is hard to speak of a fractal unless the self-similar
range extends over a reasonably wide range.

\citet{Sreenivasan09011989} found a clear power law from
two-dimensional sections of the interface, and measured a constant
dimension $D$ away from the saturation caused by the shortest and
longest scales. They concluded that the interface is a monofractal in
that range. \citet{MoisyJimenez} computed the fractal dimension of
three-dimensional enstrophy isosurfaces in homogeneous turbulence,
using the full three-dimensional field instead of cross sections and
three-dimensional boxes instead of two-dimensional ones. They found
that the self-similar range observed by \citet{Sreenivasan09011989} is
only an approximation, and defined a local dimensional exponent to
account for the dependence on the box size
\begin{equation}
D_b(r) =  -\frac{\mbox{d} \log N_b}{\mbox{d} \log r}.
\label{eq:fexponent}
\end{equation}
This definition includes the previous two. If $D_b(r)$ is constant and the
T/NT interface is a monofractal, $\dim_b=D=D_b(r)$.

\begin{figure}
\centerline{%
\includegraphics[width=0.48\textwidth, clip=true, trim=1em 0em 1em 2em]{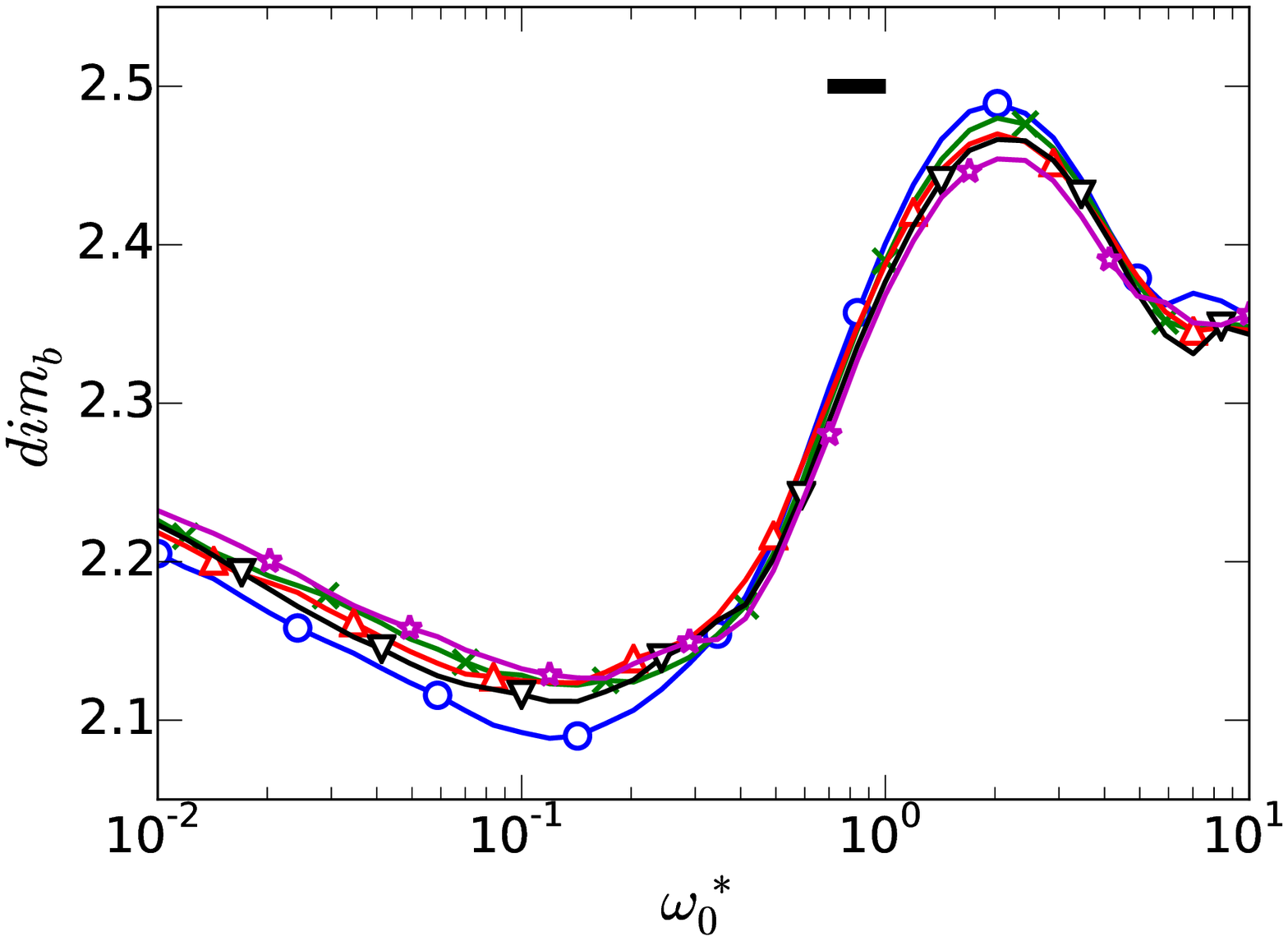}%
\mylab{-0.40\textwidth}{0.29\textwidth}{(a)}%
\hspace{2mm}%
\includegraphics[width=0.48\textwidth, clip=true, trim=1em 0em 1em 2em]{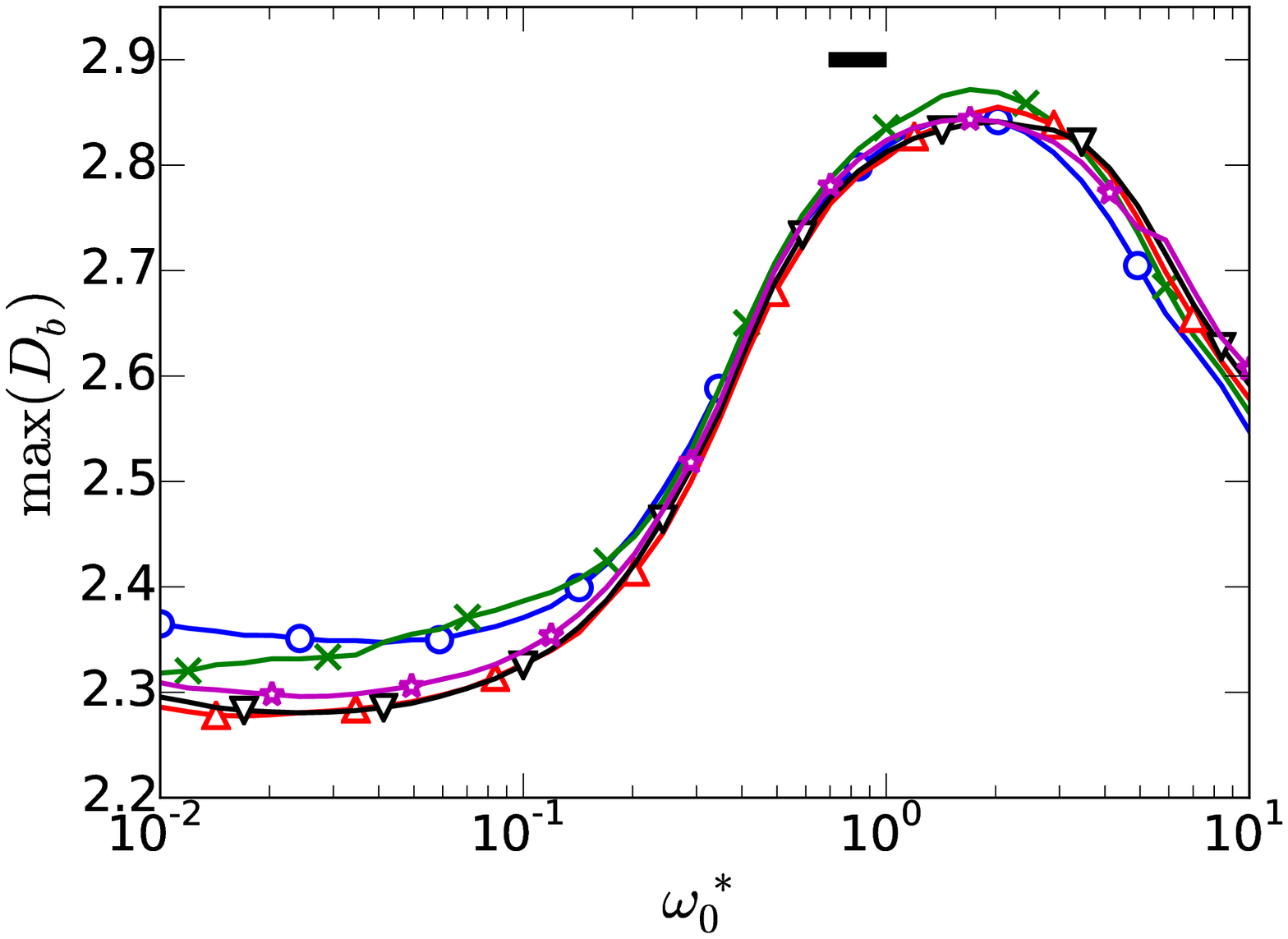}%
\mylab{-0.40\textwidth}{0.29\textwidth}{(b)}%
}
\caption{\label{fig:fdimension} Estimation of the box-counting fractal
  dimension, obtained in (a) from the local fractal exponent in the
  limit of small box size, and in (b) from the maximum of the local
  exponent over $r$. Symbols are $\delta_{99}^+$: $\circ$, 1100;
  $\times$, 1300; $\bigtriangledown$, 1500; $\bigtriangleup$, 1700;
  $\star$, 1900.
The horizontal bar is the variation of $\omega^*/\omega^+$ in our range of $\delta_{99}^+$.
}
\end{figure}

The local exponent \eqref{eq:fexponent} of our `cleaned' interface is
presented in figure \ref{fig:fdimension} as a function of $\omega_0^*$
for several Reynolds numbers. Figure \ref{fig:fdimension}(a) plots
$D_b$ for the smallest possible value of $r$, and tries to approximate
\eqref{eq:fdimension}. Figure \ref{fig:fdimension}(b) plots the
maximum value $D_b$ over the whole range of $r$. The differences
between the two figures quantify how far from a monofractal is the
T/NT interface. Note the good collapse of the different Reynolds
numbers when parametrised with $\omega_0^*$. The black horizontal bar
near the peaks of both figures is the variation of $\omega^*/\omega^+$
in our range of Reynolds number. A similar bar is included in all
later figures that make a Reynolds number comparison, and measures how
much the collapse of the different curves would deteriorate if the
data had been normalised with $\omega_0^+$ instead of with
$\omega_0^*$.

The dependence of $D_b$ on $\omega_0^*$ confirms the visual impression
from figure \ref{fig:vorticity_isocontour} that the threshold has a
dramatic effect on the interface. At low thresholds, the dimension
approaches the smooth limit $D=2$ but, at higher ones, the T/NT
interface is significantly more
convoluted. \citet{Sreenivasan09011989} predicted $D=7/3$ for the T/NT
interface, precisely the value observed by \citet{de2013multiscale} in
a more recent experiment, which is within the range of the present
results. It would correspond to $\omega_0^*\simeq 1$ in figure
\ref{fig:fdimension}(a), and to the lowest possible dimension in
figure \ref{fig:fdimension}(b).

Regardless of the differences in their absolute values, the two
estimations of the fractal dimension in figure \ref{fig:fdimension}
behave similarly with respect to $\omega^*_0$. There is a transition
between $\omega_0^*\simeq 0.2$ and $\omega_0^*\simeq 2$, across which
the geometrical complexity of the interface increases
significantly. This suggests that across this range the isosurface
moves inside the turbulent core, where it reflects the geometrical
features of the turbulent vorticity itself. It also follows from
figure \ref{fig:mean_std} that $\omega_0^* \simeq 0.2$ corresponds to
the threshold for which the average location of the interface $\bra
y_I\ket$ decreases fastest as the threshold increases.

The decrease of the dimension beyond $\omega_0^* \simeq 2$ was already
observed by \citet{MoisyJimenez}, who used thresholds equivalent to
$\omega^*_0\simeq 2-12$ to study the geometry of the \emph{volume} of
the vorticity in isotropic turbulence. There is no simple relation
between the fractal dimensions of a set and of its surface, but
\citet{MoisyJimenez} noted that in the limit of very high thresholds
the vorticity would be reduced to a discrete cloud of points for which
$D\simeq 0$. A similar argument can be applied to the interface.

\subsection{\label{sec:genus}Genus}

The geometric complexity of an object can also be characterized by its
topological properties. The genus $g$ is a topological invariant of
any connected orientable surface, and measures the number of its
`handles' (figure \ref{fig:features}a). A sphere has genus zero, a
torus has genus one, and two connected tori have genus two.  To our
knowledge, the genus was first used to characterize turbulent
structures in homogeneous turbulence by \citet{Genus}, who cite
instances of its earlier use in disciplines such as astrophysics. In
most of those cases, the genus is obtained by integrating the mean and
gaussian curvatures over the interface, which requires a careful
triangulation of the surface. This step is time consuming and prone to
errors, and we bypass it by computing the genus directly from the
Euler characteristic $\chi$ of the numerically defined contour. The
algorithm is described in \cite{GenusTOMS}, and is optimised to
exploit discrete data in structured grids.

Briefly, any numerical isosurface in a cartesian grid is a polyhedron of stacked
parallelepipeds. If $V$ is the number of vertices, $E$ the number of edges, and $F$
the number of faces, its Euler characteristic is given by the Euler--Poincar\'e
formula,
\begin{equation}
  \label{eq:euler_formula}
  \chi = V-E+F,
\end{equation}
and the genus is
\begin{equation}
  \label{eq:genus}
  g = 1-\frac{\chi}{2}.
\end{equation}
The genus is a measure of complexity, like the fractal dimension, but
the two are not equivalent. A wrinkled piece of paper has genus zero,
independently of the amount of wrinkling. A regular Brownian surface
is defined as a fractal single-valued map on the real plane. Its
fractal dimension is $D=2.5$, but it has no handles \citep{RussBook}.

\begin{figure}
\centerline{%
\includegraphics[width=0.48\textwidth]{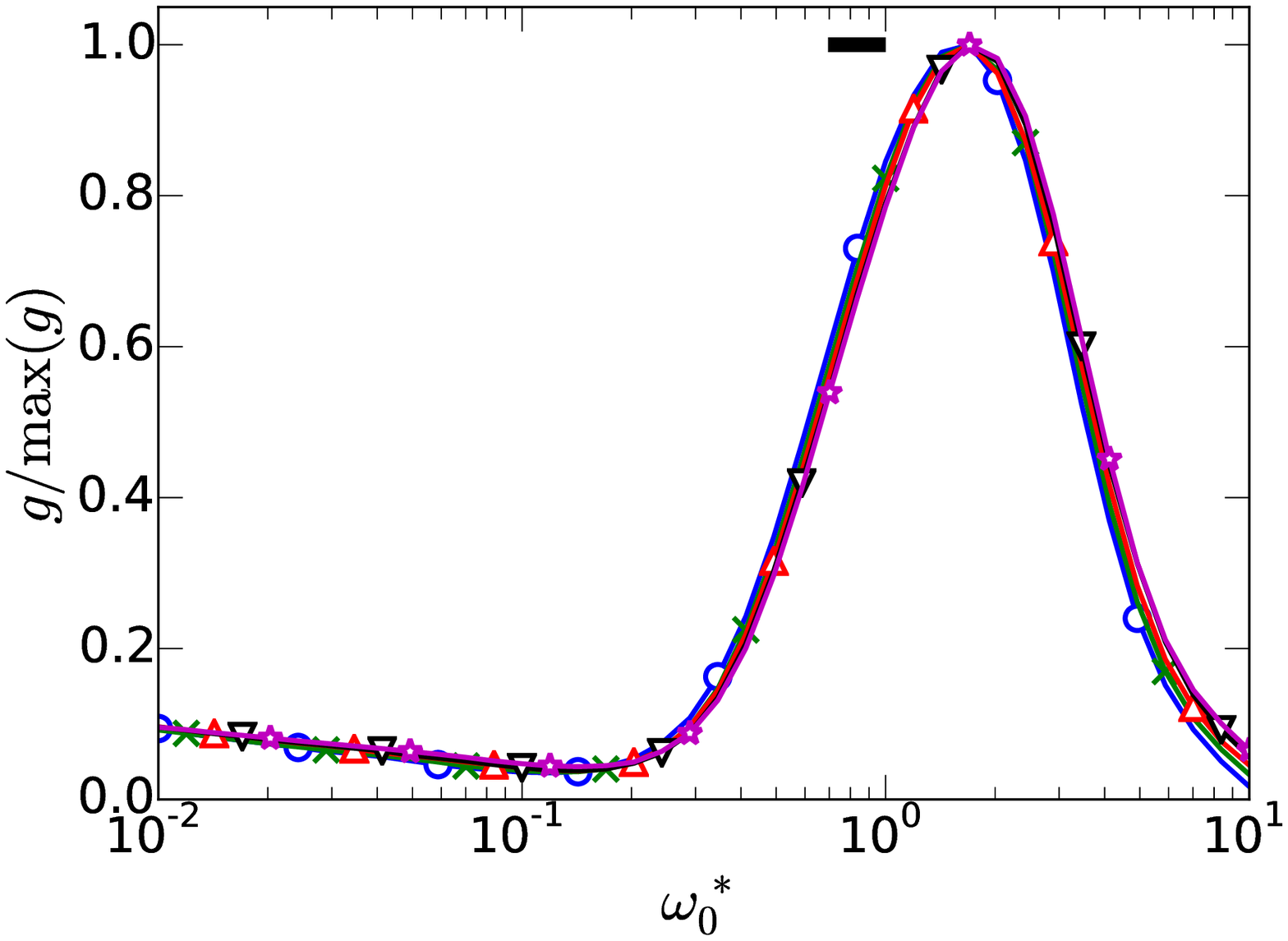}
\mylab{-0.40\textwidth}{0.29\textwidth}{(a)}%
\hspace{2mm}%
\includegraphics[width=0.48\textwidth]{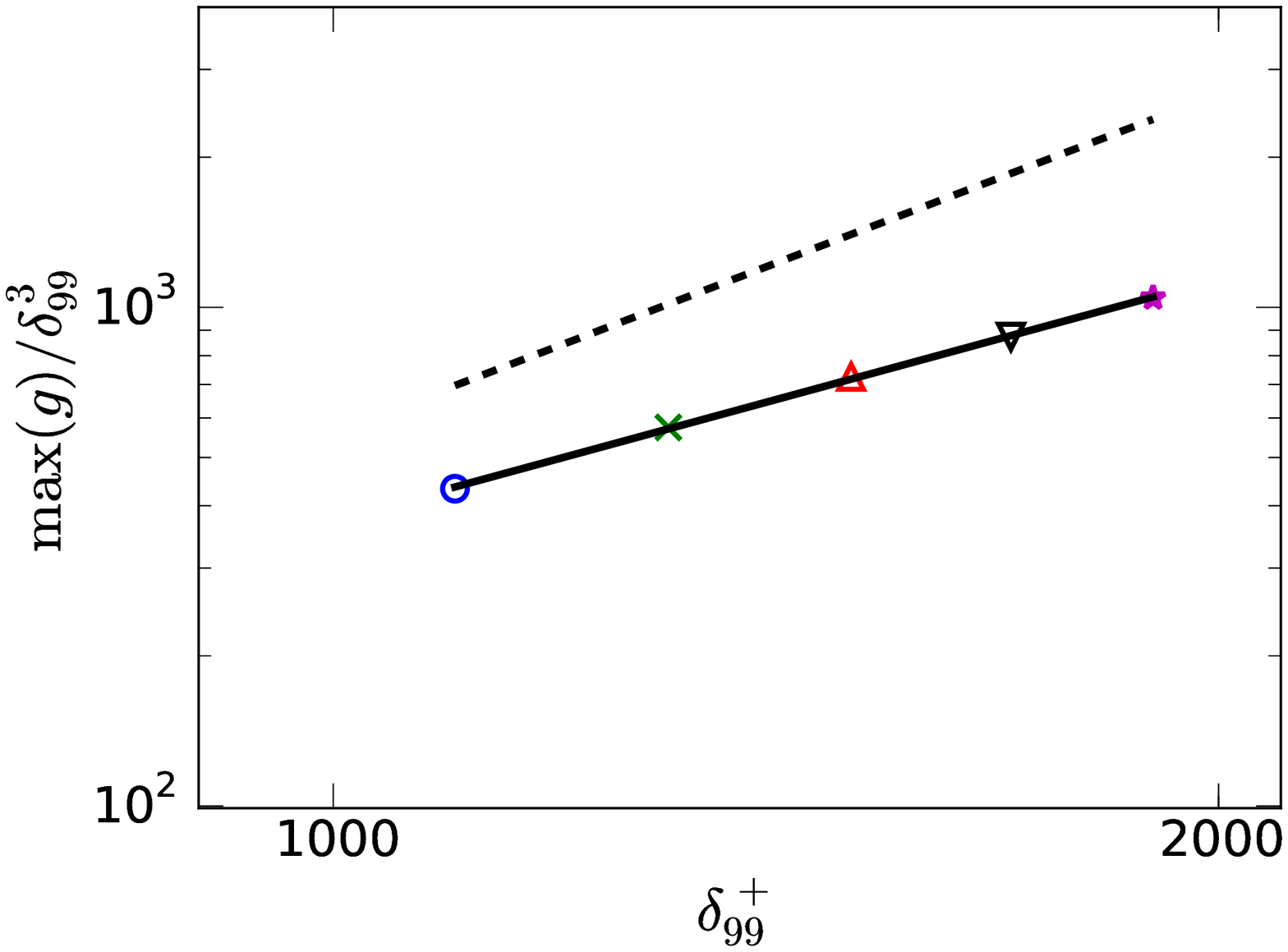}%
\mylab{-0.38\textwidth}{0.29\textwidth}{(b)}%
}
\caption{(a) Genus normalized with its maximum over $\omega_0$. The
  horizontal bar is the variation of $\omega^*/\omega^+$ in our range
  of $\delta_{99}^+$. The horizontal bar is the ratio of
  $\omega^*/\omega^+$ over the present range of $\delta_{99}^+$
(b) Maximum genus per cubic boundary-layer thickness, occurring in all
  cases at $\omega_0^* \simeq 2$. Both axes are logarithmic. The solid
  line is a power law fit, $\max(g)/{\delta_{99}^+}^3\propto
  {\delta_{99}^+}^{1.6}$. The dashed one is the Kolmogorov limit,
  $\max(g)/{\delta_{99}^+}^3\propto {\delta_{99}^+}^{9/4}$.
Symbols are $\delta_{99}^+$, as in figure \ref{fig:fdimension}.  }
\label{fig:genus}
\end{figure}

As in the previous section, we compute the genus for the largest
connected component of the vorticity isosurface, which is shown in
figure \ref{fig:genus}(a) normalized by its maximum over
$\omega_0$. There is a topological transition in which handles begin
to appear over roughly the same range, $\omega^*_0 \simeq (0.2-2)$, as
the growth of the fractal dimension. Around $\omega^*_0 \simeq 1$,
handles are the dominant feature of the surface, and there are
hundreds or thousands of them in a volume $O(\delta_{99}^3)$. We
suggested in the discussion of the fractal dimension that the T/NT
interface at these high thresholds is basically a reflection of the
internal geometry of the turbulent vorticity, and the reasons for the
decrease of the dimension beyond the end of the transition also apply
here. Some turbulent features disappear for very large thresholds,
causing the genus to decrease. The maximum genus occurs at the end of
the topological transition $\omega^*_0 \simeq 2$, and figure
\ref{fig:genus}(b) shows that it increases with the Reynolds number as
$\max(g)/{\delta_{99}^+}^3\propto {\delta_{99}^+}^{1.6}$. This
exponent is somewhat lower than for the number of Kolmogorov-size
structures per cubic integral scale ${\delta_{99}^+}^{9/4}$, which
sets an upper bound for the scaling of the possible complexity. It is
tantalisingly close to the corresponding number of $\lambda$-sized
structures, ${\delta_{99}^+}^{3/2}$. Note again the good collapse
provided by $\omega_0^*$ for the Reynolds number dependence of the
genus.

This proliferation of handles will become important for the
conditional analysis of the flow in the next section. When the
analysis of a surface with handles is carried out using a
lower-dimensional section, such as a two-dimensional plane or a line,
the results can be subject to interpretation artefacts. For example,
the planar section of a torus across its principal axis is two
circles, giving the impression of two disconnected geometrical
objects. Up to a point, the same is true for pockets such as those in
figure \ref{fig:features}(b). For example, the interface shown below
in figure \ref{fig:distancefield}(b) is a section of a
singly-connected isosurface, although it appears to contain many
unconnected irrotational bubbles within the turbulent region.  Another
effect of the handles has to do with values conditioned to the
direction normal to the interface. The usual assumption in this case
is that a normal defined from high towards low values of the vorticity
points into the the free stream. In a handle, or in a narrow pocket,
this is only true over distances of the order of the feature
thickness, and becomes an issue if handles and pockets are
dominant. The problem is less pressing when the threshold is chosen
below $\omega^*_0 \simeq 0.2$, where the T/NT interface is smoother,
but figure \ref{fig:mean_std} shows that a lot of the published work
uses thresholds within the topological transition, characterised by
non-trivial fractal dimensions and, presumably, large genera.

The main conclusion from this section is that the properties of the
fully turbulent flow appear gradually in the geometry of the interface
as the threshold traverses the topological transition, and that the
handles, folds, and high fractal dimensions are probably the
reflection of the internal structure of the flow.

\section{\label{sec:conditional} Conditional analysis of the vorticity field.}

In this section we study the properties of the vorticity field as a
function of the distance to the T/NT interface. Given the geometrical
complexity of the interface, it is to be expected that different
definitions of distance produce different conditional results. To
allow us to differentiate between genuine flow properties and possible
measurement artefacts, we will pay especial attention to the cases in
which the results of two alternative distance definitions are not
equivalent.

\begin{figure}
\centering
\subfloat[]{
  \includegraphics[width=0.48\textwidth]{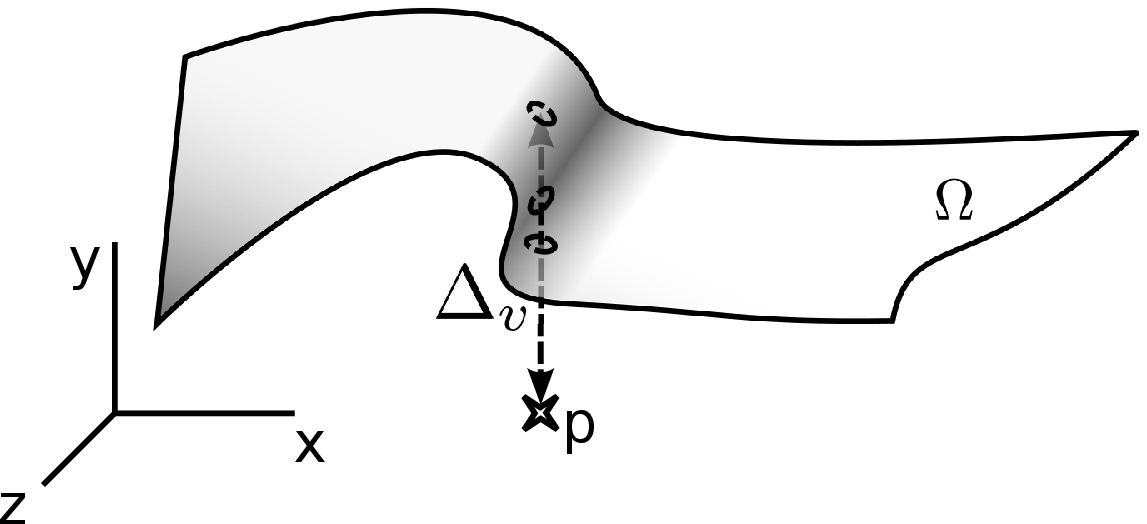}
}
\subfloat[]{
  \includegraphics[width=0.48\textwidth]{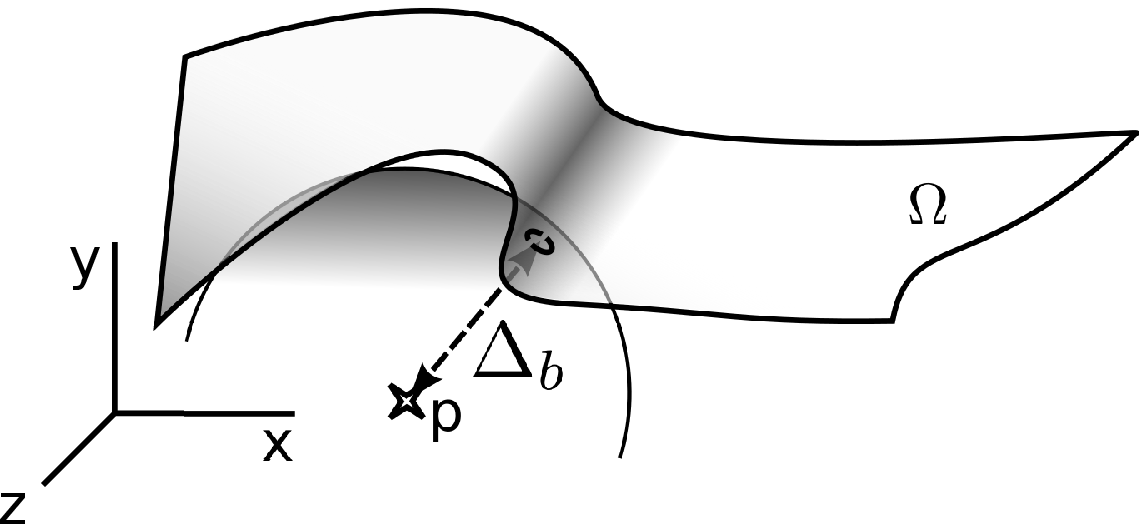}
}
\caption{(a) Sketch of the vertical distance $\Delta_v$, and (b) the
  ball distance $\Delta_b$, between a point $p$ and a surface
  $\Surf$. In the case of $\Delta_v$, the wall-normal line may
  intersect $\Surf$ multiple times, but only the top intersection is
  kept.  Here, the surface has a pocket and the line crosses it three
  times. In the case of the ball distance, there is usually only one
  point where the sphere centred at $p$ with radius $\Delta_b$ is
  tangent to $\Surf$, marked here with a small circle.  }
\label{fig:distance_sketch}
\end{figure}

Consider first the vertical distance $\Delta_v$. Given a surface
$\Surf$, $\Delta_v$ is the distance between a point $p$ and the
topmost intersection with $\Surf$ of a line normal to the wall going
through $p$. A sketch is given in figure \ref{fig:distance_sketch}(a),
emphasizing that even if the line used to measure distance crosses the
interface multiple times, only the highest intersection is used.  Note
that discarding the lower intersections hides part of the complexity
of the interface, and that most handles and pockets are not captured.
This criterion has been used to study the T/NT interface in boundary
layers by \citet{Chauhan:14} using normals to the wall, and in jets by
\citet{West:etal:09} and \citet{SilvaTaveira} using normals to the
symmetry plane.

Our second definition of distance is the separation between the point
$p$ and its closest point in $\Surf$. We will call it the ball (or
minimum) distance $\Delta_b$, and has a simple geometrical
interpretation as the radius of the sphere tangent to the interface
and centred at $p$. It is sketched in figure
\ref{fig:distance_sketch}(b). Some properties of this distance are
particularly convenient for a conditional analysis. Regardless of the
complexity of the surface, there is always a closest surface point to
any point in space, and the ball distance is always uniquely
defined. If the point $p$ is infinitesimally close to the interface,
$\Delta_b$ is equivalent to the distance measured along the local
normal. It also has a relatively simple mathematical formulation,
since it satisfies the Eikonal equation $|\nabla (\Delta_b)| = 1$ with
$\Delta_b=0$ at the interface. This equation has a solution regardless
of the complexity of the boundary condition, and can be integrated by
several fast methods \citep{jones20063d}.

\begin{figure}
\centering
\subfloat[]{
  \includegraphics[width=0.48\textwidth]{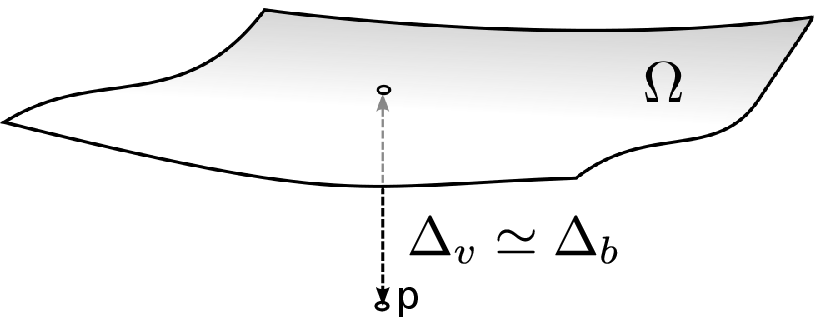}
}
\subfloat[]{
  \includegraphics[width=0.48\textwidth]{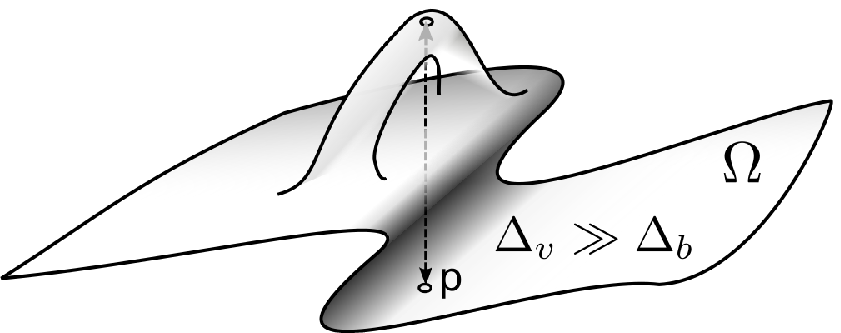}
}
\caption{(a) Sketch of simple almost horizontal surface for which
  $\Delta_v\simeq \Delta_b$.
(b) Example of a case in which both distances are very
  different. Here, point $p$ is very close to the interface and
  $\Delta_b\simeq 0$, but lies underneath a pocket and a handle, and
  $\Delta_v\gg \Delta_b$ }
\label{fig:distance_behavior}
\end{figure}

The relation between the two distance definitions depends on the local
orientation and complexity of the surface. Referring to figure
\ref{fig:distance_behavior}(a), when the T/NT interface is mostly
horizontal, simple and smooth, the two definitions produce similar
results. When the interface is more complex or not parallel to the
wall, as in figure \ref{fig:distance_behavior}(b), the two results are
different. For example, point $p$ in figure
\ref{fig:distance_behavior}(b) is very close to the interface in terms
of $\Delta_b$, but relatively deep into the turbulent side in terms of
$\Delta_v$.

Other authors have introduced alternative definitions of conditional
distance.  \citet{watanabe2015turbulent} and \citet{da2008invariants}
use the local normal to the interface, obtained respectively in two
and three dimensions. This is similar to the ball distance close to
$\Delta=0$, particularly in the three-dimensional case, but farther
away normals may intersect each other, and the two definitions are not
comparable. There have been efforts to study the conditional
properties of the interface using lagrangian trackers in jets
\citep{FLM:1757608, taveira2013lagrangian}, but the trajectories soon
get complicated away from the interface, and
\citet{atkinson2014numerical} showed that tracking them close to the
edge of the boundary layer is significantly more difficult than in
jets, where the free stream velocity is very low.

\begin{figure}
\centering
\subfloat[]{
  \includegraphics[width=0.35\textwidth]{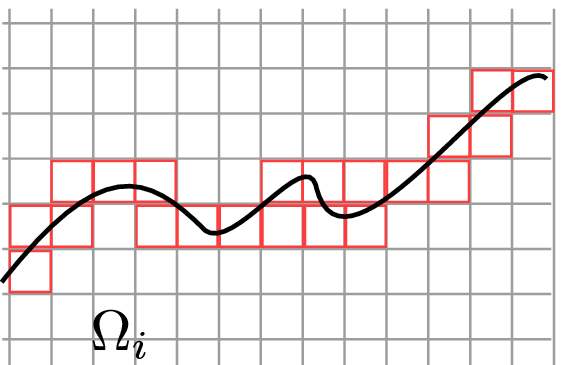}
}
\subfloat[]{
  \includegraphics[width=0.35\textwidth]{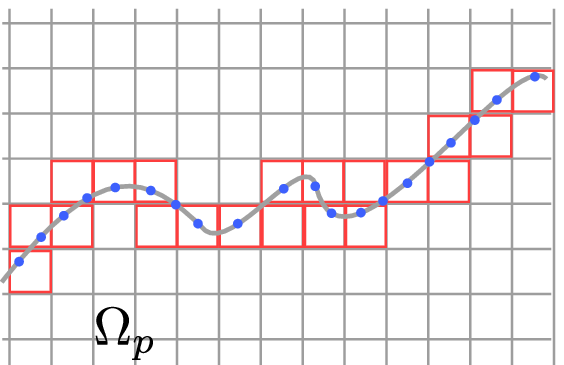}
}
\caption{(a) Set  $\Surf_i$ of voxels that contain the interface.
(b) Set $\Surf_p$ of points used to approximate the T/NT interface.  }
\label{fig:voxeltopoint}
\end{figure}

Our algorithm to obtain the ball distance starts from the set
$\Surf_i$ of interface voxels defined in
\eqref{eq:interface_algorithm}. The vorticity within each voxel is
approximated by a trilinear interpolation of the values at the
vertices, so that the T/NT interface is approximated by a polyhedron
of which each interface voxel contains a planar face. The interface is
approximated by the set $\Surf_p$ of the points which are closest to
the centre of the voxel in each of those faces. The sets $\Surf_i$ and
$\Surf_p$ are illustrated in figure \ref{fig:voxeltopoint}.

The ball distance between $p$ and $\Surf$ is approximated by the
distance between $p$ and its nearest neighbour in $\Surf_p$. The
nearest-neighbour search (NNS) is a common problem in optimization. If
$N_p$ is number of elements in the set $\Surf_p$, a fast solution
requiring $O(\log N_p)$ computations was found by \cite{Arya:1998}.
Most data analysis packages and toolkits provide implementations of
some variant of NNS, and free libraries are available
\citep{muja2014scalable}. By convention, the distance to the interface
is defined as positive or negative depending on whether the point $p$
has been classified as being in the turbulent or in the non-turbulent
region. Note that, because the distance is computed with respect to
the cleaned interface defined in \S\ref{sec:drops}, turbulent and
non-turbulent points refer to the smoothed flow regions. Bubbles are
counted as turbulent, and drops as non-turbulent.

In our analysis, the ball distance is treated as a field and computed
for all the collocation points in the computational grid. Assuming a
total number $N$ of grid points, obtaining the field of ball distances
requires $O(N \log N_p)$ operations.  For our data, $N_p$ is of the
order of $10^8$, and $N$ is of the order of $10^{9}$ for each
snapshot. The same procedure is followed for the vertical distance
$\Delta_v$, with a computational cost $O(N)$.

\subsection{The signed distance field}\la{sec:disfield}

\begin{figure}
\centering
\subfloat[]{
  \includegraphics[width=0.48\textwidth, clip=true, trim=0em 5em 0em
  11em]{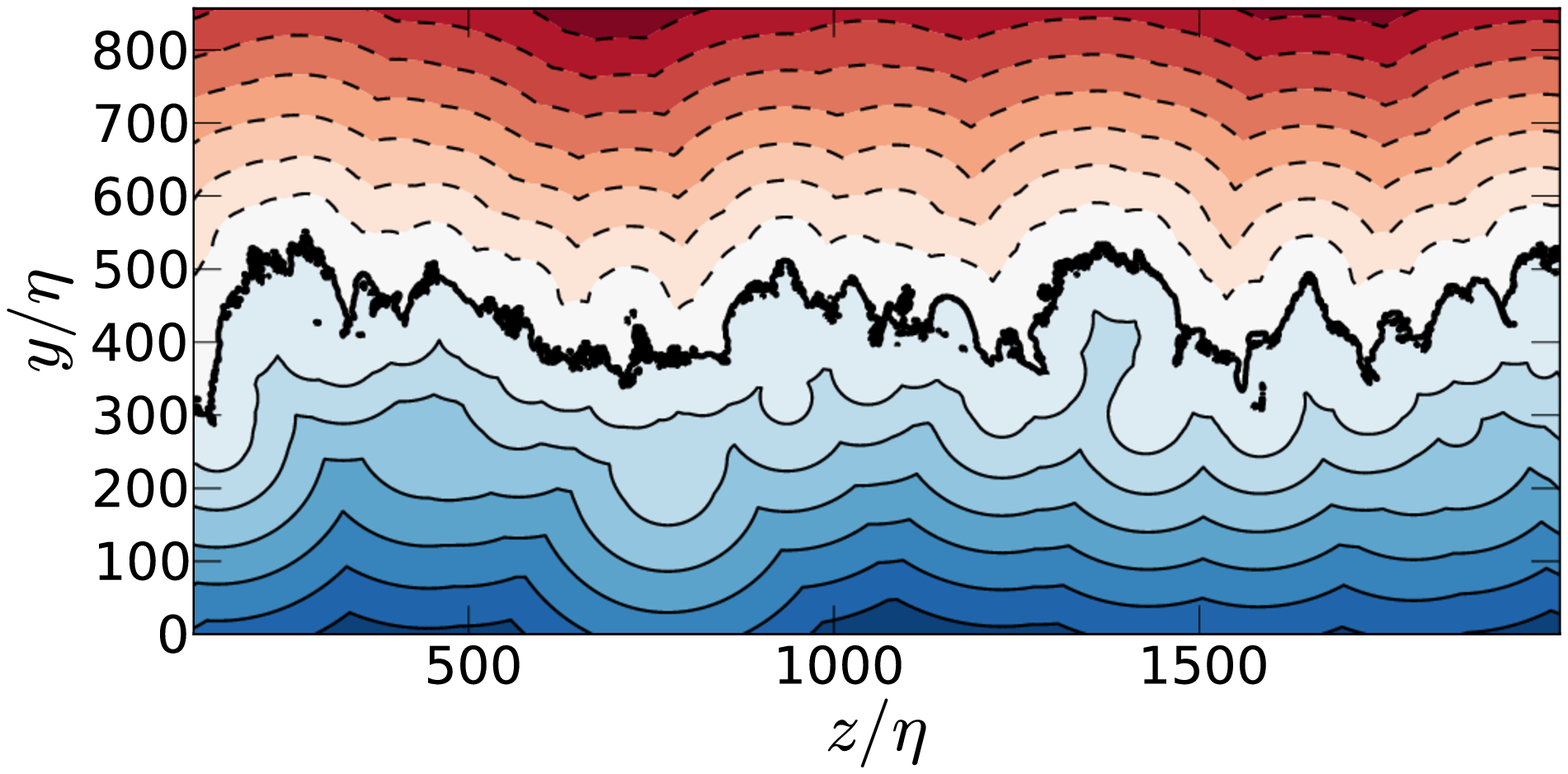}
}
\subfloat[]{
  \includegraphics[width=0.48\textwidth, clip=true, trim=0em 5em 0em
  11em]{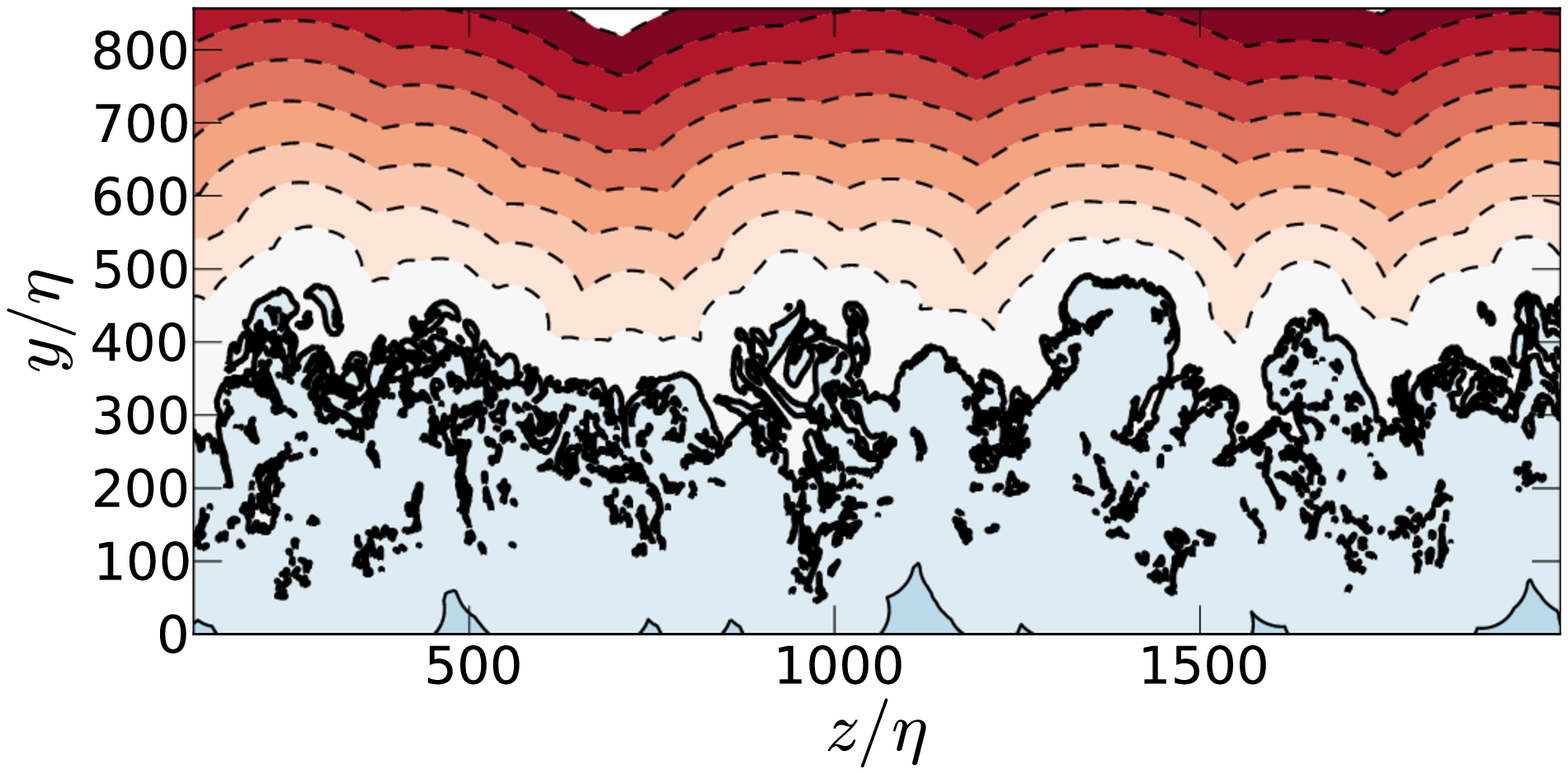}
}

\subfloat[]{
  \includegraphics[width=0.48\textwidth, clip=true, trim=0em 0em 0em
  0em]{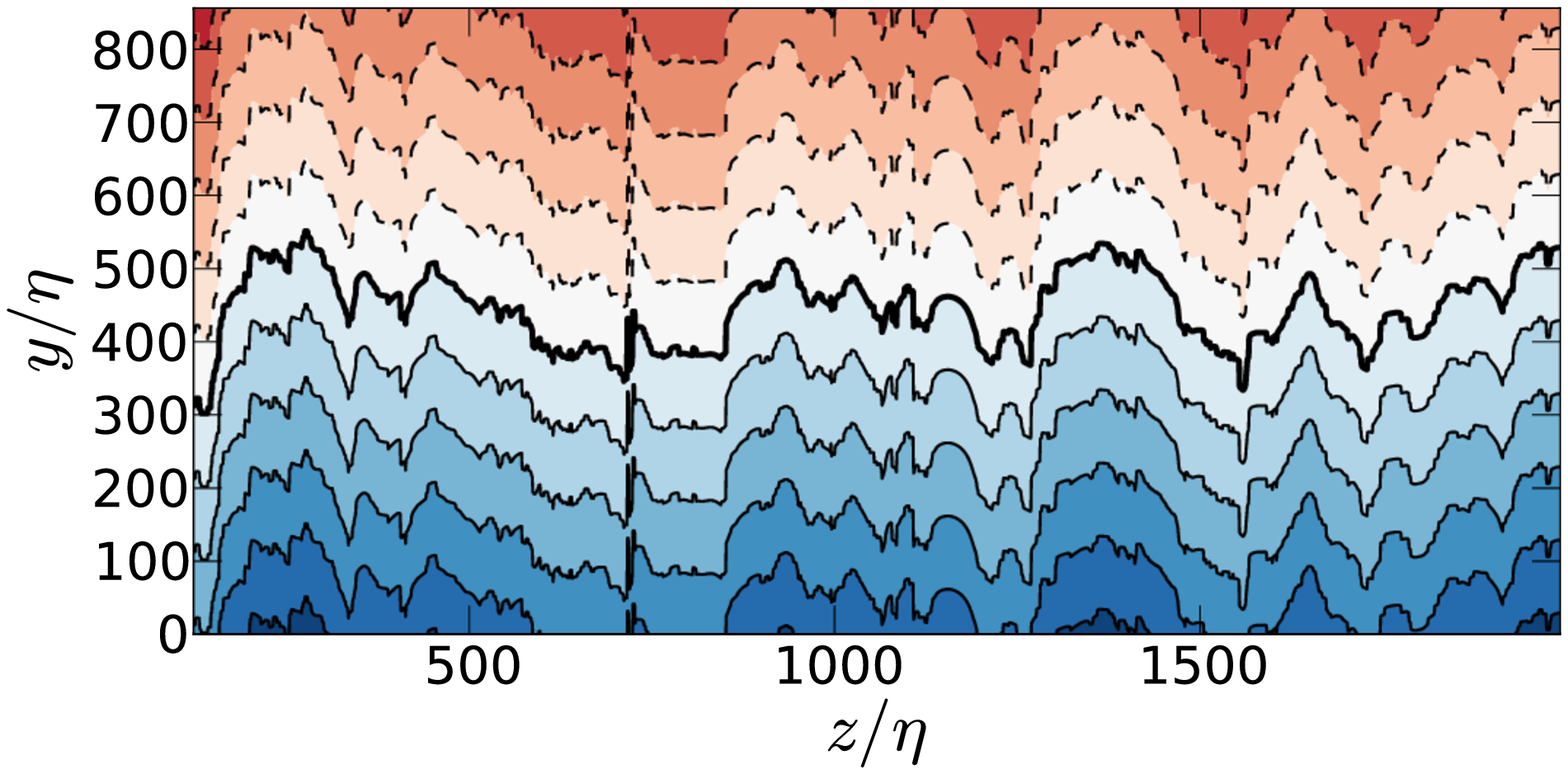}
}
\subfloat[]{
  \includegraphics[width=0.48\textwidth, clip=true, trim=0em 0em 0em
  0em]{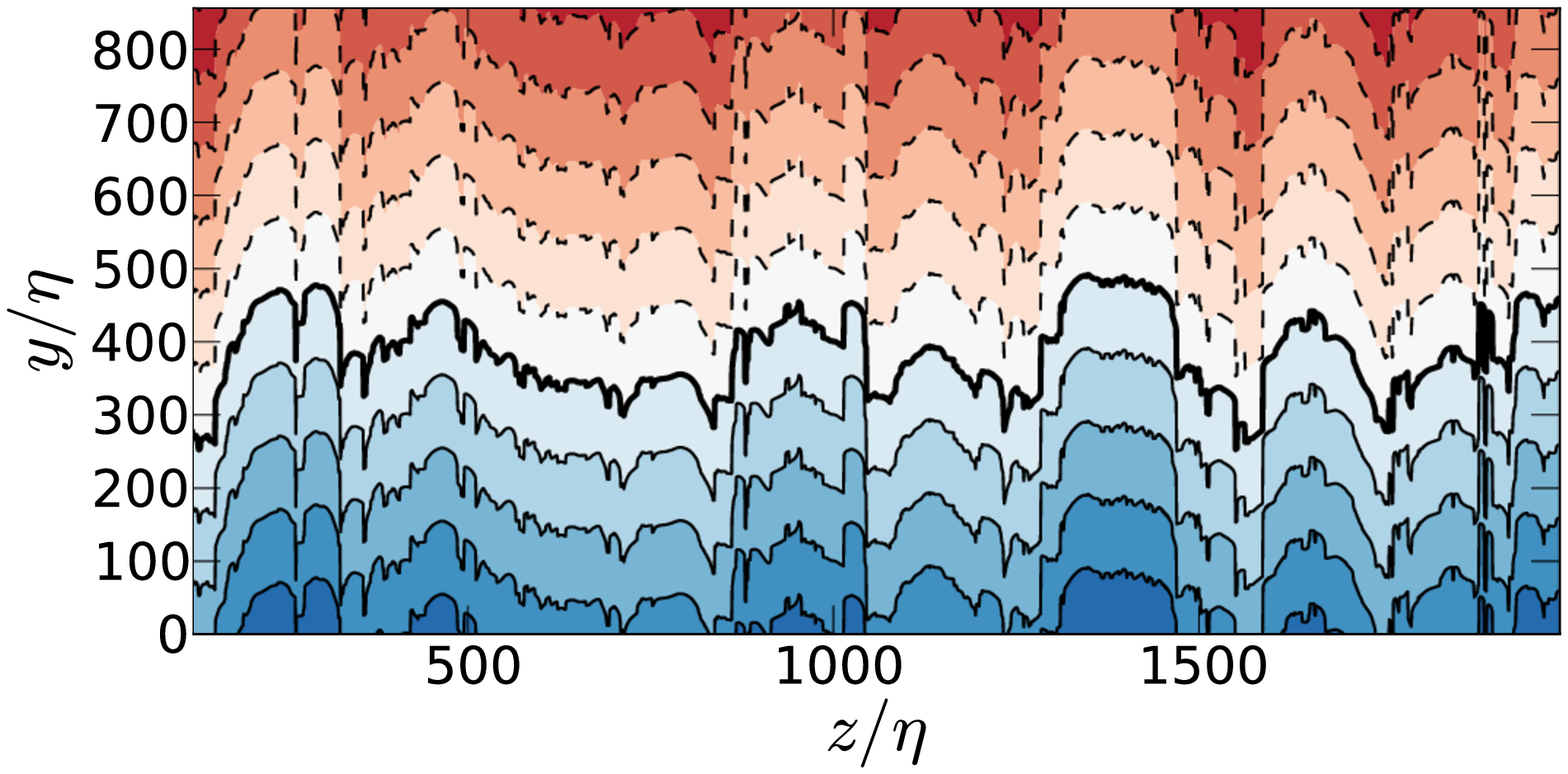}
}
\caption{Cross-stream sections of the signed distance fields for the
  lowest and highest available thresholds, and for the two definitions
  of distance. (a) $\Delta_b$ and $\omega_0^* = 0.01$; (b) $\Delta_b$
  and $\omega_0^* = 0.5$; (c) $\Delta_v$ and $\omega_0^* = 0.01$; (d)
  $\Delta_v$ and $\omega_0^* = 0.5$. All sections correspond to the
  same flow field at $\delta_{99}^+=1500$. The thicker solid line
  represents the T/NT interface for each distance definition, and
  always corresponds to a single connected surface. The isolated spots
  are due to three-dimensional contortions. Other contour levels are
  separated by $50\eta$ for $\Delta_b$, and by $100\eta$ for the
  $\Delta_v$. Negative contours are dashed.  }
\label{fig:distancefield}
\end{figure}

The discrete fields obtained with the minimum and vertical distance are called the
ball-distance field $\Delta_b(x,y,z)$, and the vertical-distance field
$\Delta_v(x,y,z)$, respectively. The symbol $\Delta$ denotes distance regardless of
the particular definition.

The concept of a distance field is also found in \citet{FLM:5533652}, who use the length
of the trajectories along lines of maximum gradient of an advected scalar to measure the
distance with respect to the interface. While their definition can also be used regardless
of the complexity of the surface, the gradient lines of the vorticity magnitude become
very contorted away from the interface in the turbulent side, and less suitable for
conditional analysis than any of the two definitions mentioned above.

Sections of the two distance fields of the same snapshot of the flow are shown in
figure \ref{fig:distancefield}, each one computed for two different thresholds. The
isosurface $\Delta=0$ is our effective definition of the interface, but note that
the two distance definitions generate different surfaces. The first observation is
that the two distances give fairly different results in the turbulent side,
particularly for the higher vorticity thresholds. In the non-turbulent side, where
the interface is more convex, the differences are not as important. When the
threshold is within the topological transition, such as $\omega_0^*=0.5$ in figures
\ref{fig:distancefield}(b,d), the contortions of the `ball' interface are so intense
that there are very few points in the turbulent side for which $\Delta_b>100\eta$. We
emphasize that $\Delta_b=0$ in figure \ref{fig:distancefield}(b) corresponds to a
single connected surface from which bubbles have been removed, and that the
apparently isolated contours within the turbulent side are artefacts of the
two-dimensional section. Comparison of the results of the two thresholds for each
distance definition shows that the vertical distance field in figures
\ref{fig:distancefield}(c,d) is less sensitive to the contortions than the ball
distance in figures \ref{fig:distancefield}(a,b), and also less sensitive to the
choice of the threshold. Because of this, it misses most of the interface complexity
and the existence of the topological transition.

Away from the wall, the two distance definitions also behave differently. Because of
its connection with the Eikonal, the ball distance can be visualised as a wavefront
moving away from the interface at a uniform velocity. As is does, the interface
irregularities are eliminated by successive mergings of caustics, and the $\Delta_b$
isosurface becomes smoother. Roughly speaking, a $\Delta_b$ isosurface only retains
wavelengths larger than $O(|\Delta_b|)$. The vertical distance does not share this
smoothing property. Because the $\Delta_v$ isosurfaces are vertical translations of
the interface, they retain its irregularities at all distances. Note also that neither
distance is additive. Because a $\Delta\ne 0$ isosurface is not an iso-vorticity
surface, even if the $\Delta=0$ surface is defined as one, it is impossible to define a
distance between interfaces with different thresholds such that
$\Delta(p\rightarrow\omega_0) =\Delta(p\rightarrow\omega_1) +
\Delta(\omega_1\rightarrow\omega_0)$. This will later lead to ambiguities in the
definition of the thickness of the T/NT interface layer.

The angle $\theta$ between the normal to the $\Delta_b$ interface and the vertical
can be estimated by $\left. \dr\Delta_v/\dr\Delta_b\right|_{\Delta_b=0} =
1/\cos\theta$, but there is no simple way to evaluate the orientation of the
$\Delta_v$ interface in this manner. 

\begin{figure}
\centerline{%
\includegraphics[width=0.48\textwidth]{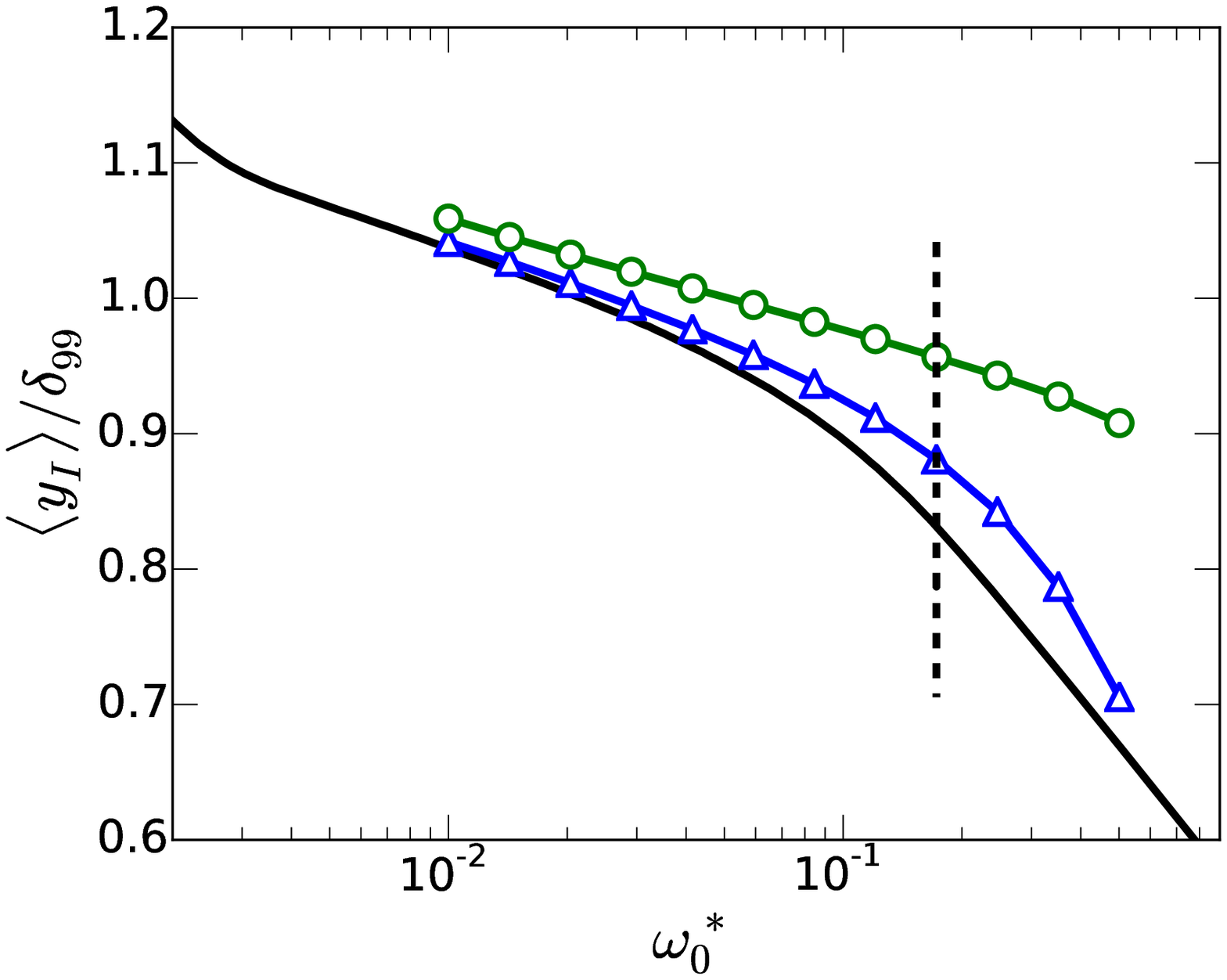}%
\mylab{-0.15\textwidth}{0.29\textwidth}{(a)}%
\hspace{1mm}%
\includegraphics[width=0.48\textwidth]{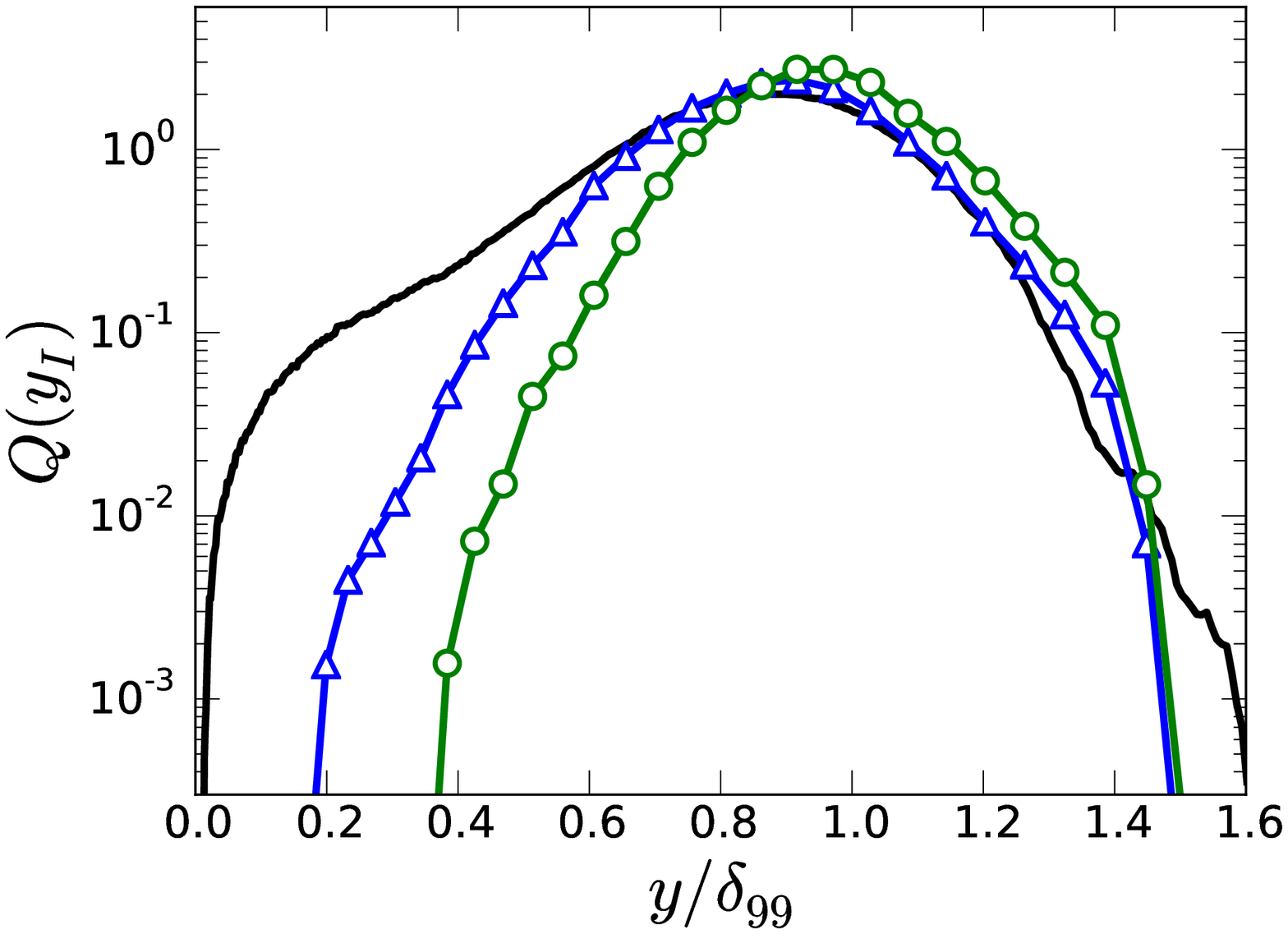}%
\mylab{-0.37\textwidth}{0.29\textwidth}{(b)}%
}
\caption{(a) Mean position of the T/NT interface as a function of the vorticity
threshold. No symbols, vorticity isosurface as in figure \ref{fig:mean_std};
\trian, interface defined as $\Delta_b=0$; \circle, $\Delta_v=0$.
The threshold in (b) is the vertical dashed line.
(b) PDFs of the vertical position of the three isosurfaces for $\omega_0^*=0.19$.
Lines as in (a). $\delta^+=1500$.
}
\label{fig:intdist}
\end{figure}

None of the interfaces defined by the above distance criteria exactly
coincides with a vorticity isosurface. In the case of $\Delta_b$ the
only difference is the absence of the bubbles and drops discarded in
the smoothing step, and the deviations are relatively minor.  The
vertical distance misses substantial parts of the isosurface, and may
deviate a lot from it. Figure \ref{fig:intdist}(a) shows the mean
position of the two interfaces as a function of $\omega_0^*$, compared
with the mean position of the vorticity isosurface. The mean $\bra y_I
(\Delta_b) \ket$ deviates little from the position $\bra y_I \ket$ of
the vorticity isosurface (figure \ref{fig:mean_std}), but $\bra y_I
(\Delta_v) \ket$ remains close to the edge of the boundary layer even
when the vorticity isosurface moves closer to the wall. This is
confirmed by the PDFs of the height of the three isosurfaces, given in
figure \ref{fig:intdist}(b). For low thresholds (not shown), the PDFs
of the two interfaces and of the vorticity isosurface roughly
coincide, and are approximately gaussian \citep{NACA:1244}.  But for
the higher threshold in figure \ref{fig:intdist}(b), $y_I (\Delta_b)$
follows the isosurface into the turbulent core of the boundary layer
substantially better than $y_I (\Delta_v)$. As a consequence, $y_I
(\Delta_b)$ results in a much better representation of the
intermittency parameters of the boundary layer, such as $\gamma$. Note
that the vorticity threshold used in figure \ref{fig:intdist}(b),
$\omega_0^*=0.19$, although relatively high, is below the topological
transition, and in the range of most of the studies collected in table
\ref{table:gamma}.

\subsection{Conditional analysis of distance and vorticity.}\la{sec:distandvor}

\begin{figure}
\centerline{%
\includegraphics[width=0.48\textwidth, clip=true,trim=0em 1em 1em 4em]{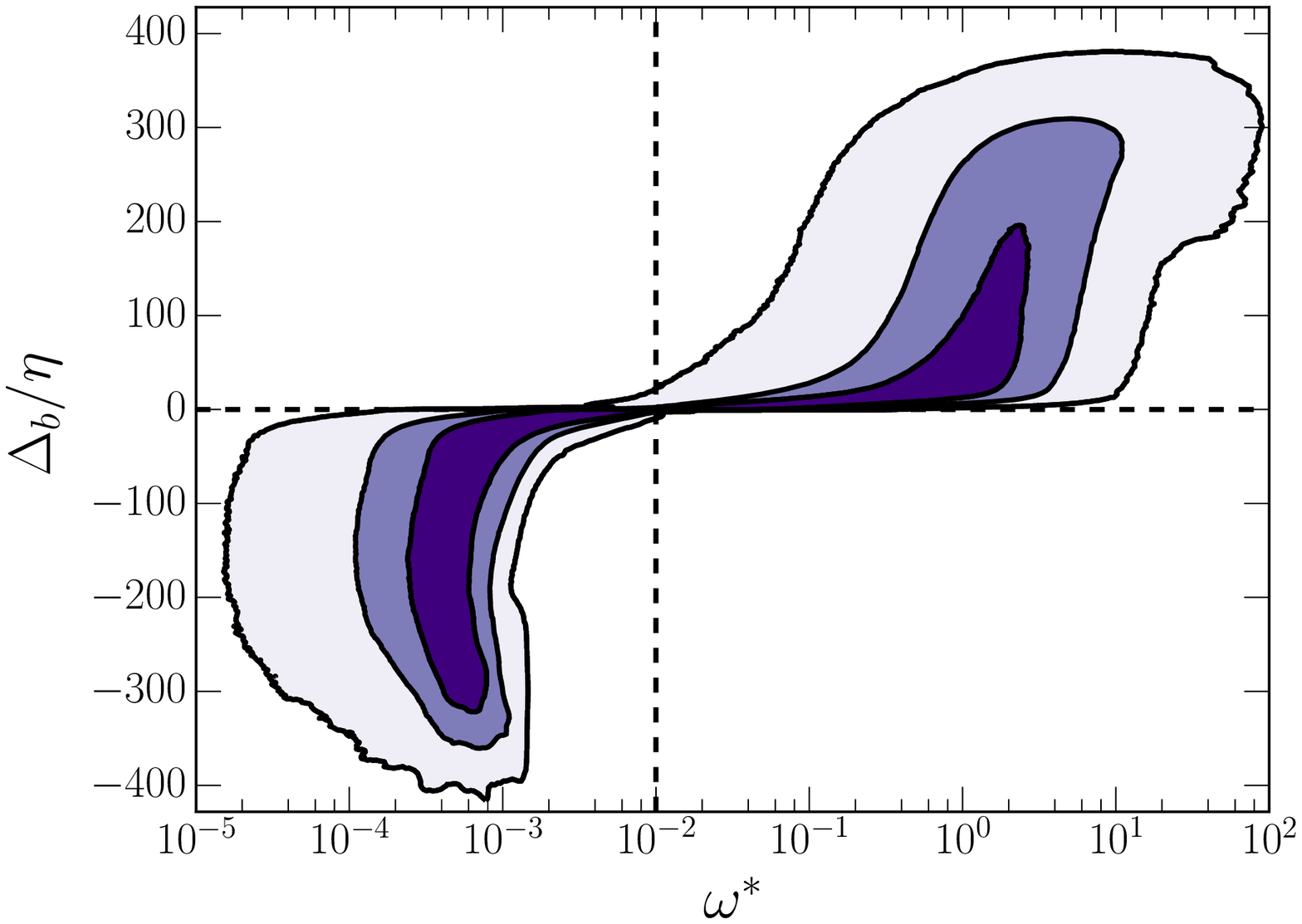}%
\mylab{-0.37\textwidth}{0.29\textwidth}{(a)}%
\hspace{1mm}%
\includegraphics[width=0.48\textwidth, clip=true, trim=0em 1em 1em 4em]{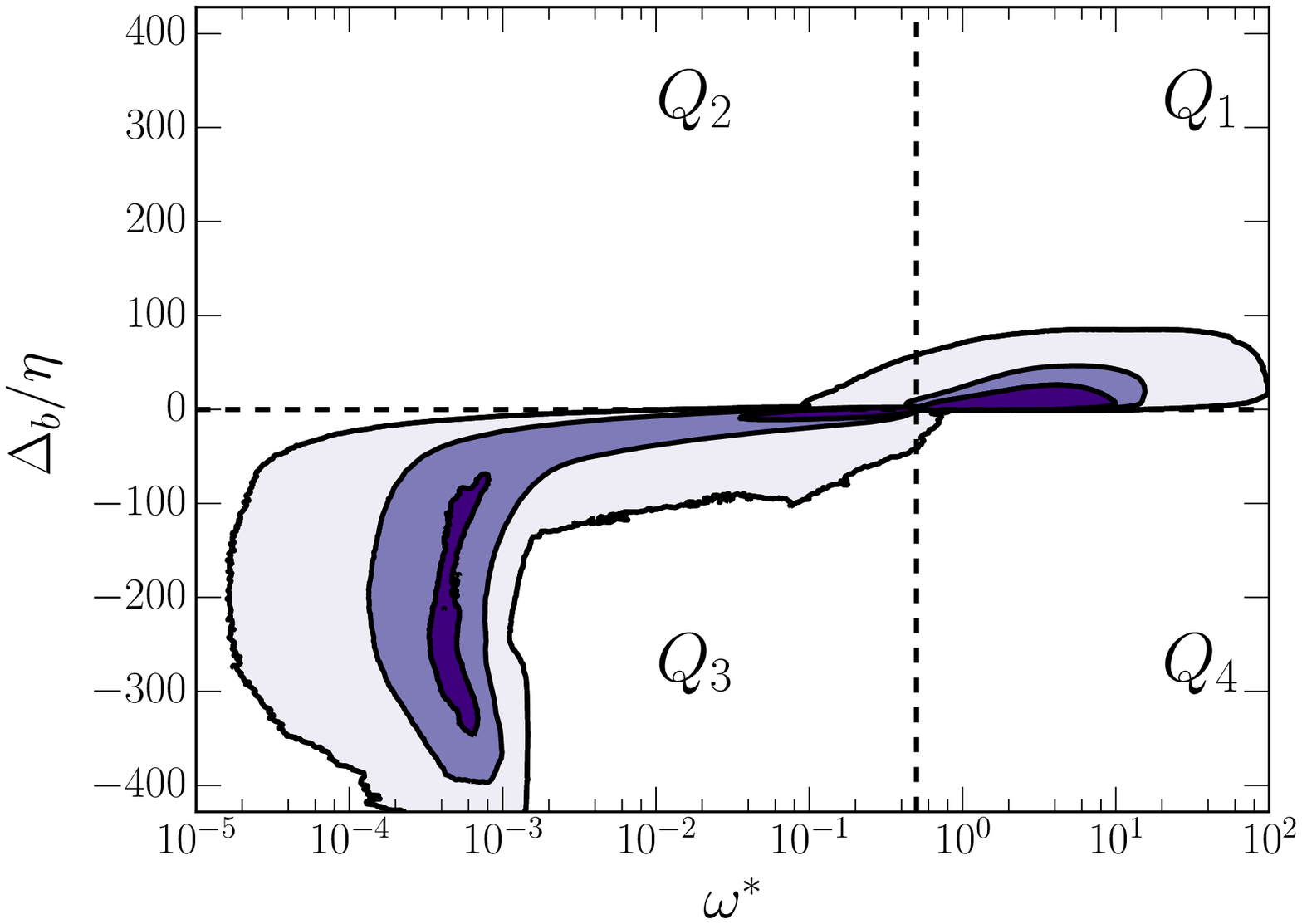}%
\mylab{-0.37\textwidth}{0.29\textwidth}{(b)}%
}
\vspace{1ex}%
\centerline{%
\includegraphics[width=0.48\textwidth, clip=true,trim=0em 1em 1em 4em]{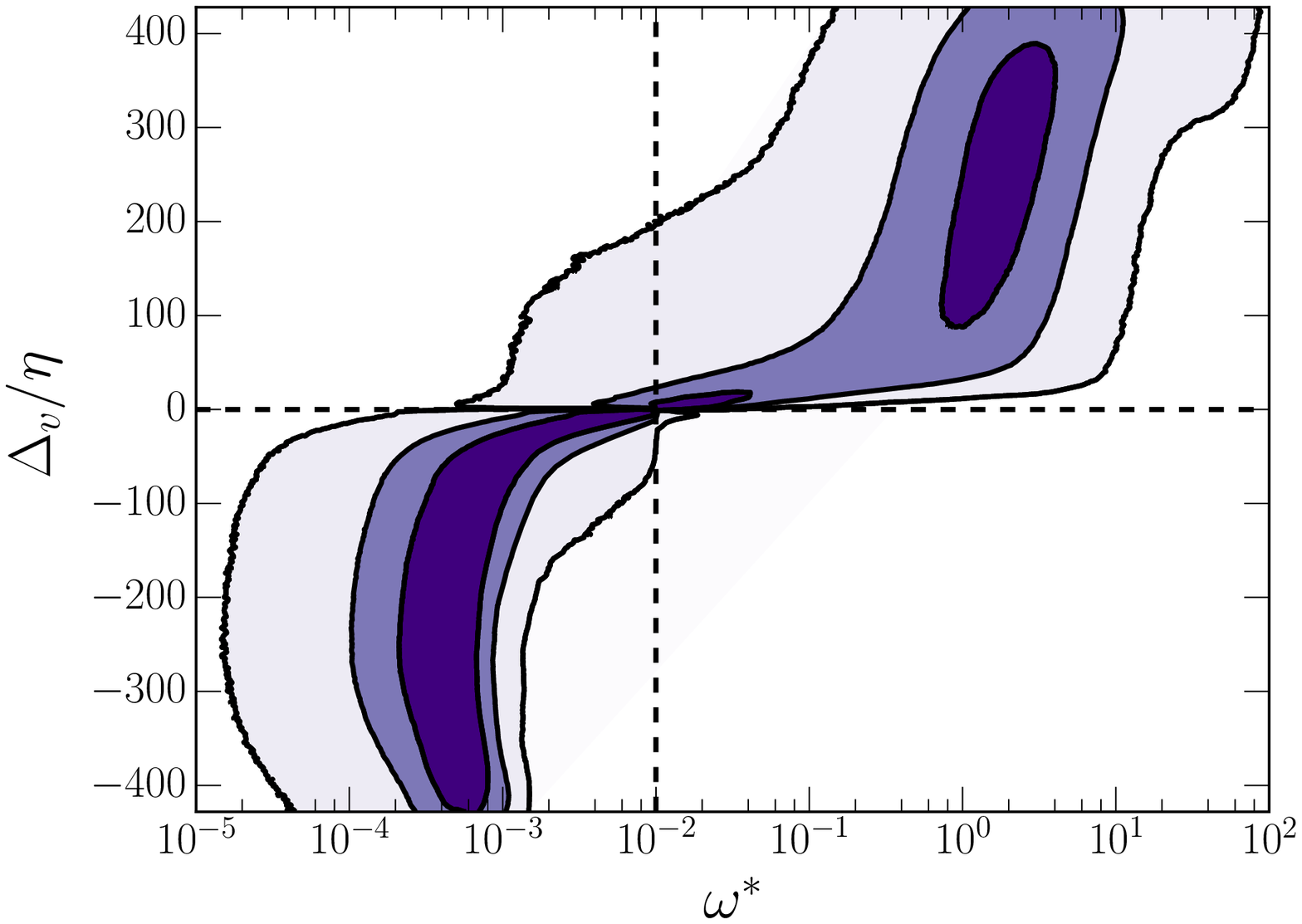}%
\mylab{-0.37\textwidth}{0.29\textwidth}{(c)}%
\hspace{1mm}%
\includegraphics[width=0.48\textwidth, clip=true,trim=0em 1em 1em 4em]{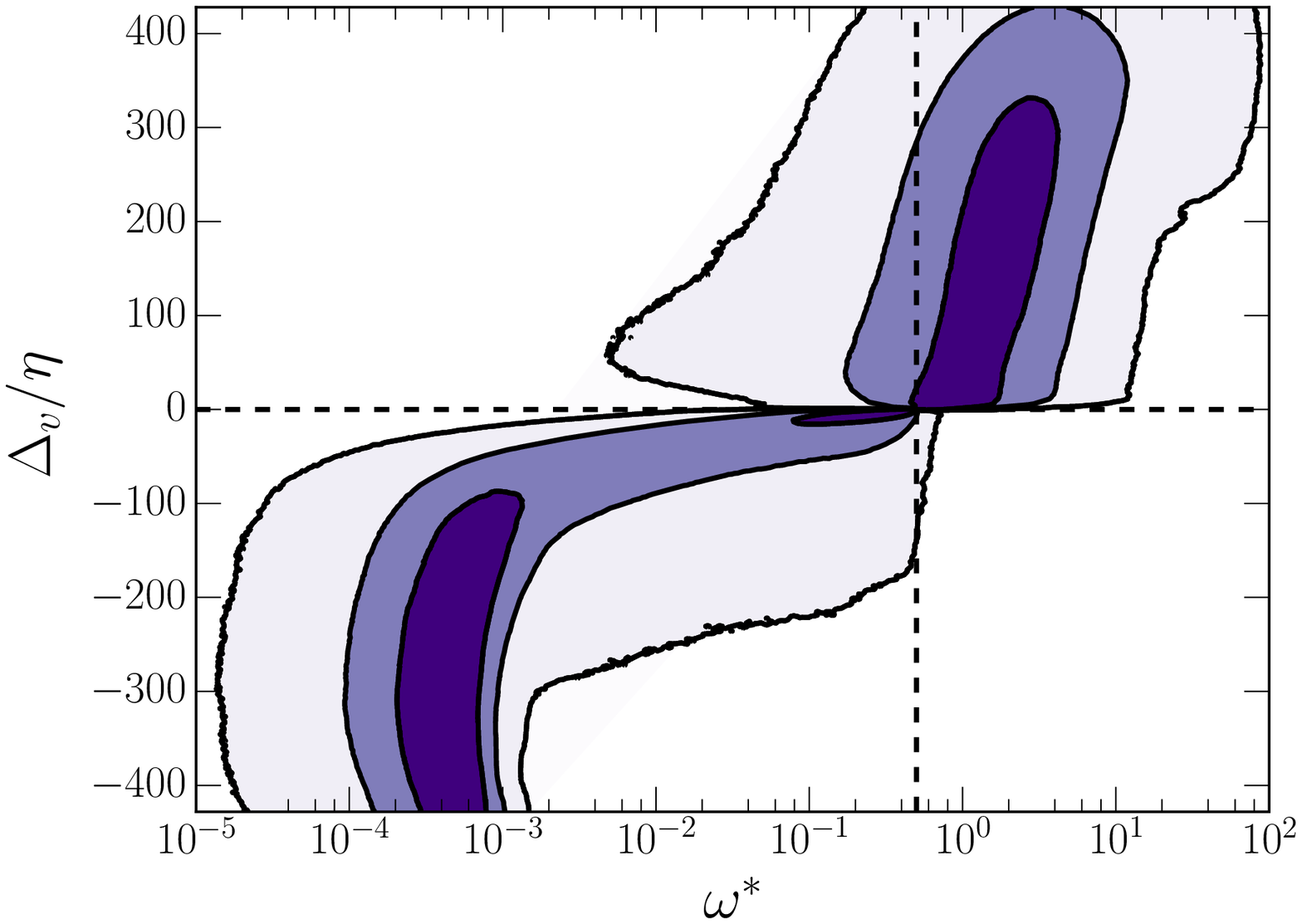}%
\mylab{-0.37\textwidth}{0.29\textwidth}{(d)}%
}
\caption{Premultiplied joint probability density function of vorticity and distance,
$\omega F_{\omega,\Delta}$. Each subplot corresponds to the thresholds, Reynolds
number, and distance definitions in figure \ref{fig:distancefield} (top row
$\Delta_b$; bottom row $\Delta_v$; left column $\omega_0^*=0.01$; right column
$\omega_0^*=0.5$). Contours contain 50\%, 90\%, and 99\% of points, respectively. 
}
\label{fig:jointpdf}
\end{figure}

The properties of the vorticity conditioned to its position with
respect to the interface can be analysed using the joint PDF of the
vorticity magnitude and of the distance, $F_{\omega,\Delta}$. Figure
\ref{fig:jointpdf} shows four examples corresponding to the thresholds
and distance definitions in figure \ref{fig:distancefield}. Similar
PDFs were obtained by \cite{Tav:Sil:PF14} using $\Delta_v$ in a planar
jets, and by \cite{silva2014characteristics} for jets, a shearless
interface and a subset of the present boundary layer. The analysis in
the present paper has been carried out for five Reynolds numbers in
the range $\delta_{99}^+ \in (1100$ -- 1900), and ten thresholds in
$\omega_0^* \in (0.01$ -- 0.5), each of them computed for the two
distance definitions mentioned above.
 
The joint PDF can be divided into four quadrants (figure \ref{fig:jointpdf}b), separated
by the axes $\Delta=0$ and $\omega=\omega_0$, marked with dashed lines in figure
\ref{fig:jointpdf}. Given that the flow field is the same in the four figures, the
differences in the joint PDF are due to the different distance definitions and thresholds.

The first quadrant, which contains points classified as turbulent and with a
relatively high vorticity, represents the core turbulent flow. As already seen in
figure \ref{fig:distancefield}, the minimum and vertical distances behave similarly
for low thresholds (figures \ref{fig:jointpdf}a,c), but very differently for
thresholds within the topological transition. The field of vertical distances depends
only slightly on the threshold (figures \ref{fig:jointpdf}c,d), but there are few
points at distances beyond $\Delta_b = 100\eta$ for the higher threshold in figure
\ref{fig:jointpdf}(b).

The second quadrant contains different geometrical objects depending on the distance
definition. It contains bubbles for $\Delta_b$, and bubbles, handles, and pockets for
$\Delta_v$. For the ball distance, the weight of the second quadrant is always small
compared with the first one, and contributes little to the averaged vorticity in the
free-stream side of the interface (figures \ref{fig:jointpdf}a,b). This is not the
case for the vertical distance, and it is clear from figures \ref{fig:jointpdf}(c,d)
that the weight of this quadrant increases as the interface becomes more complex at
high thresholds. This quadrant, with especial reference to the properties of the
pockets, will be studied in more detail in \S\ref{sec:pockets}.

The third quadrant contains points of low vorticity classified as non-turbulent. It
represents the bulk of the free stream which, in the case of $\Delta_b$, also
includes the irrotational pockets. It depends only weakly on the threshold and on the
distance definition, except for $\omega\approx \omega_0$.

The fourth quadrant, with $\omega>\omega_0$ and negative distances,
corresponds to the objects defined in \S\ref{sec:drops} as drops. It
is almost empty for all the cases considered in this study, confirming
that the smoothing of the free stream described in \S\ref{sec:drops}
does not affect the results of the conditional analysis.

\begin{figure}
\centerline{%
\raisebox{0.1\textwidth}[0\textwidth][0\textwidth]{%
\includegraphics[width=0.4\textwidth, clip=true,trim=0em 0em 0em -6em]{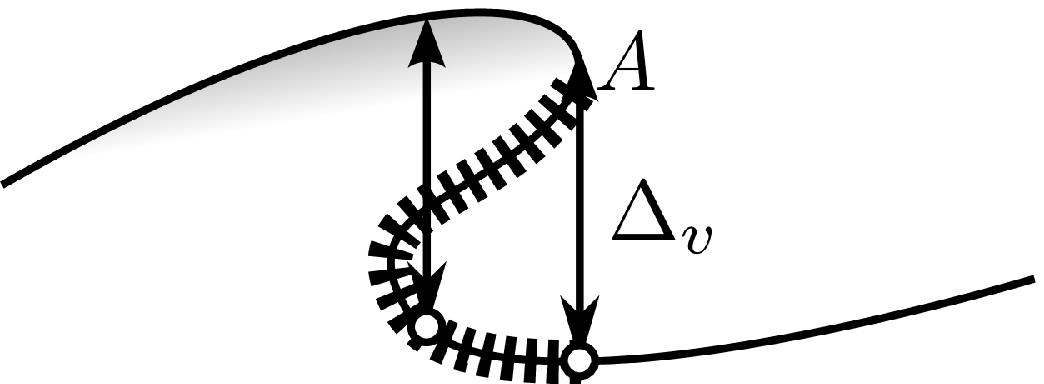}%
\mylab{-0.20\textwidth}{0.19\textwidth}{(a)}}%
\hspace{5mm}%
\includegraphics[width=0.5\textwidth, clip=true, trim=1em 0em 1em 2em]{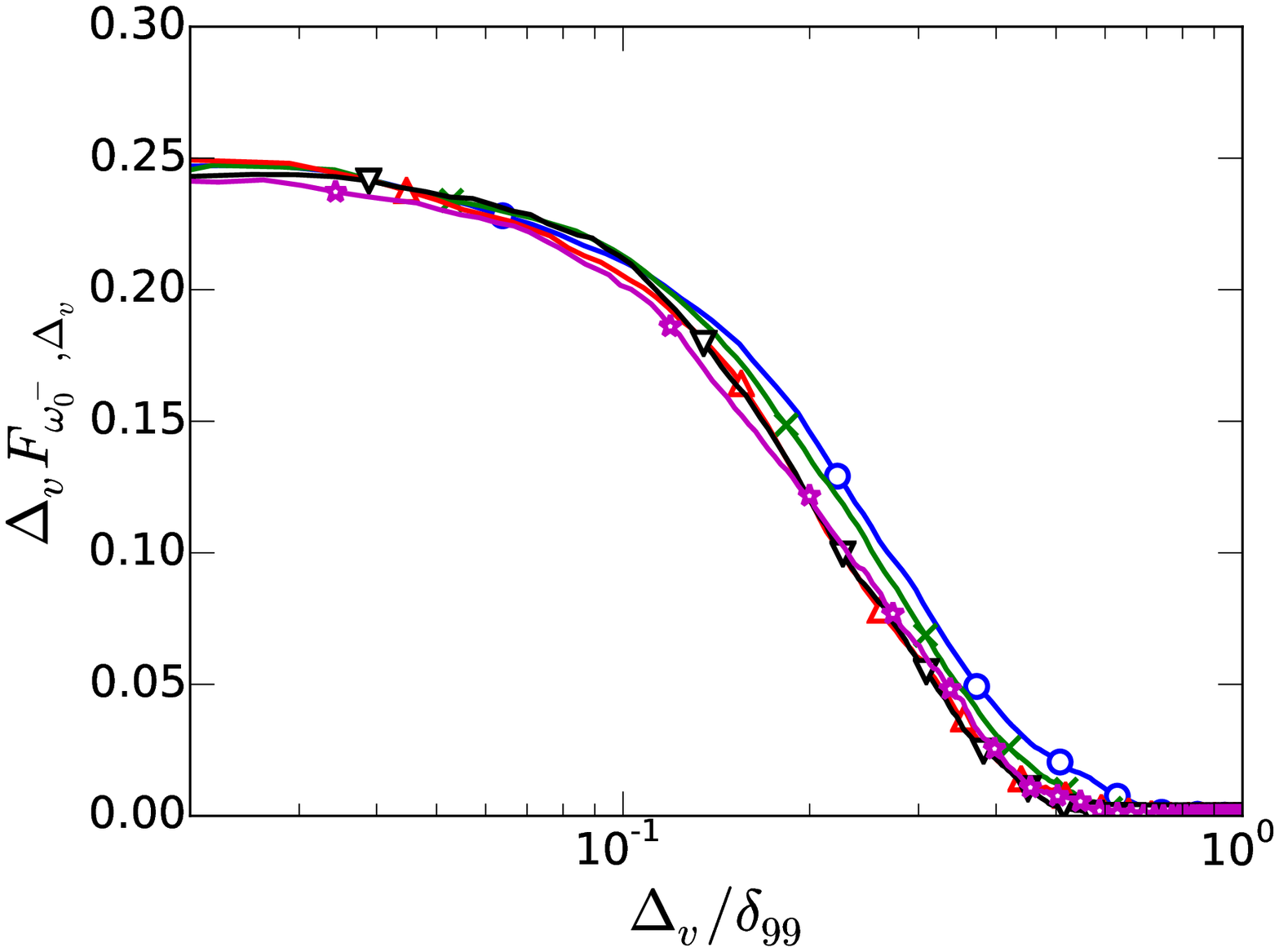}%
\mylab{-0.12\textwidth}{0.29\textwidth}{(b)}%
}
\caption{(a) Sketch of the discontinuity of the vertical distance at the edge of a
pocket. The hatched line marks points where $\omega = \omega_0$ and
$\Delta_b=0$, but $\Delta_v > 0$. The non-turbulent region just outside A has
$\Delta_b\simeq 0$ but $\Delta_v < 0$. 
(b) Premultiplied PDF, $\Delta_v F_{\omega_0^-,\Delta_v}$, of the vertical distance
of  the non-turbulent points with $\omega^*=0.25$--0.5, whose vorticity is close to the threshold
$\omega_0^*=0.5$. $\circ$ (blue), $\delta_{99}^+=1100$;
$\times$ (green), $1300$; $\triangle$ (red), $1500$; $\triangledown$ (black), $1700$;
$\star$ (magenta), $1900$ .
}
\label{fig:deltav_ambiguity}
\end{figure}

The influence of the distance definition on the joint PDF is most
visible far from the horizontal axis in the neighbourhood of
$\omega=\omega_0$. These are points in which the vorticity is close to
the identification isosurface, but that are incorrectly identified as
being far from the interface. The range of possible ball distances for
$\omega\simeq\omega_0$ (figure \ref{fig:jointpdf}a,b) is very narrow,
$|\Delta_b|< 100\eta$, especially in the second quadrant, and can be
interpreted as a typical distance to the interface of the irrotational
bubbles that have been labelled as turbulent by the smoothing
process. On the other hand, the vertical distances in the same region
can be as large as 250$\eta$ at both sides of the interface (figure
\ref{fig:jointpdf}c,d). Denote by $\omega_0^-$ the vorticities just
below the threshold. The wide $\Delta_v$ tails of
$F_{\omega_0^-,\Delta_v}$ have several causes, sketched in figure
\ref{fig:deltav_ambiguity}(a). On the positive side, $\Delta_v>0$ in
Q$_2$, all the points represented with a hatched line in that figure
are on the $\omega=\omega_0$ and $\Delta_b=0$ isosurfaces, but not on
the $\Delta_v$ interface, which is only the top of the overhang.
Points near the hatched line have vorticities close to $\omega_0$, but
are counted as being deep within the turbulent region by
$\Delta_v$. The $\Delta_v<0$ tail of $F_{\omega_0^-,\Delta_v}$ in
Q$_3$ contains points whose vorticity is slightly below than the
threshold, but which are classified by $\Delta_v$ as being far within
the irrotational region. They correspond to points such as A in figure
\ref{fig:deltav_ambiguity}(a), where the orientation of the interface
is vertical and induces a discontinuity in the height of the
$\Delta_v$ interface. Such discontinuities are clearly visible in
figures \ref{fig:distancefield}(c,d). These tangencies are less common
than the overhung surfaces, and the mass in the negative tail of
$F_{\omega_0^-,\Delta_v}$ is typically smaller than in the positive
one, especially in the convoluted interfaces at the higher thresholds
(only 15\% as many in figure \ref{fig:distancefield}d).

It is clear from figure \ref{fig:deltav_ambiguity}(a) that the negative tail of
$F_{\omega_0^-,\Delta_v}$ contains information about the `depth' of the pockets,
rather than about the thickness of the interface. The premultiplied probability
distribution $\Delta_v F_{\omega_0^-,\Delta_v}$, integrated over the band
$\omega\in(\omega_0/2,\omega_0)$, is presented in figure
\ref{fig:deltav_ambiguity}(b) for a relatively high threshold. It is well
approximated by a power law $F_{\omega_0^-,\Delta_v}\propto \Delta_v^{-1}$ for
$\Delta_v\lesssim 0.2\delta_{99}$. Although the reason for this particular power is
not completely clear, it suggests a regular structure for the $\Delta_v$ interface.
That interface has no overhangs, and renders pockets as holes with steep sides. If
we assume that the interface is covered with pockets of size $\Delta$, the
contribution of each hole to the PDF in figure \ref{fig:deltav_ambiguity}(b) would be
proportional to the $O(\Delta)$ length of its lip. Their number would be proportional
to $\Delta^{-2}$ and the total lip length would be proportional to $\Delta^{-1}$, as
in the figure.

Even if this explanation turns out to be oversimplified, the fact that the
distribution of pocket heights satisfies a power law is consistent with a fractal
interface, and suggests that the discontinuities represent a
self-similar hierarchy of overhangs. For the threshold in figure
\ref{fig:deltav_ambiguity}(b), the self-similar range ends around $\Delta_v\approx
0.2\delta_{99}$, and the probability of finding pockets deeper than that limit is
very low. This is about three times the standard deviation of the position of the
vorticity isosurface for this threshold (figure \ref{fig:mean_std}). At lower
thresholds such as those in figures \ref{fig:jointpdf}(a,c), the self similar range
disappears, and the `pocket' distribution in concentrated around
$\Delta_v=10\eta$.

\subsection{Conditional averages}\la{sec:condav}

\begin{figure}
\centerline{%
\includegraphics[width=0.48\textwidth, clip=true,trim=1em 2em 1em 0em]{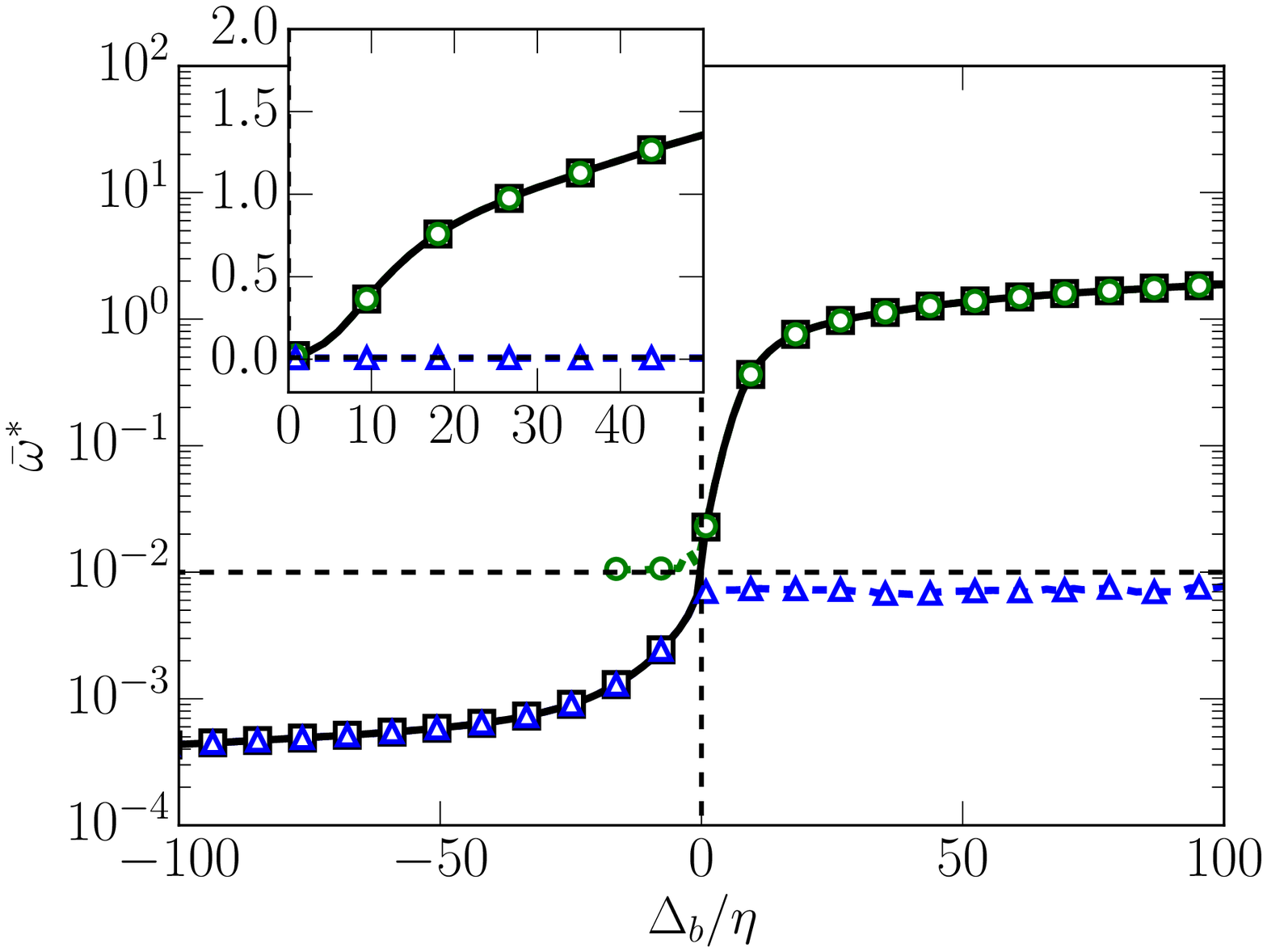}%
\mylab{-0.10\textwidth}{0.29\textwidth}{(a)}%
\hspace{2mm}%
\includegraphics[width=0.48\textwidth, clip=true,trim=1em 2em 1em 0em]{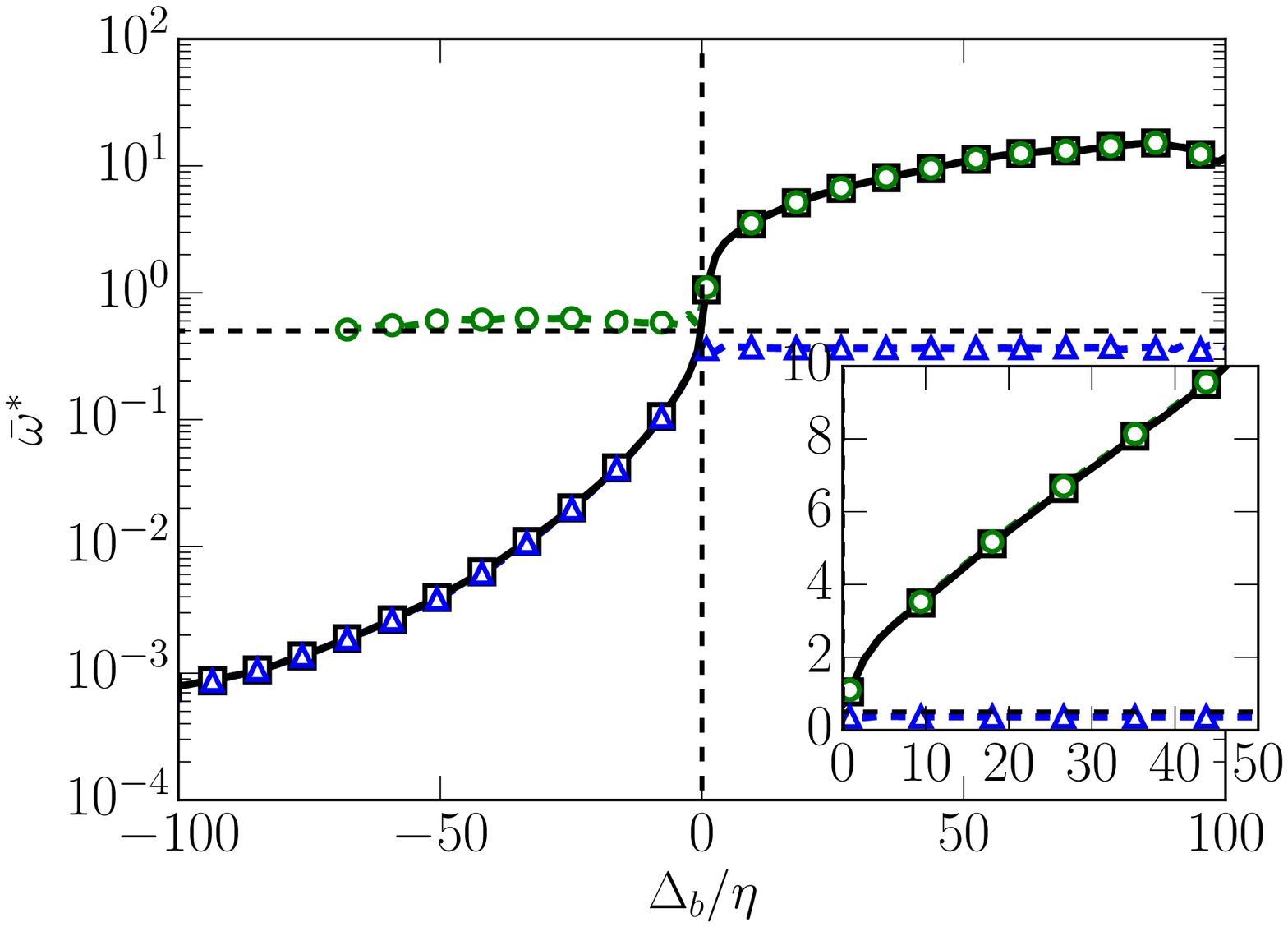}%
\mylab{-0.37\textwidth}{0.29\textwidth}{(b)}%
}
\vspace{1ex}%
\centerline{%
\includegraphics[width=0.48\textwidth, clip=true,trim=1em 2em 1em 0em]{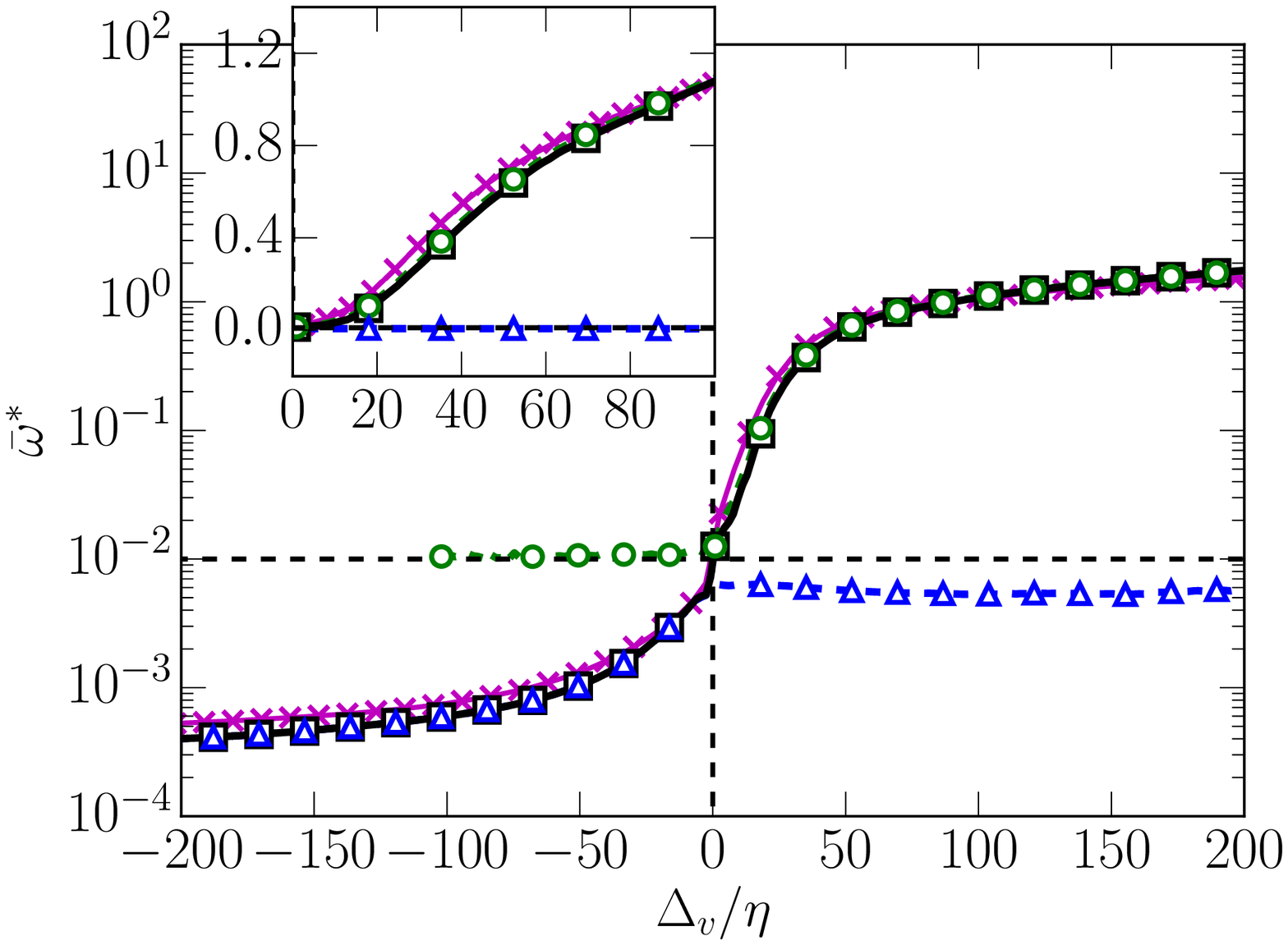}%
\mylab{-0.10\textwidth}{0.29\textwidth}{(c)}%
\hspace{2mm}%
\includegraphics[width=0.48\textwidth, clip=true,trim=1em 2em 1em 0em]{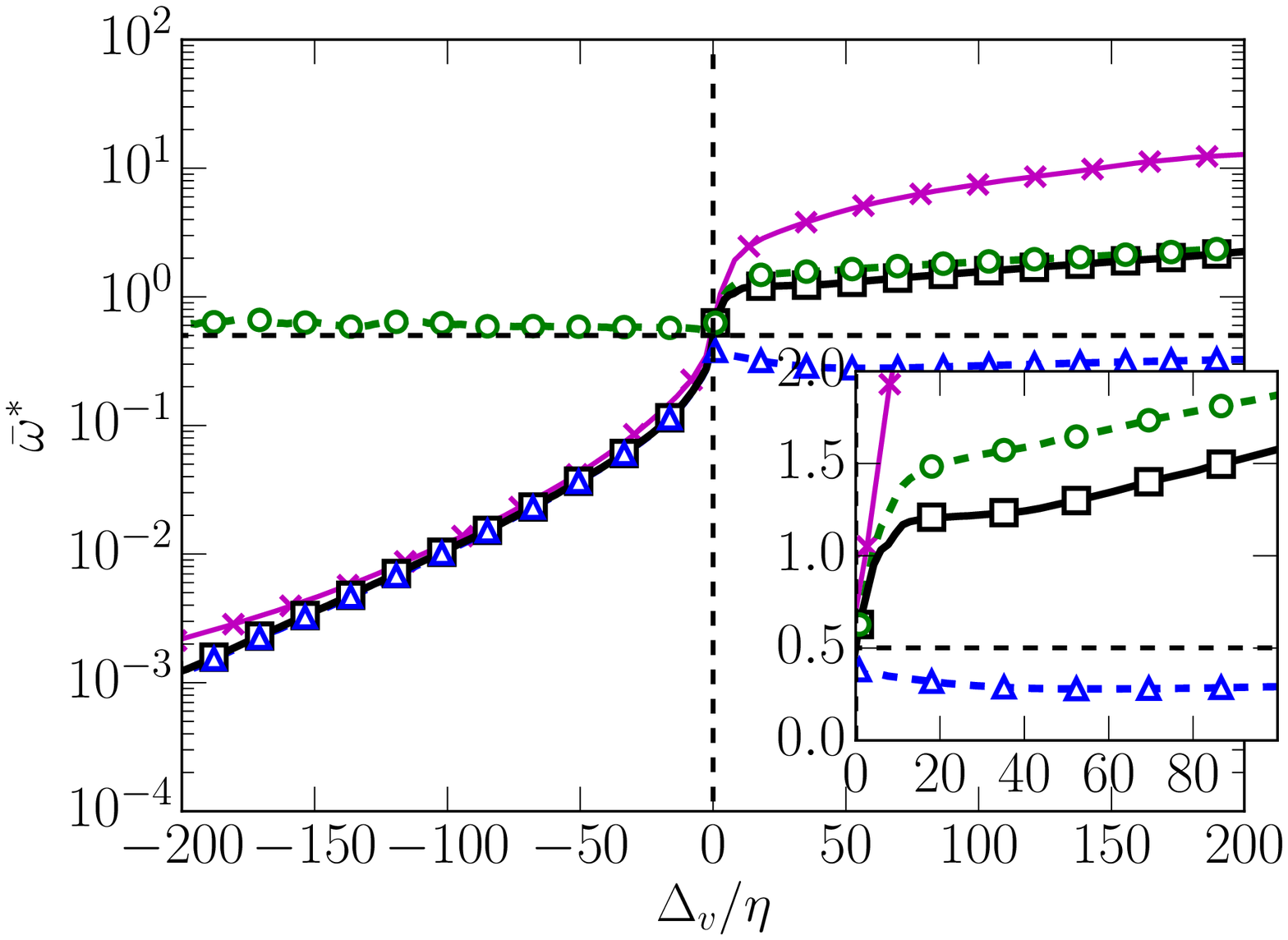}%
\mylab{-0.37\textwidth}{0.29\textwidth}{(d)}%
}
\caption{
  $- -\circ- -$, $\overline{\omega_1}$; $- - \triangle - - $,
  $\overline{\omega_2}$. The figures follow the same arrangement as in
  figure \ref{fig:distancefield} (top row $\Delta_b$; bottom row
  $\Delta_v$; left column $\omega_0^*=0.01$; right column
  $\omega_0^*=0.5$). The black dashed lines correspond to the value of
  the threshold (vertical) and zero distance (horizontal).  The inset
  in each figure correspond to the same plot, using linear coordinates
  for the vorticity magnitude. The two solid lines with crosses in (c)
  and (d) correspond to $ \overline{\omega}(3.1\Delta_b)$ (see \S
  \ref{ballvsvertical}).  }
\label{fig:condaverage} 
\end{figure}

The averaged vorticity conditioned to the distance to the interface
can be computed from $F_{\omega,\Delta}$ as
\begin{equation}
  \label{eq:total_average}
  \overline{\omega}(\Delta) = \frac{ \int_0^\infty \omega
    F_{\omega,\Delta}\  \mbox{d}\omega}
  { \int_0^\infty
    F_{\omega,\Delta}\  \mbox{d}\omega}.
\end{equation}
It is given by the solid lines with squares in figures
\ref{fig:condaverage}(a-d), and is equivalent to the conditional
vorticity profiles in \citet{WesterweelEiF}. Note the use of the bar
over the symbol to distinguish \eqref{eq:total_average} from the more
usual mean profile $\langle \omega \rangle$ at a given distance from
the wall, defined as
\begin{equation}
  \label{eq:total_average_wall}
  \langle \omega \rangle(y) = \frac{ \int_0^\infty \omega
    \Gamma_{\omega,y}\  \mbox{d}\omega}
  { \int_0^\infty
    \Gamma_{\omega,y}\  \mbox{d}\omega}.
\end{equation}
We will use the notation $\overline{\omega}(\Delta_b)$ and
$\overline{\omega}(\Delta_v)$ to distinguish between conditional
profiles obtained with each definition of distance.

The conditional vorticity in all the panels of figure \ref{fig:condaverage} increases
monotonically to its expected fully turbulent level, $\omega^*=O(1)$, within a few
Kolmogorov lengths from the interface, except for the plateau at
$\Delta_v/\eta=15\mbox{--} 40$ in figure \ref{fig:condaverage}(d).

The existence of a plateau or of a maximum in the conditional
vorticity profile near the T/NT interface has been mentioned in wakes
\citep{TownsendBook,FLM:95049} and reported in jets
\citep{West:etal:09,FLM:8400021}. Its presence has sometimes been used
to define the thickness of the interface layer \citep{SilvaTaveira},
and taken as the basis for theoretical models in which the interface
is maintained by the presence of a strong localised shear
\citep{HuntDurbin}. Similar models have been used to suggest
similarities between the T/NT interface in jets \citep{West:etal:09}
and strong internal vortex layers in homogeneous turbulence
\citep{Ishihara}. \cite{Chauhan:14} report a strong conditional
vorticity peak in boundary layers, but their interface is defined in
terms of the velocity fluctuations, and we will argue below that it is
probably different from the one discussed here. Moreover, not all
these papers use the same definition of the interface or even the same
thresholded scalar. In fact, when \cite{SilvaAR14} compiled
conditional vorticity statistics for a variety of flows, the only
obvious peak is found in the early stages of the evolution of a
shearless mixing layer \citep{SilvaTaveira}. \cite{FLM:95049} also
find strong vorticity peaks for some high vorticity thresholds in
their wake, but attribute them to the presence of isolated vorticity
patches, and discard them in favour of a lower threshold
$(\omega^*\approx 0.1)$ for which the maximum is barely
noticeable. Although we will find and discuss below comparable peaks
in other variables, we note at this stage that, if the vorticity
magnitude were particularly intense close to the interface, a plateau
analogous to the one in figure \ref{fig:condaverage}(d) should also
appear in the $\overline{\omega}(\Delta_b)$ profile in figure
\ref{fig:condaverage}(b), but this is not the case. An alternative
explanation is that the vorticity close to the interface is not
particularly intense but that, when the conditional profiles are
obtained as a function of $\Delta_v$ at a sufficiently high threshold,
some non-turbulent flow is counted as being turbulent within the inner
part of the interface, lowering the local average vorticity
\citep{FLM:95049}.

To differentiate between the two hypotheses we split the conditional profile
$\overline \omega (\Delta_v)$ into contributions from the high-vorticity first
quadrant, $Q_1$, and the mislabeled non-turbulent points in $Q_2$. Equation
\eqref{eq:total_average} is split into
\begin{equation}
  \label{eq:contributions}
  \overline{\omega} =  W_1
  \overline{\omega}_1 + W_2 \overline{\omega}_2 ,
\end{equation}
where
\begin{equation}
  \label{eq:conditioned_average}
  \overline{\omega}_1 = \frac{\int_{\omega_0}^\infty \omega
    F_{\omega,\Delta}\  \mbox{d}\omega}{\int_{\omega_0}^\infty
    F_{\omega,\Delta}\  \mbox{d}\omega}
  ,\qquad 
\overline{\omega}_2 = \frac{ \int_0^{\omega_0} \omega
    F_{\omega,\Delta}\  \mbox{d}\omega}
  { \int_0^{\omega_0}  F_{\omega,\Delta}\  \mbox{d}\omega} ,
\end{equation}
are the conditional averages for $Q_1$ and $Q_2$, and
\begin{equation}
  \label{eq:quadrants}
  W_1= \frac{ \int_{\omega_0}^{\infty} 
    F_{\omega,\Delta}\  \mbox{d}\omega}
  { \int_0^{\infty}  F_{\omega,\Delta}\  \mbox{d}\omega}
  ,\qquad  
W_2 = \frac{ \int_0^{\omega_0}
    F_{\omega,\Delta}\ \mbox{d}\omega} { \int_0^{\infty}
    F_{\omega,\Delta}\ \mbox{d}\omega} ,
\end{equation}
are the corresponding weights. The profiles of $\overline{\omega}$,
$\overline{\omega}_1$, and $\overline{\omega}_2$ are given in figure
\ref{fig:condaverage}. In the case of low thresholds (left column of the figure),
$\overline{\omega}_1\simeq\overline{\omega}$, and the contribution of the second
quadrant is small, regardless of the distance definition.

The only case in which $\overline{\omega}_1$ is clearly different from
the overall average is figure \ref{fig:condaverage}(d), in which the
contribution of the handles and pockets is significant. In this
figure, the maximum relative weight of $Q_2$ is $W_2 \simeq W_1/4$ at
$\Delta_v=20\eta$. At the even higher thresholds at which the
interface reaches its maximum geometrical complexity near the end of
the topological transition, the weights of the two quadrants are
comparable. This has a noticeable effect on the conditional profiles,
and it is clear from figure \ref{fig:condaverage}(d) that the plateau
is a consequence of the negative contribution from
$\overline{\omega}_2$. If we consider this contribution as a spurious
effect of $\Delta_v$, the `true' conditional vorticity
$\overline{\omega}_1$ in figure \ref{fig:condaverage}(d) increases
monotonically near the the interface. In essence, the conditional
vorticity remains constant or decreases away from the interface
because $\Delta_v$ misclassifies some weakly vortical pockets as part
of the turbulent flow.

\begin{figure}
\centerline{%
\includegraphics[width=0.48\textwidth, clip=true,trim=1em 2em 1em 0em]{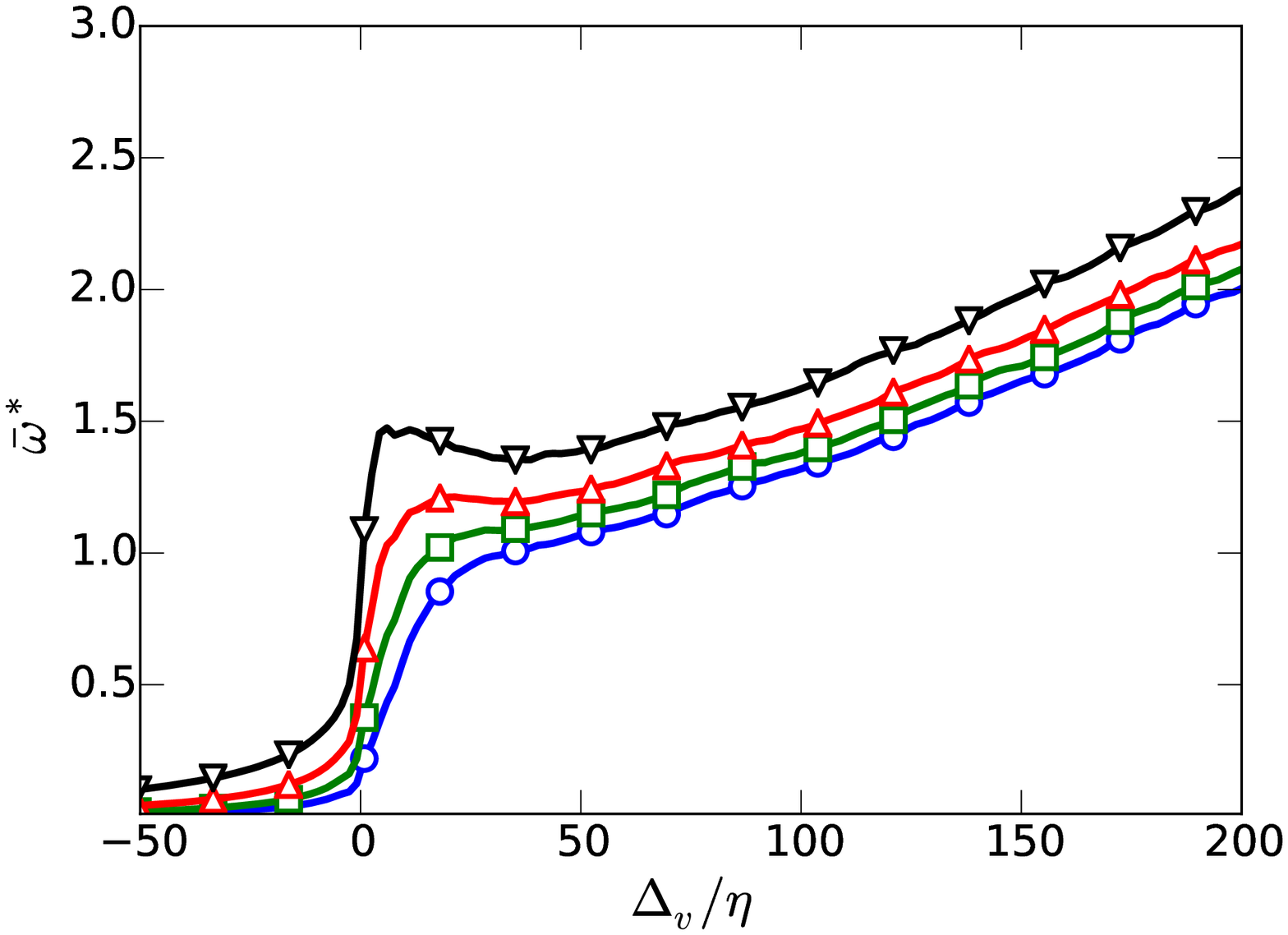}%
\mylab{-0.40\textwidth}{0.29\textwidth}{(a)}%
\hspace{2mm}%
\includegraphics[width=0.48\textwidth, clip=true,trim=1em 2em 1em 0em]{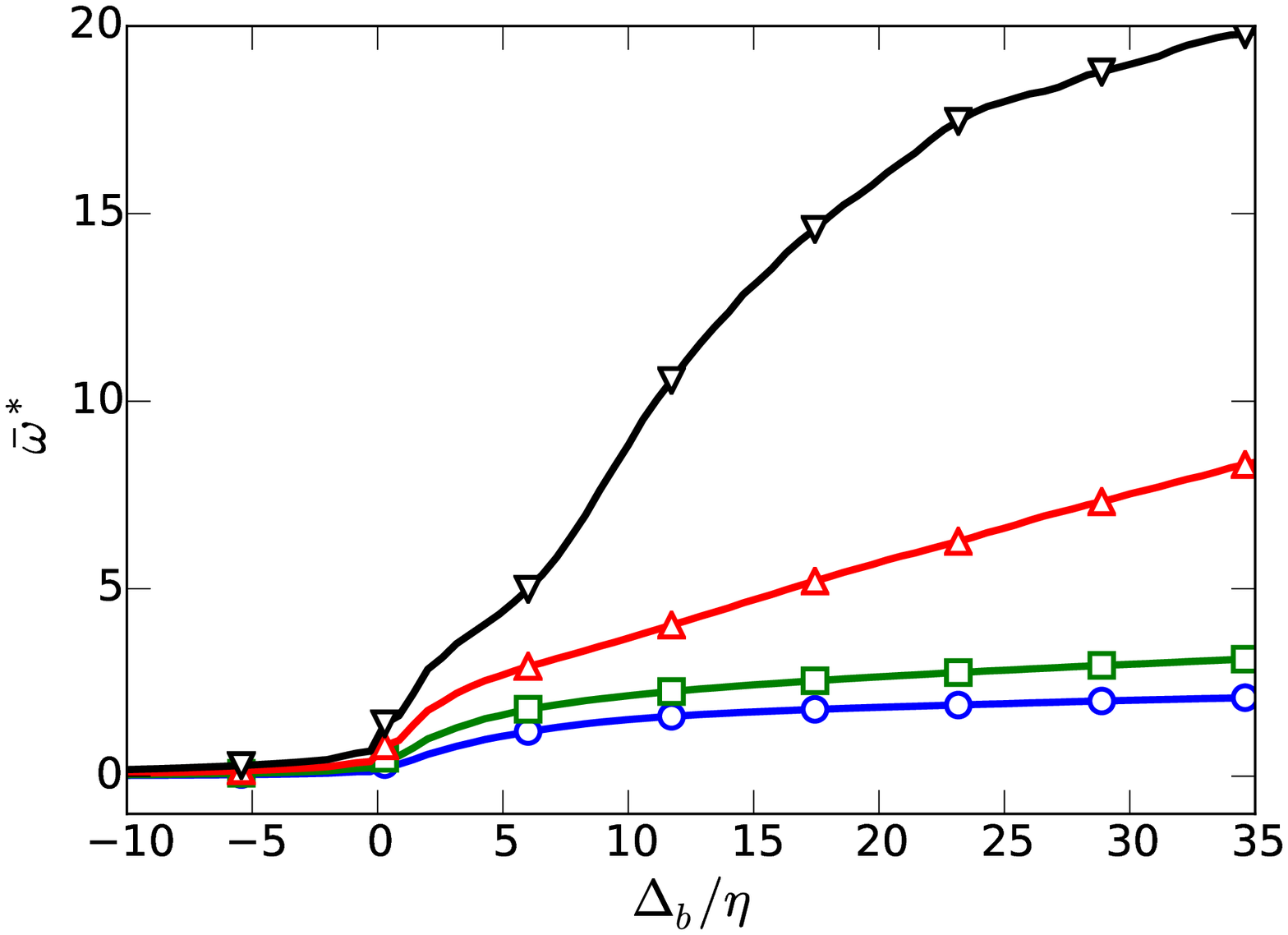}%
\mylab{-0.40\textwidth}{0.29\textwidth}{(b)}%
}
\caption{(a) Conditional vorticity profiles for $\delta_{99}^+=1900$, computed as in
figure \ref{fig:condaverage}, as functions of the threshold.
\circle, $\omega_0^*=0.17$; \squar, 0.29; \trian, 0.52; \dtrian, 0.88.
(a) $\Delta_v$.  (b) $\Delta_b$. 
}
\label{fig:condaverage_seepeak}
\end{figure}

This effect is clearer in figure \ref{fig:condaverage_seepeak}, which
presents conditional vorticities for several interface
thresholds. Figure \ref{fig:condaverage_seepeak}(a) is computed with
$\Delta_v$, and develops a plateau and eventually a peak as the
threshold increases. As in figure \ref{fig:condaverage}(d), it can be
shown that this is a due to the increasingly negative contribution
from the pockets as the complexity of the interface increases.  Figure
\ref{fig:condaverage_seepeak}(b) presents the same cases computed for
$\Delta_b$, and shows no trace of an interface peak.

Note that the distances in figure \ref{fig:condaverage_seepeak}(b) are
much lower than in figure \ref{fig:condaverage_seepeak}(a), while the
conditional vorticities are higher. In fact, similar conditional
vorticities are found when the horizontal axis of figure
\ref{fig:condaverage_seepeak}(a) is extended to $\Delta_v\simeq 400$,
carrying the plot to the neighbourhood of the wall. The plot of
$\overline{\omega}(\Delta_v)$ for these large distances is very
similar to a shifted version of $\omega'(y)$ (figure
\ref{fig:intermittency}a). The vorticity isosurface at these high
thresholds permeates the whole boundary layer, and occasionally comes
very close to the wall. The ball distance recognises this fact and
brings the strong near-wall vorticity closer to the interface, while
the vertical distance misses that complexity.

This discussion suggests that the apparent strongly vortical interface
layer observed in some of the studies mentioned above is either an
artefact of how a one-dimensional definition of distance interacts
with a fully three-dimensional geometry, or only manifests itself in
variables different from the vorticity magnitude. In the present case,
the above arguments show that the interface peak is due to the neglect
of the effect of irrotational pockets on the conditional
quantities. We next discuss the relevance of these pockets in the
entrainment process.

\subsubsection{\label{ballvsvertical}An approximate relation between $\Delta_v$ and $\Delta_b$}

A property of the ball distance that can be used to approximately
relate it to the more common vertical definition $\Delta_v$ is that,
close enough to the interface, it corresponds to the distance along
the local normal. In consequence, both definitions are related by
  \begin{equation}
    \label{eq:angle}
    \lim_{\Delta_b\to 0} \Delta_b/\Delta_v =  \cos \theta,
  \end{equation}
where $\theta$ is the angle between the local normal and the
  vertical direction. Equation \eqref{eq:angle} can be averaged,
  giving quantitative relationship between the conditional profiles,
  \begin{equation}
    \label{eq:average_angle}
    \overline{\frac{\Delta_b}{\Delta_v}} \sim \overline {\cos \theta}.
  \end{equation}
showing that conditional profiles obtained in terms of the ball
distance are $\overline {\cos \theta }$ narrower than those expressed
in terms of the vertical distance. The profile $\overline{\omega}
(\Delta_b/\overline{\cos \theta})$ is represented in figures
\ref{fig:condaverage}(c,d) as a magenta line with crosses. Whenever
$\overline{\omega} (\Delta_b/\overline {\cos \theta}) \simeq
\overline{\omega}(\Delta_v)$, the projection of one distance onto the
other is quantitatively valid. The results suggest that the two
measures are comparable if the geometry of the interface is only
moderately complex. For example, figure \ref{fig:condaverage}(c) shows
that the range of validity of \eqref{eq:angle} for low vorticity
thresholds extends to a substantial portion of the boundary layer
thickness, using $\overline{\cos \theta} = 0.32$. On the other hand,
when the threshold approaches the topological transition, the two
measurements are only comparable within a region very close to the
interface (figure \ref{fig:condaverage}d).

Although this relationship between $\Delta_b$ and $\Delta_v$ is only a
gross approximation, it contains information about the shape of the
interface. Lower values of $\overline{\cos \theta}$ imply that the
interface is steeper on average, so that the local normal is less
likely to be aligned with the vertical axis. To give some perspective
on the the empirical $0.32$ factor mentioned above, the same result
for a hemisphere yields $\overline{\cos \theta}=0.5$.

\subsection{The relevance of pockets}\la{sec:pockets}

We saw in figure \ref{fig:deltav_ambiguity} that pockets form a
self-similar hierarchy of many different sizes, and it has been
conjectured that their formation signals the large-scale engulfment of
irrotational fluid before it is finally entrained by small-scale
`nibbling'. Their abundance has been used to quantify the relative
importance of the two processes \citep{mathew:2065,CTRSandham}.

\begin{figure}
\centerline{%
\includegraphics[width=0.48\textwidth, clip=true,trim=1em 2em 1em 0em]{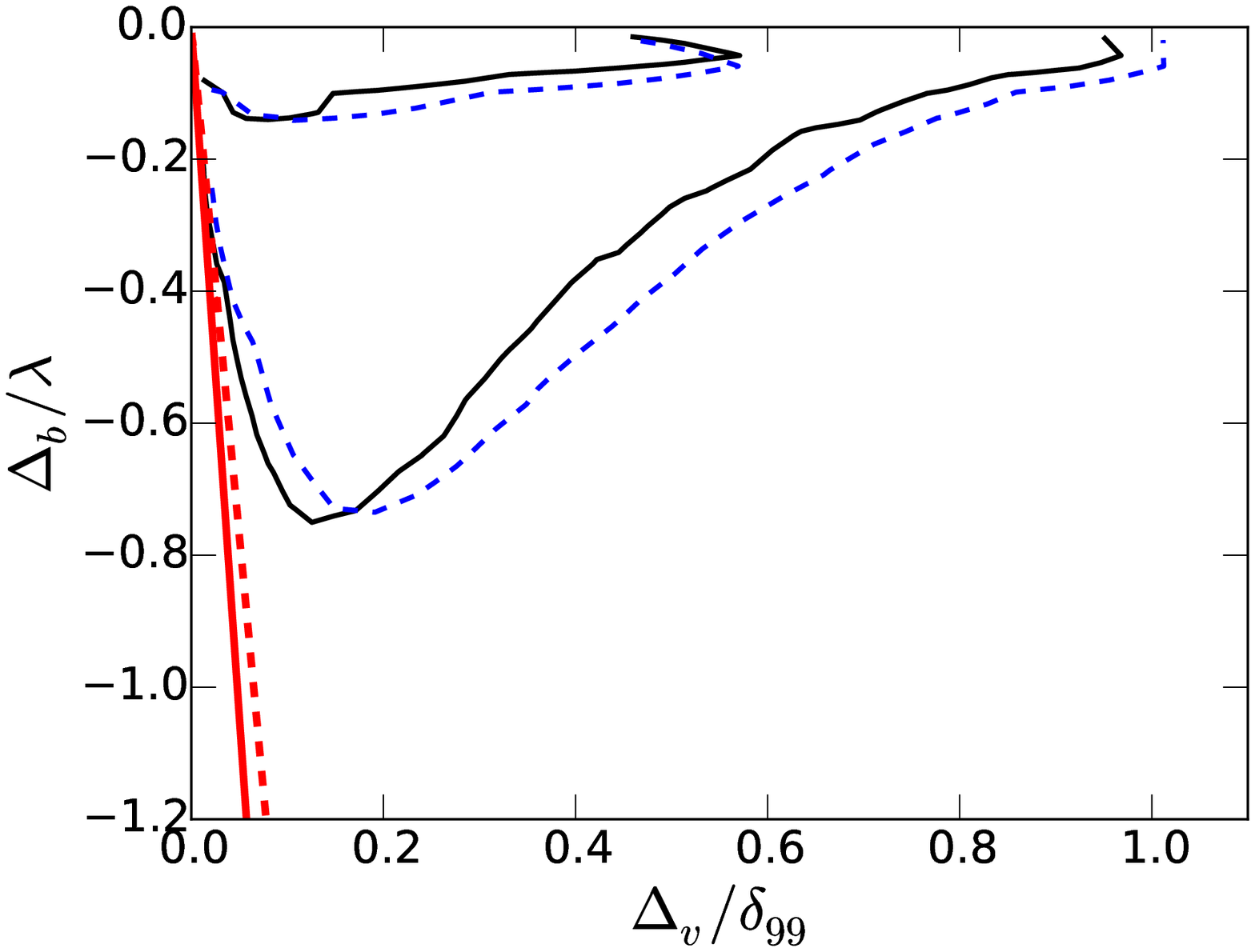}%
\mylab{-0.10\textwidth}{0.25\textwidth}{(a)}%
\hspace{2mm}%
\includegraphics[width=0.48\textwidth, clip=true,trim=1em 2em 1em 0em]{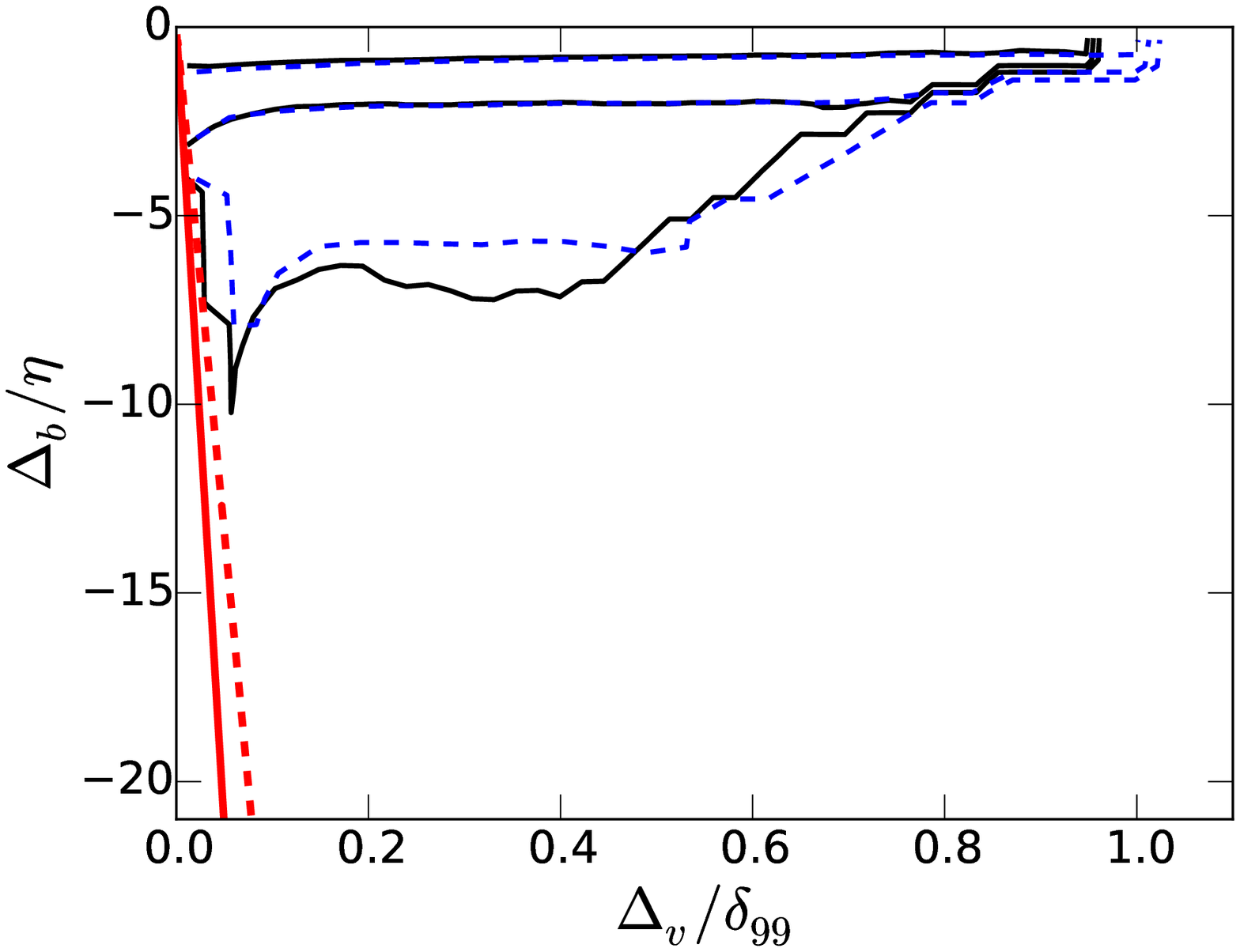}%
\mylab{-0.10\textwidth}{0.25\textwidth}{(b)}%
}
\caption{%
Flow within pockets of the interface at $\omega^*_0=0.5$. \dashed,
$\delta_{99}^+=1100$; \solid, $\delta_{99}^+=1900$
(a) Joint PDF of $\Delta_v$ and $\Delta_b$ within the region. The two
contours for each Reynolds number contain 60\% and 98\% of the points,
respectively.
(b) Average vorticity within the pockets as a function $\Delta_v$ and
$\Delta_b$ within the region that contains 98\% of the
points. Contours are $\omega^* = 0.15$, 0.25, 0.35.
The two red diagonals are $\Delta_v=\Delta_b$.  }
\label{fig:pockets}
\end{figure}

We can define pockets as regions identified by the ball distance as
part of the free stream, $\Delta_b<0$, and by the vertical distance as
turbulent, $\Delta_v>0$. For the purpose of this section, they include
the underside of handles as well as simple folds of the
interface. Figure \ref{fig:pockets}(a) shows the joint PDF of the two
distances in the range corresponding to pockets. The figure is drawn
for the relatively high vorticity threshold of figures
\ref{fig:distancefield}(b,d), guaranteeing both the presence of
abundant pockets and the possibility of observing how the vorticity
diffuses into the irrotational flow. It includes the two extreme
Reynolds numbers in our simulation, allowing some scaling
comparisons. For instance, the agreement between the two profiles of
$\bra y_I \ket$ suggests that the thickness of the intermittent region
is not expected to change substantially with higher Reynolds
numbers. It turns out that the size of the pockets, as measured by the
maximum $\Delta_b$, scales best in terms of the Taylor microscale,
while their depth within the layer, as measured by $\Delta_v$, scales
better with the boundary-layer thickness. The joint PDF is roughly
triangular. It is bounded on the left by the trivial limit
$\Delta_b\le \Delta_v$, plotted for each Reynolds number as a thick
inclined straight line, and on the right by a roughly hyperbolic curve
that can be interpreted to mean that deeper pockets (large $\Delta_v$)
tend to be smaller (small $\Delta_b$), presumably because they have
been broken down by the turbulence while being entrained.

The question of whether being entrained into a pocket also promotes the diffusion of
vorticity is tested in figure \ref{fig:pockets}(b), which shows the distribution of the
conditionally averaged vorticity in the same parameter space as figure \ref{fig:pockets}(a).
Note that all the vorticity levels in this figure are below the interface threshold, so that
the band of higher vorticities along the top of figure \ref{fig:pockets}(b) portrays how
vorticity diffuses into the irrotational fluid. Notice that the size of the pockets in this
figure is normalised with $\eta$. Comparison with figure \ref{fig:pockets}(a) shows that the
difference between scaling it with $\eta$ or $\lambda$ is not great, but the collapse of the
vorticity band at the top of figure \ref{fig:pockets}(b) is considerably better with the
$\eta$ than with $\lambda$. Its width, approximately 5--$10\eta$, strongly suggest a
viscous origin \citep{Ree:Holz:14}, and it is clear from the figure that the vorticity is
correlated with the ball distance, but not with the vertical position with respect to the
interface. The only exceptions are points near the line $\Delta_v=\Delta_b$, where both
measures coincide.

The implication is that the fluid within pockets is sensitive to how close it is to
the interface, but not to how deep it is within the turbulent layer. If engulfment
were an important mechanism to promote the diffusion of vorticity into the
irrotational fluid, for example by preferentially straining it, one would expect some
correlation between $\Delta_v$ and the width of the diffusion band at the top of
figure \ref{fig:pockets}(b), but there is little evidence for that. Apparently,
whether the fluid is within a pocket or not is immaterial to its behaviour, although
the break-up of the deeper pockets into smaller sizes should enhance the overall
effect of viscous diffusion. We will only use $\Delta_b$ from now on in our analysis.

\subsection{\label{sec:Layer}The interface layer.}

\begin{figure}
\centerline{%
\includegraphics[width=0.48\textwidth, clip=true,trim=0em 0em 4.5em 2.5em]{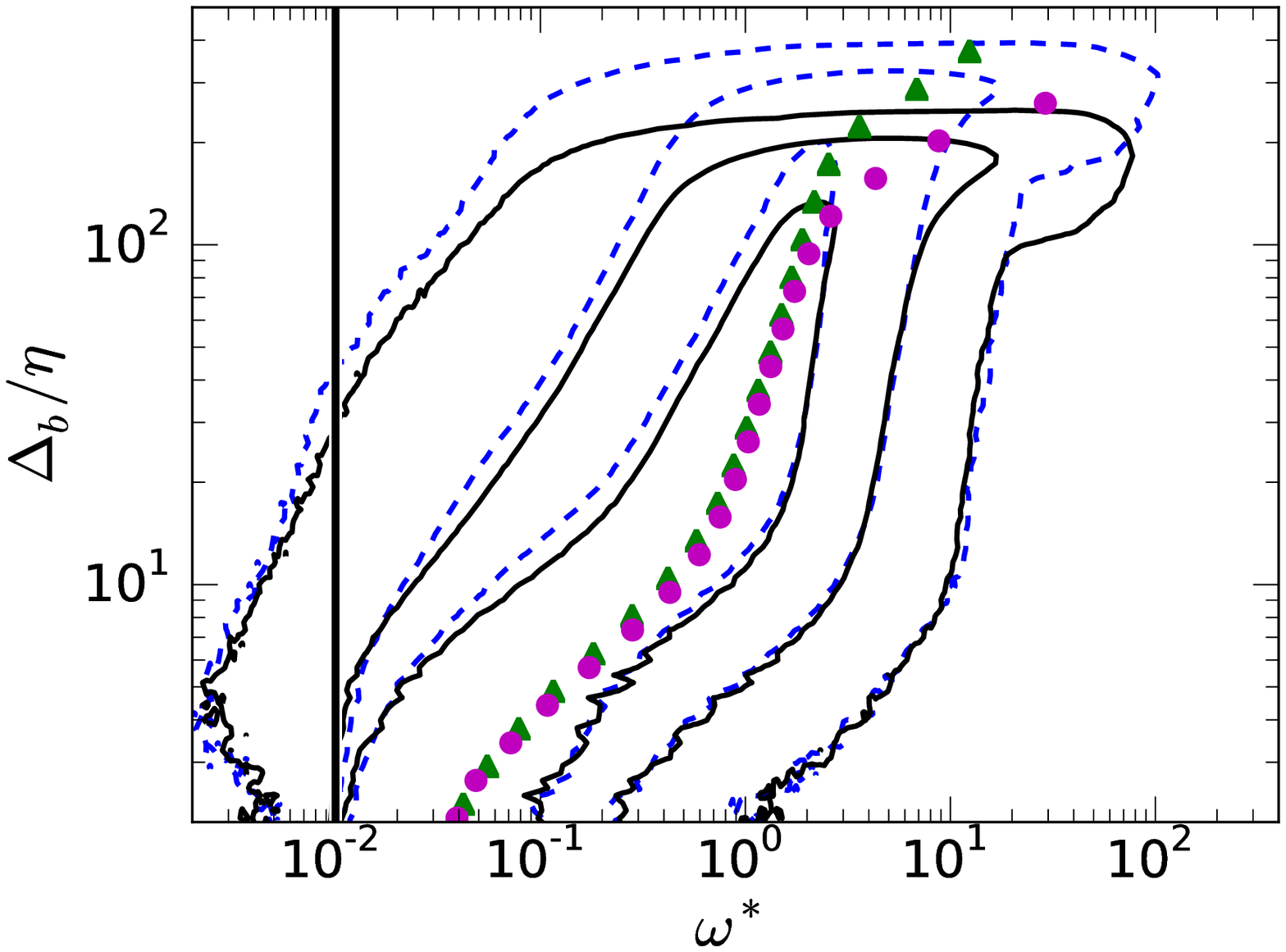}%
\mylab{-0.385\textwidth}{0.31\textwidth}{(a)}%
\hspace{2mm}%
\includegraphics[width=0.48\textwidth, clip=true,trim=0em 0em 4.5em 2.5em]{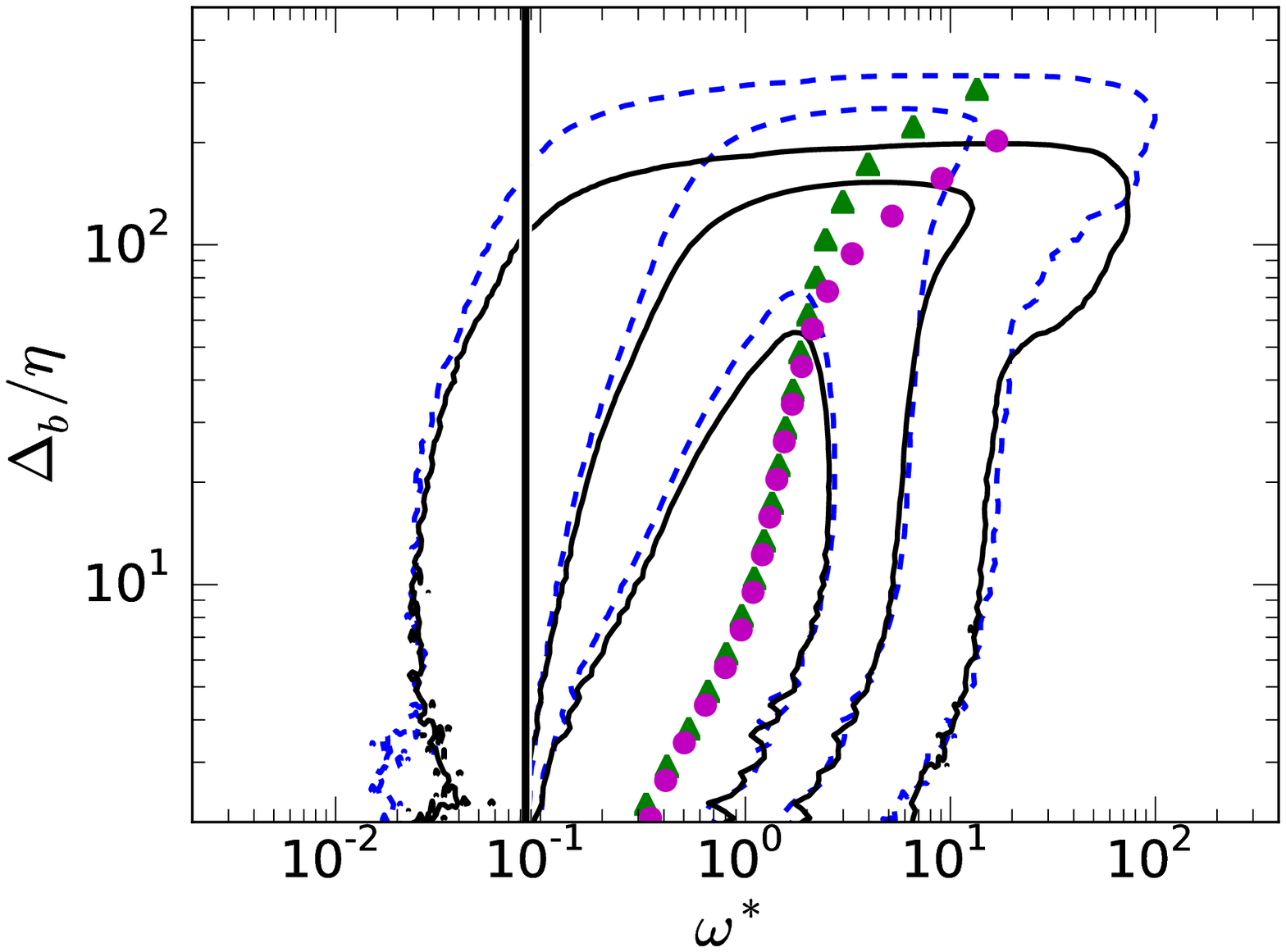}%
\mylab{-0.385\textwidth}{0.31\textwidth}{(b)}%
}
\vspace{1ex}%
\centerline{%
\includegraphics[width=0.48\textwidth, clip=true,trim=0em 0em 4.5em 2.5em]{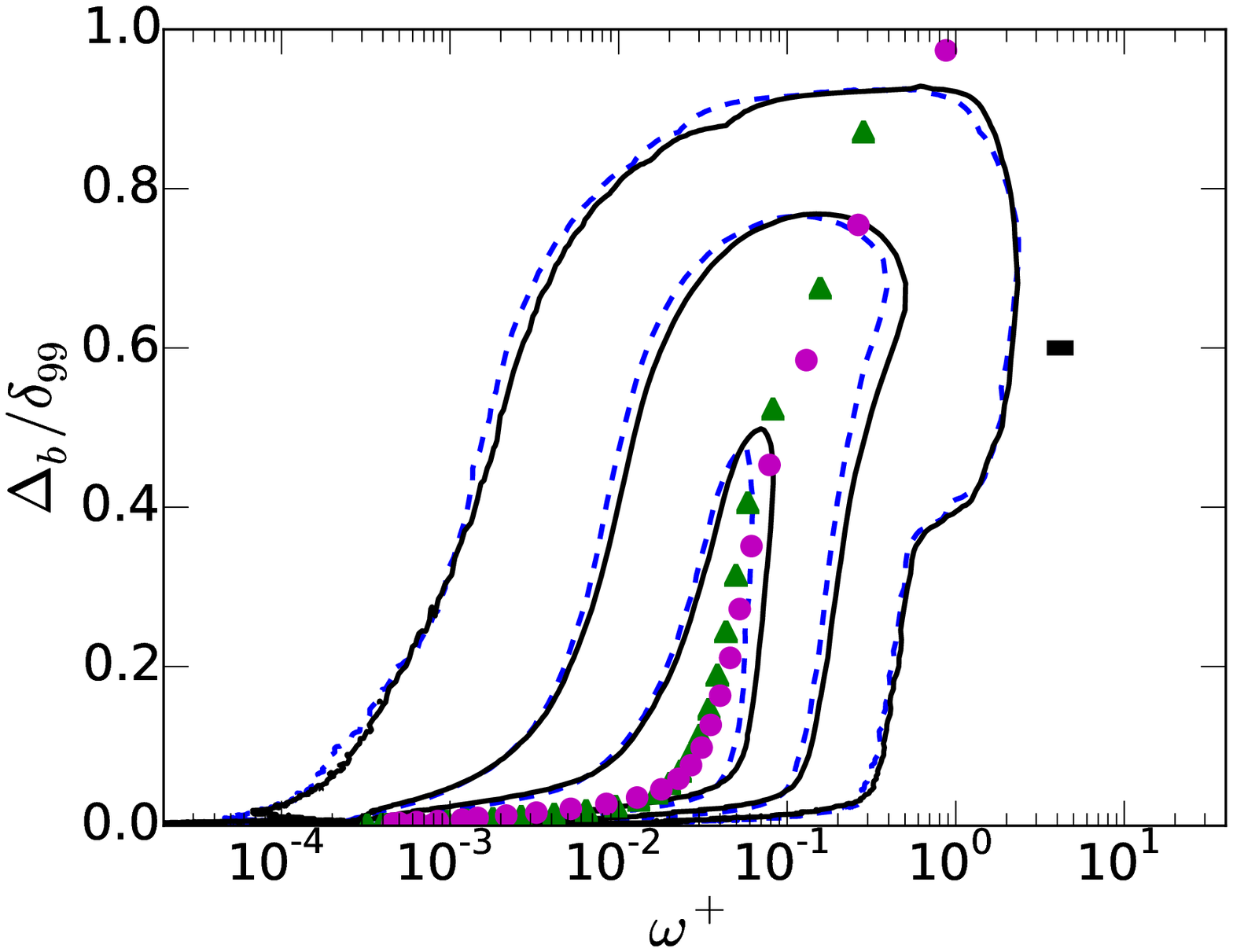}%
\mylab{-0.385\textwidth}{0.31\textwidth}{(c)}%
\hspace{2mm}%
\includegraphics[width=0.48\textwidth, clip=true,trim=0em 0em 4.5em 2.5em]{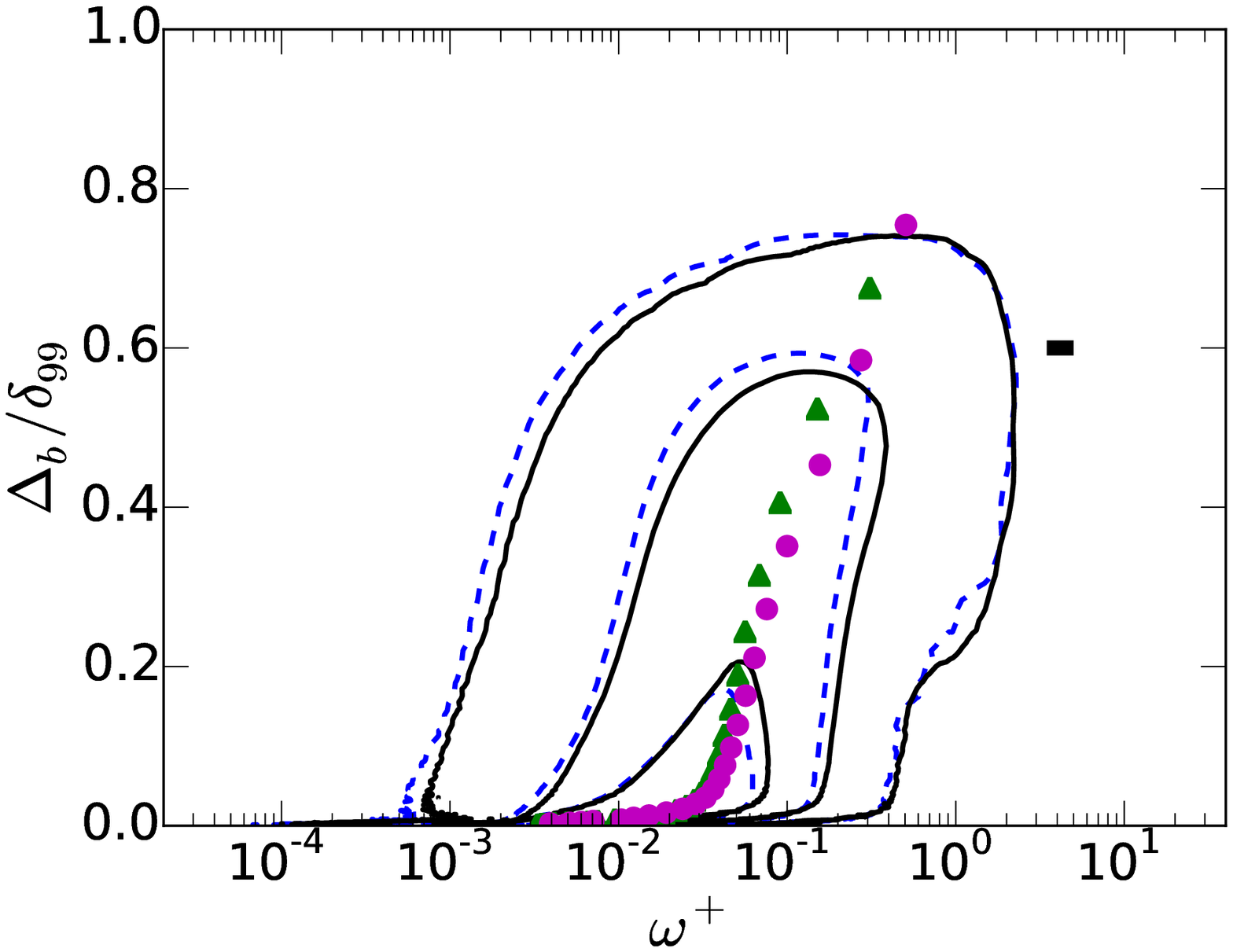}%
\mylab{-0.385\textwidth}{0.31\textwidth}{(d)}%
}
\caption{Premultiplied joint probability density function, $\omega
F_{\omega,\Delta_b}$, of the vorticity and ball distance in the turbulent side of
the interface for: (a) a low threshold, $\omega_0^*=0.01$, and (b) a moderate one at
the beginning of the topological transition, $\omega_0^*=0.09$. Two Reynolds numbers
are presented in each figure, $\delta_{99}^+=1100$ ($\solid$ black), and
$\delta_{99}^+=1900$ ($\dashed$ blue). The vertical solid line is $\omega^*_0$.
(c,d) Same as (a,b), but with the vorticity in wall units and the distance normalized
with the boundary layer thickness. The horizontal bar is the variation of
$\omega^*/\omega^+$ in our range of $\delta_{99}^+$.
The curves with markers correspond to the average vorticity magnitudes
for each Reynolds number, $\delta_{99}^+=1900$ (green $\vartriangle$),
and $\delta_{99}^+=1100$ (magenta $\circ$). Contours
contain 50\%, 90\%, and 99\% of points, respectively.  
}
\label{fig:short}
\end{figure}

While the previous sections deal with the properties of the interface {\em surface},
it is also interesting to characterise the properties of the interface {\em layer},
understood as the part of the turbulent flow that is directly influenced by its
proximity to the free stream. As a first step, figures \ref{fig:short}(a,b) reproduce
the first and second (turbulent) quadrants of the joint PDFs of the vorticity and
distance in figures \ref{fig:jointpdf}(a,b). The distance axis is now logarithmic, to
emphasize the region close to the interface, and each figure includes the two extreme
Reynolds numbers in our data set.
 
Three regions can be distinguished in order of increasing distance
from the interface. The first and closest to the interface contains
the strongest vorticity gradients. A precise definition of the
thickness of this layer will be given in \S\ref{sec:Thickness} but, if
we define the limit of this layer by the intersection of two straight
lines tangent to the probability isocontours near and far from the
interface, its thickness is of the order of a few Kolmogorov units for
the different Reynolds numbers, suggesting a viscous origin.  However,
this thickness depends on the identification threshold. It is
approximately $10\eta$ in figure \ref{fig:short}(a)
$(\omega_0^*=0.01)$, $5\eta$ in figure \ref{fig:short}(b)
$(\omega_0^*=0.09)$, and almost vanishes at the beginning of the
topological transition, $\omega_0^*=0.2$ (not shown). In the cases in
which this region can be identified in the joint PDF, its limit is
$\overline{\omega}^*\approx 1$, which we have seen above to be the
level of fully developed turbulence.

The viscosity-dominated region just outside the interface has been
recently studied by \citet{Ree:Holz:14} and \citet{Tav:Sil:PF14} in
temporally evolving turbulence fronts. They identify it with the
`superlayer' conjectured by \citet{NACA:1244}, and find that its
characteristic thickness is the Kolmogorov microscale computed with
the energy dissipation rate of the core flow. The enstrophy level in
this viscous layer depends somewhat on the definition, but is
typically very low. The viscous region in figure \ref{fig:short} is
probably not the superlayer, whose observation requires a higher
numerical resolution and a quieter free stream than those in our
simulation \citep{Ree:Holz:14}. We will see below that both the rate
of strain and the vortex stretching remain high in the viscous layer
of figure \ref{fig:short}, and that that region is probably best
interpreted as part of the `buffer layer' defined by
\cite{Ree:Holz:14} in the range $\omega^*\in (0.1-1)$. In analogy to
the similarly named layer in wall-bounded turbulence, both nonlinear
and viscous effect are important that region. It is interesting that
such a hybrid mechanism was proposed by \cite{TownsendBook}, who noted
that viscous diffusion of vorticity and its tangential transport
should be comparable near the interface. On the assumption of
homegeneity, the magnitude of the rate of strain is proportional to
the enstrophy, and the predicted result of this mechanism is also a
thickness $O(\eta)$.

The region beyond the buffer interface layer is self similar, in the
sense that both the conditionally averaged vorticity and the
probability isocontours follow power laws in $\Delta_b$. That implies
that there is no intrinsic length or vorticity scale, which is
consistent with the intuitive idea of an inertial layer but difficult
to interpret dynamically. In the first place, different parts of the
PDF scale differently. The strong-vorticity isocontours to the right
of figure \ref{fig:short}(a) are almost vertical $(\omega\propto
\Delta_b^0)$, but those corresponding to weak vorticity on the left of
the figure follow $\omega\propto \Delta_b^1$, and the conditional mean
enstrophy approximately satisfies $\omega\propto \Delta_b^{1/2}$. In
the second place, those slopes change with the detection threshold,
and it is hard to distinguish any power law in figure
\ref{fig:short}(b) or at higher thresholds. We mentioned in
\r{eq:omp}--\r{eq:epsilon_y} that $\bra\omega\ket \propto y^{-1/2}$ is
a consequence of the local energy equilibrium and the logarithmic
profile in the logarithmic layer, but the same argument cannot be used
here. The interface is not an impermeable boundary that limits the
size of eddies as the wall does, although it could be argued that the
size of the eddies defines the position of the interface. The trend in
figure \ref{fig:short}(a), that larger eddies have more intense
vorticity, is contrary to the inertial relation of homogeneous
turbulence, $\omega \propto \Delta^{-2/3}$ \citep{kol41}, and the most
plausible explanation is that larger eddies reach closer to the wall
and are therefore stronger. The apparent self-similarity in figure
\ref{fig:short} may be coincidental.

The width of this intermediate region depends on the identification threshold, but
scales with the boundary layer thickness. It extends to the hockey-stick at the top
of the PDFs, which contains the points with the highest vorticity and farthest from
the interface. This last region is mostly formed by points near the wall. When
$\omega$ is scaled in wall units and $\Delta_b$ is normalized with the boundary-layer
thickness, as in figures \ref{fig:short}(c, d) the two Reynolds numbers collapse well
for long distances and high vorticities. At the two Reynolds numbers in figure
\ref{fig:short}, $\delta_{99}/\eta \approx 250$ and 450, respectively.

\subsection{Other velocity gradients}\la{sec:Strain}

\begin{figure}
\centering
\vspace*{5mm}
  \includegraphics[width=0.55\textwidth, clip=false, 
  trim=1em 1em 1em 4em]{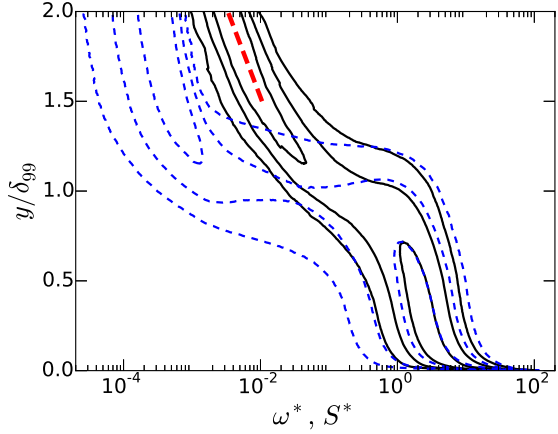}
\caption{\label{fig:strainscaling} Premultiplied joint PDFs: \solid (black), $S
\Gamma_{y,S}$; \dashed (blue), $\omega \Gamma_{y,\omega}$. Contours contain 50\%,
90\%, and 99\% of points, respectively. The dashed diagonal is the exponential decay
of the Fourier modes of irrotational strain with a wall-parallel wavelength
$2\delta_{99}$. 
}
\end{figure}

In the previous sections, we have discussed the properties of the vorticity field
near a vorticity isosurface, and it is perhaps not surprising that they may be
special. For example, an interesting question is whether the vorticity within the
interface layer has different properties from the core of the turbulent flow, such as
perhaps being weaker because it is less strained, but such questions are hard to
answer if the interface is defined by the magnitude of the vorticity itself. It is
useful for that purpose to determine the conditional properties of other quantities
besides the one being thresholded. In this section we study the properties of the
strain rate tensor $\Smat$ in the neighbourhood of the vorticity interface, as well
as the behaviour of the vorticity in the neighbourhood of an interface defined in
terms of the strain. Define $S$ as the euclidean norm of the rate-of-strain tensor,
$S=\|\Smat\|$. In analogy to equation \eqref{eq:vortouter}, and taking into account
that
\begin{equation}
  \label{eq:Somega}
  \bra\omega^2\ket = 2\bra S^2\ket
\end{equation}
in homogeneous flows, the star units  for the $S$ are defined as
\begin{equation}
  \label{eq:starstrain}
  S^* = S \frac{\nu \sqrt{2 \delta_{99}^+}}{u_\tau^2}.
\end{equation}
Equation \r{eq:Somega} then becomes $\bra\omega^2\ket^* = \bra S^2\ket^*$, and 
suggests that $\omega^*$ and $S^*$ should be of the same order.
The joint PDFs of $S$ and $y$, and of $\omega$ and $y$, are presented in figure
\ref{fig:strainscaling}.

Both PDFs agree within the turbulent region in the right-lower corner of figure
\ref{fig:strainscaling}, supporting the normalisation \eqref{eq:starstrain}, but the
vorticity in the free stream on the left-hand side of the figure is almost two orders of
magnitude lower than the rate of strain. This is not unexpected in a nominally
irrotational part of the flow but, since \eqref{eq:Somega} is a kinematic relation whose
only condition is spacial homogeneity, the mismatch between the two magnitudes implies
that the strain in the free stream is an inhomogeneous residual effect of the vortical
flow within the boundary layer.

\begin{figure}
\centerline{%
\includegraphics[width=0.48\textwidth, clip=true,trim=1em 1em 1em 2em]{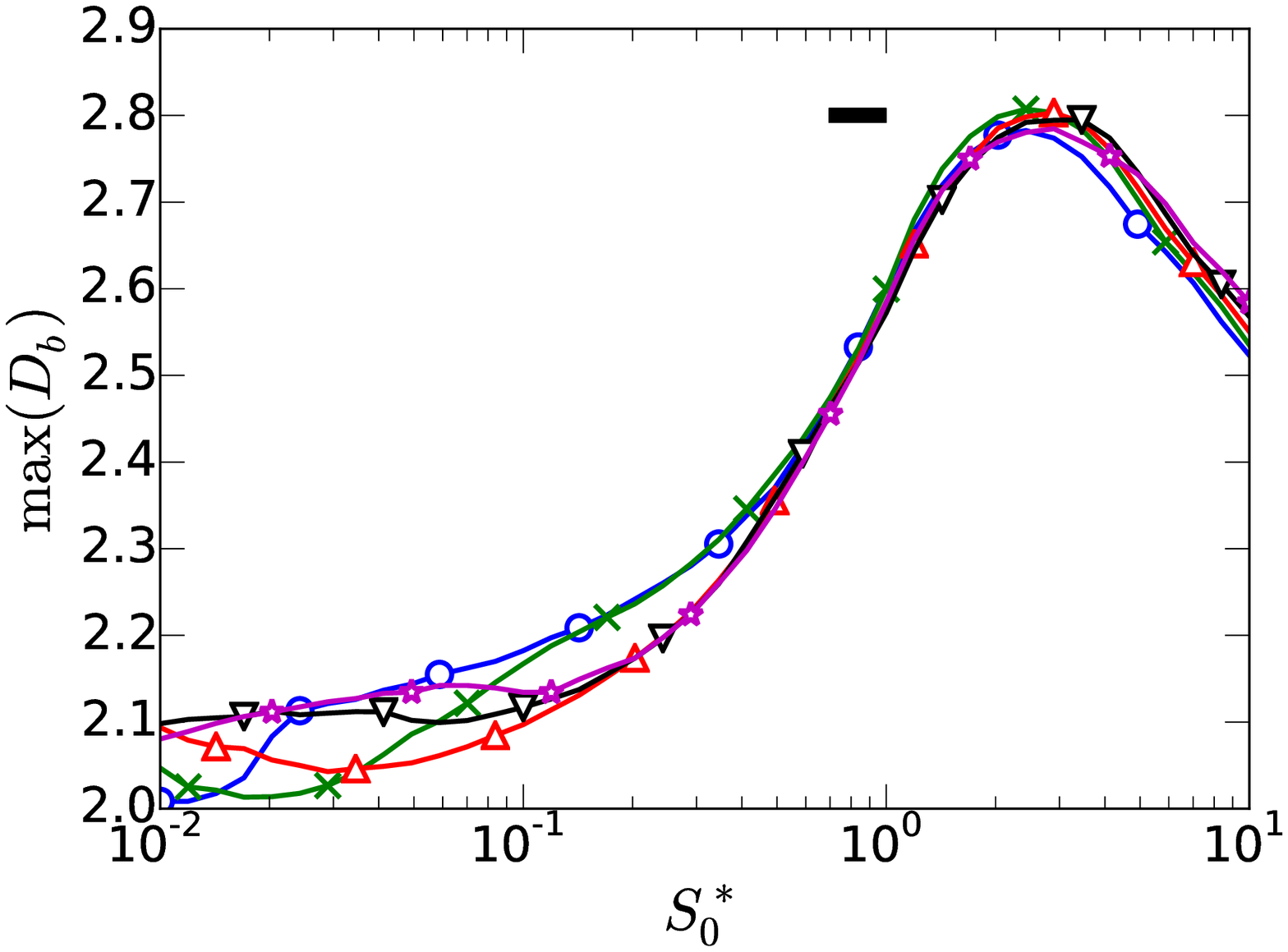}%
\mylab{-0.40\textwidth}{0.29\textwidth}{(a)}%
\hspace{2mm}%
\includegraphics[width=0.48\textwidth, clip=true,trim=1em 1em 1em 2em]{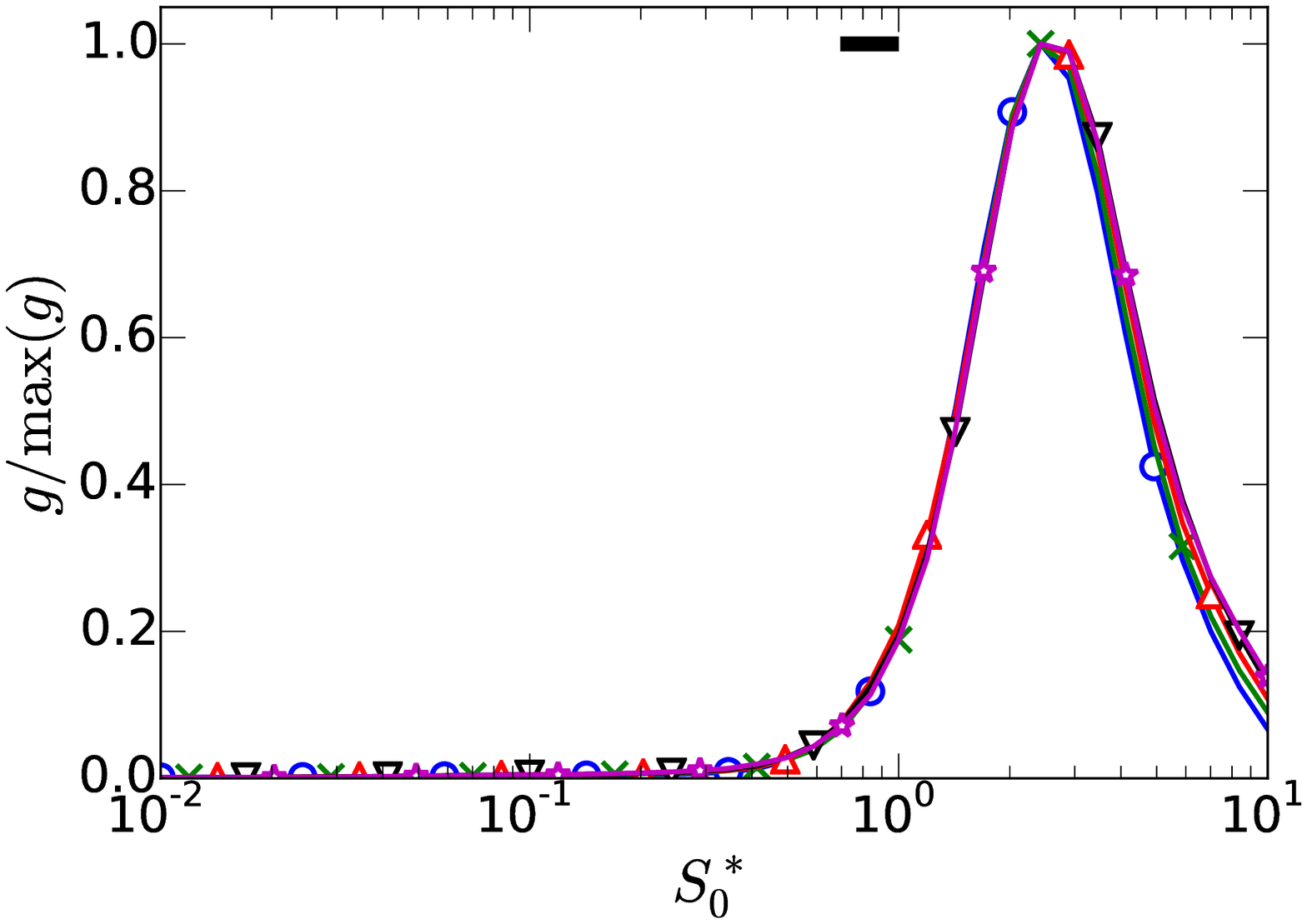}%
\mylab{-0.40\textwidth}{0.29\textwidth}{(b)}%
}
\caption{\label{fig:strain_surface} (a) Fractal dimension and (b)
  genus of an interface defined by thresholding the norm of the
  strain-rate tensor. $\circ$, $\delta_{99}^+=1100$; $\times$, 1300;
  $\triangledown$, 1500; $\vartriangle$, 1700; and $\star$, 1900. The
  horizontal bar is the variation of $\omega^*/\omega^+$ in our range
  of $\delta_{99}^+$.  }
\end{figure}

Any solenoidal velocity field can be written as
\beq
\uvec=\bnabla\wedge \Bvec + \nabla \phi,
\la{eq:potential}
\eeq
where the potentials satisfy, $\nabla^2\phi=0$ and $\bnabla^2
\Bvec=-\omvec$ \citep{bat67}.  In the irrotational free stream, both
potentials satisfy Laplace's equation and, if they are expanded in
terms of wall-parallel Fourier harmonics, decay away from the wall as
$\exp(-k y)$, where $k^2=k_x^2+k_z^2$ is the magnitude of the
wall-parallel wave vector. All the velocity components and the
rate-of-strain tensor decay exponentially at the same rate, and the
slowest decay corresponds to the largest horizontal wavelengths. It is
known that this results in an algebraic decay of the velocity
fluctuations for $y\gg\delta_{99}$, because different distances are
dominated by different wavenumbers in the long-wavelength end of the
spectrum \citep{phillips1955irrotational,stewart1956irrotational}, but
the near field is controlled by the peak of the $v$ spectrum in the
turbulent region. The thick dashed diagonal in figure
\ref{fig:strainscaling} is $S\propto \exp(-\pi y/\delta_{99})$,
corresponding to the decay of irrotational velocity fluctuations due
to structures within the boundary layer whose shortest dimension is
$O(2\delta_{99})$. This is the order of magnitude of the largest
structures in boundary layers \citep{sillero14}.

The vorticity is unrelated to the velocity potentials, and decays much
faster than the rate of strain as it enters the free stream. In fact,
this was one of the reasons why we originally chose vorticity over
other quantities to characterise the T/NT interface.

Note that the vorticity also decays exponentially with $y$ in the free
stream, although at a much lower absolute level than $S$. This is not
a kinematic result, but a consequence of the numerical inflow
conditions, which determine the three velocities at the inflow but not
their derivatives with respect to $x$. The result is that there is a
residual vorticity in the free stream due to terms like $\partial_x v$, which
inherits the exponential decay of the velocity potentials at the
inflow plane.

A consequence of the relatively high strain levels in the free-stream
is that the separation between its characteristic values in the
turbulent and the non-turbulent sides is not as clear-cut as in the
case of the vorticity. Even so, the complexity transition happens at
comparable thresholds. The geometrical properties of the strain
interface are presented in figure \ref{fig:strain_surface}. The
fractal dimension in figure \ref{fig:strain_surface}(a) should be
compared to figure \ref{fig:fdimension}(b) for the vorticity
interface. The strain isosurface is smoother, with minimum values
close to the non-fractal value, $D\approx 2$. The maximum dimension is
also somewhat lower than for the vorticity, in agreement with the
observation by \cite{MoisyJimenez} that strong dissipation structures
are less fractal (plate-like) than those of vorticity
(string-like). The evolution of the genus in figure
\ref{fig:strain_surface}(b) is also similar to the case of the
vorticity, although the maximum genus and fractal dimension are
reached for slightly higher thresholds, $S_0^*\approx 2.5$ instead of
$\omega_0^*\approx 1.5$. The topological transition is also narrower
for the strain interface, especially for the genus in figure
\ref{fig:strain_surface}(b), which starts to increase at $S_0^*\approx
1$ instead of at $\omega_0^*\approx 0.3$, as it did in figure
\ref{fig:genus}(a). The reason is probably that, while the maximum
dimension and genus mark the threshold for which the interface has
fully moved into the core turbulent flow, the slower decay of the
strain fluctuations with $y$ means that low-strain isosurfaces are
farther from the wall than for similar enstrophy thresholds, and the
corresponding interfaces becomes regular much faster.

\begin{figure}
\centering
\includegraphics[width=0.6\textwidth, clip=true, trim=0em 0em 0em
0em]{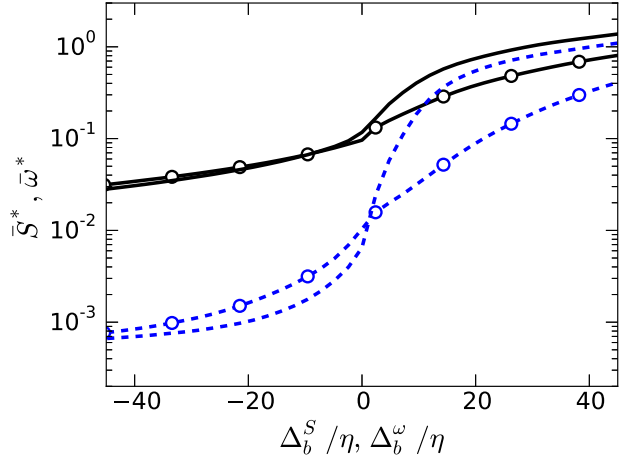}
\caption{
  strain, as functions of the distance $\Delta_b^\omega$ to the
  $\omega_0^*=0.01$ interface, or of the distance $\Delta_b^S$ to
  $S_0^*=0.1$. $\delta_{99}^+=1900$. \solid, $\overline{S}$; \dashed,
  $\overline{\omega}$. Lines without symbols are with respect to
  $\Delta_b^\omega$.  Those with symbols are with respect to
  $\Delta_b^S$.
}
\label{fig:interfaces}
\end{figure}

The evolution of the conditional statistics of the flow across the
vorticity and strain isosurfaces are compared in figure
\ref{fig:interfaces}. The two thresholds chosen are $S_0^*=0.1$ and
$\omega_0^*=0.01$, both of which are within the plateau that separates
the values of turbulent and non-turbulent flow in their joint PDFs,
and well below the beginning of the respective topological
transitions. The average height of the resulting interfaces is $\bra
y_I\ket \approx 1$ in both cases. Figure \ref{fig:interfaces} shows
the conditionally averaged enstrophy and strain for each of the two
interfaces. They are plotted as functions of the respective ball
distances, which are denoted by $\Delta_b^\omega$ and $\Delta_b^S$,
respectively.

The interesting question to be answered is whether the sharp vorticity
gradient across the vorticity interface is a statistical artefact of
the thresholding procedure, or a true physical effect. The former is a
possibility, because enstrophy is fixed at the interface, and moving
slightly away from the geometrically complex isosurface could sample
flow regions that are unrelated to it and representative of the bulk
of the turbulent and irrotational regions. In a related example,
\citet{Chauhan:14} find a sharp velocity jump across an interface
defined in terms of the velocity magnitude, raising similar
questions. In both cases, the sharp jump at the interface is what
makes the criterion useful, and the reason that an interface can be
defined at all \citep{NACA:1244}. The lines without symbols in figure
\ref{fig:interfaces} refer to the vorticity interface. Within
$\Delta_b =O(10\eta)$ of $\Delta_b^\omega=0$ the conditional vorticity
(dashed) drops by three orders of magnitude, and a similarly sharp
gradient is seen for the conditional strain (solid).

The behaviour is different for the strain interface, represented by the lines with
circles in figure \ref{fig:interfaces}, whose vorticity and strain cross
relatively smoothly the level $\Delta_b^S=0$. The difference between the two
behaviours strongly suggests that while a sharp vorticity jump is a dynamically
significant feature separating distinct regions of the flow, that of the strain is
not. 

We are now ready to define a `natural' interface as an enstrophy
isosurface below the topological transition, such as
$\omega_0^*\approx 0.01$. This threshold is somewhat lower than most
of those compiled in table \ref{table:gamma}, with the result that the
turbulent region contains part of the buffer and viscous superlayers
defined by \citet{Ree:Holz:14}.

\begin{figure}
\centerline{%
\includegraphics[width=0.48\textwidth, clip=true,trim=1em 1em 1em 0em]{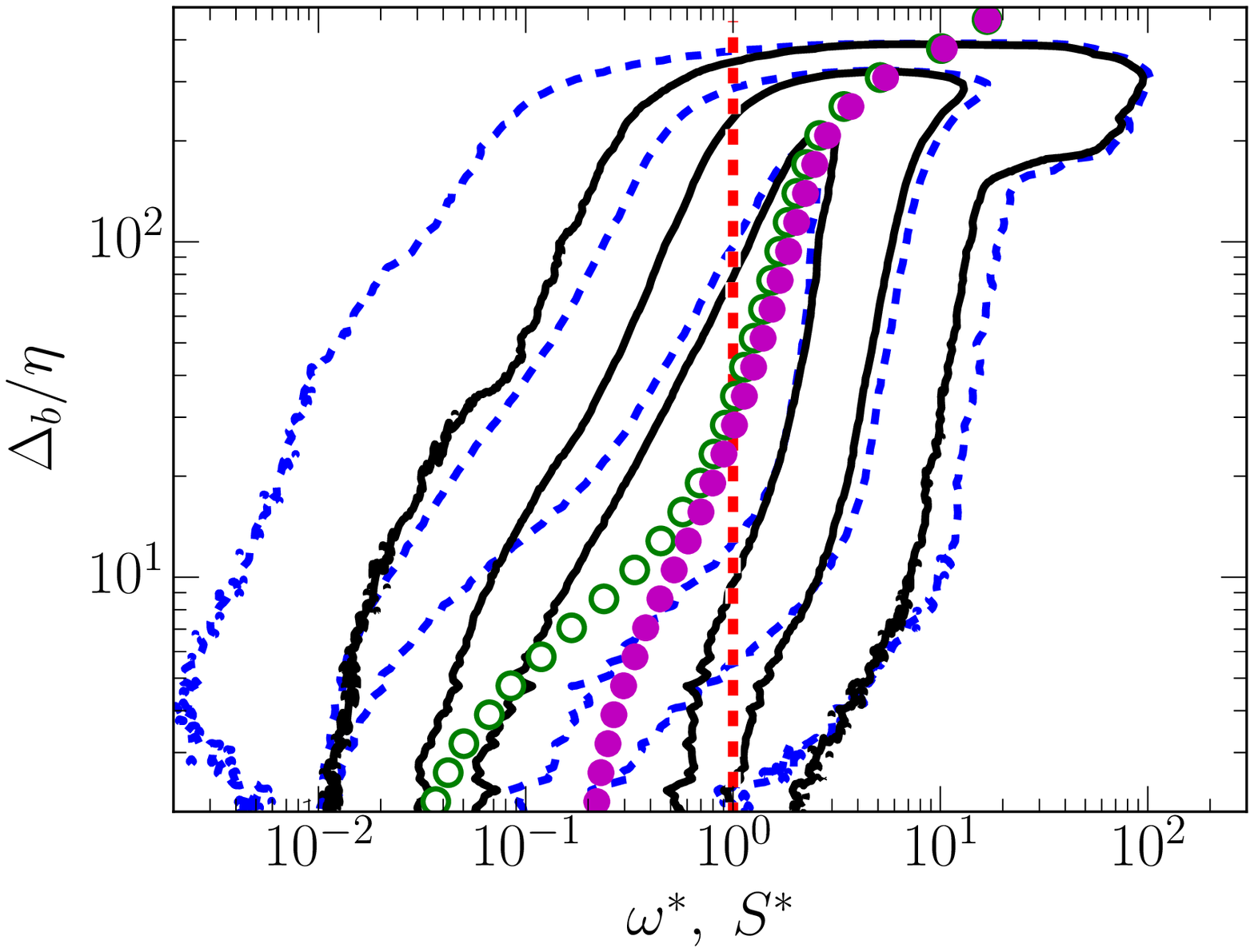}%
\mylab{-0.37\textwidth}{0.29\textwidth}{(a)}%
\hspace{2mm}%
\includegraphics[width=0.48\textwidth, clip=true,trim=1em 1em 1em 0em]{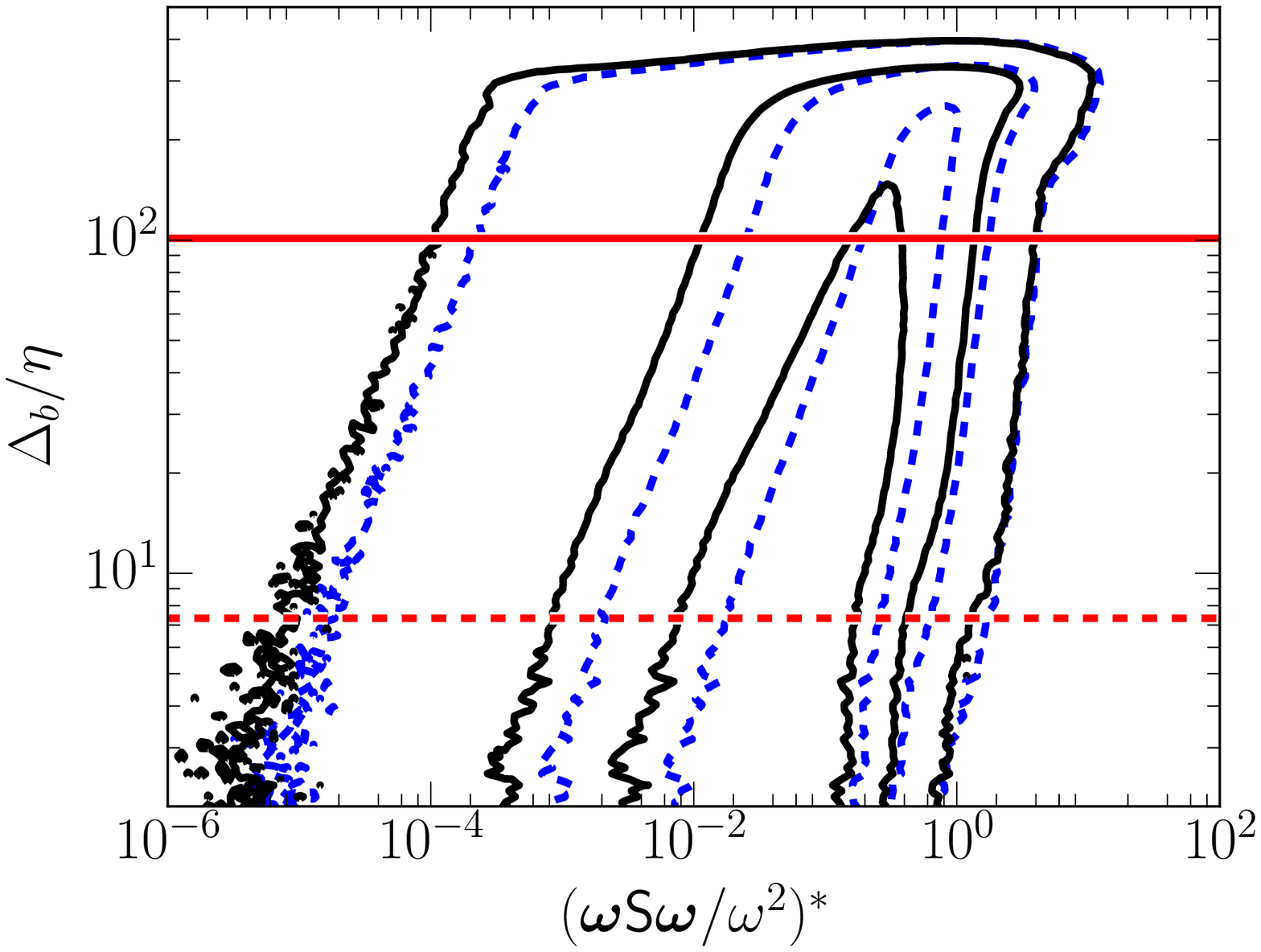}%
\mylab{-0.37\textwidth}{0.29\textwidth}{(b)}%
}
\vspace{1ex}%
\centerline{%
\includegraphics[width=0.48\textwidth, clip=true,trim=1em 1em 1em 2em]{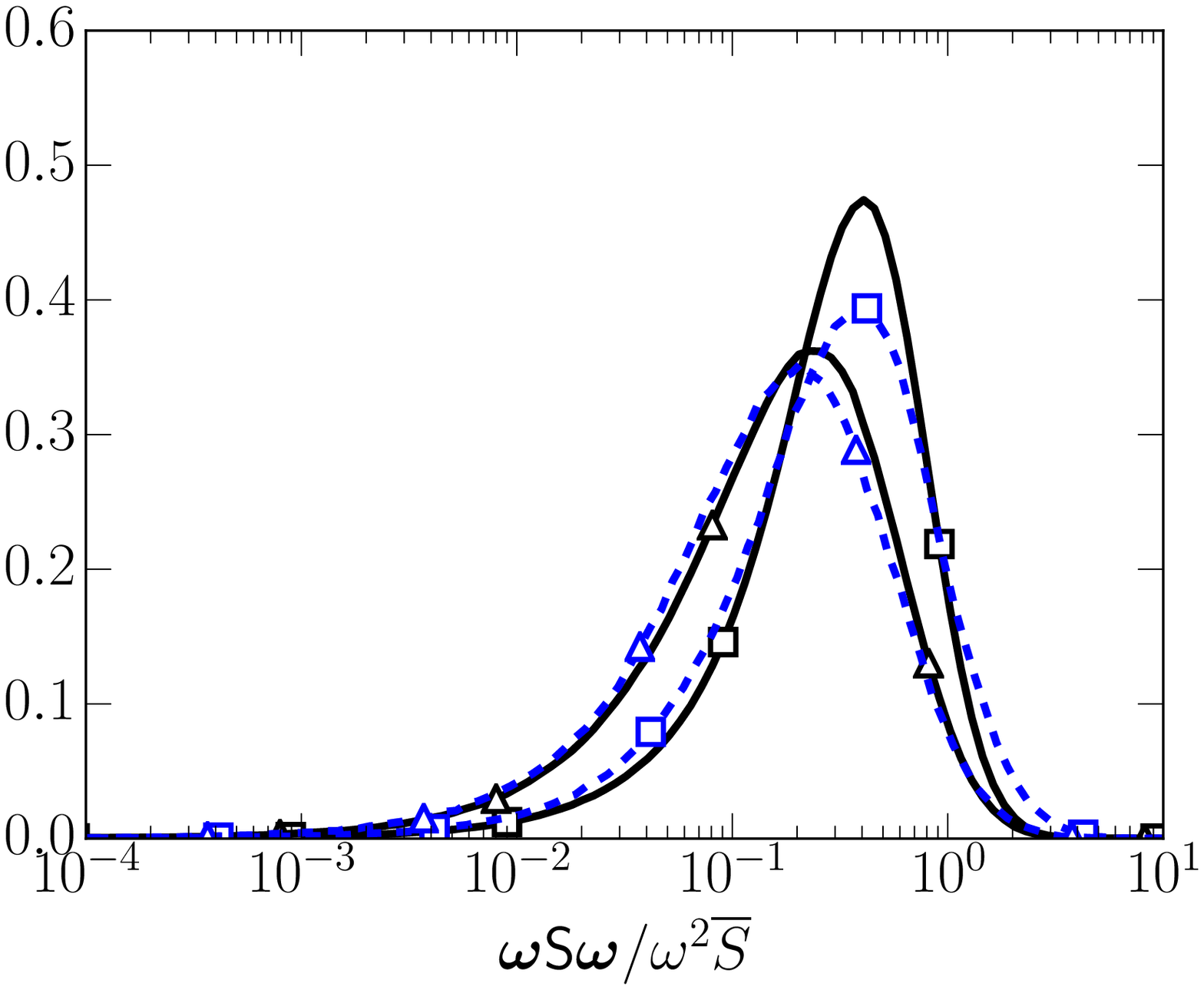}%
\mylab{-0.37\textwidth}{0.29\textwidth}{(c)}%
\hspace{2mm}%
\includegraphics[width=0.48\textwidth, clip=true,trim=1em 1em 1em 2em]{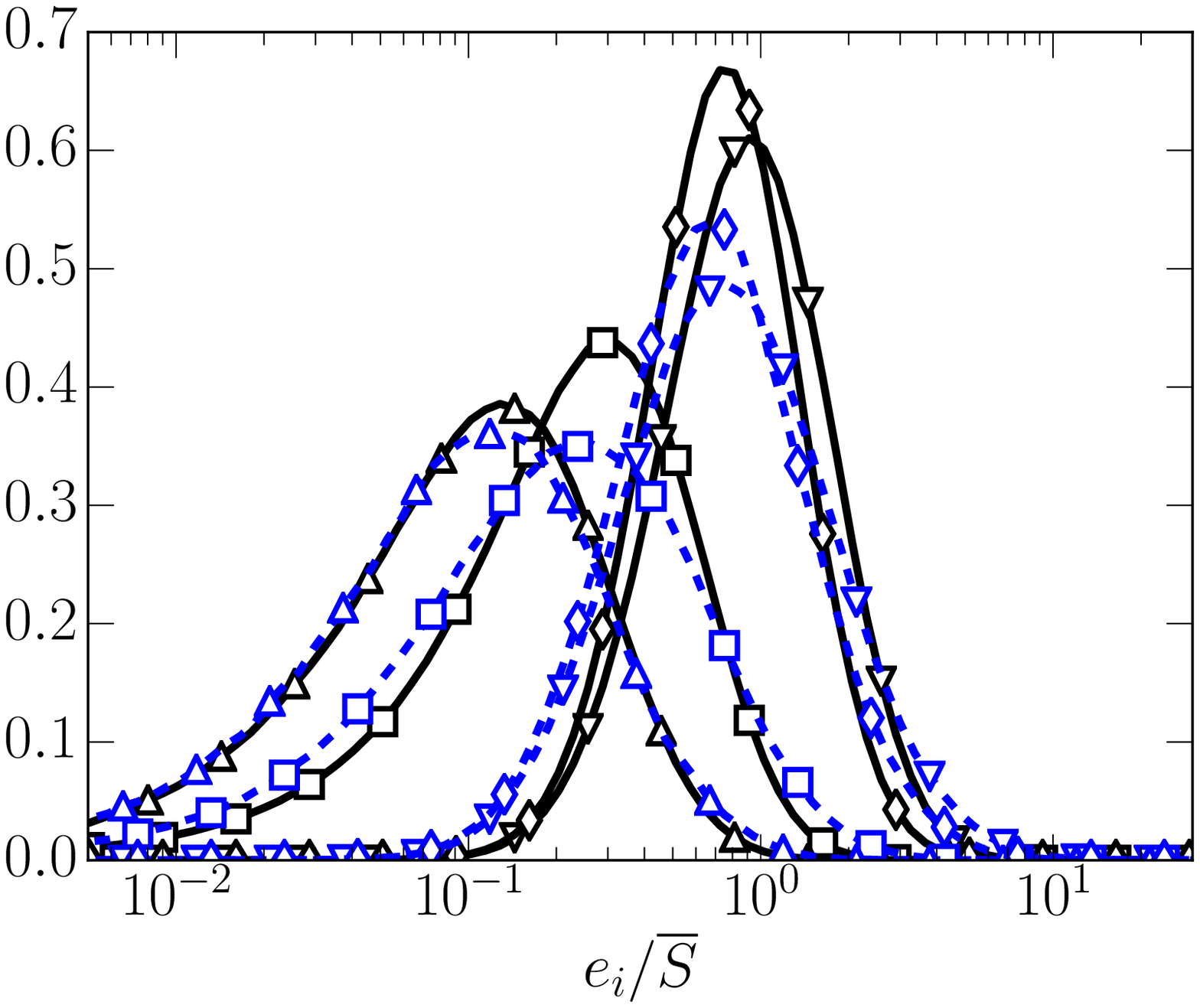}%
\mylab{-0.37\textwidth}{0.29\textwidth}{(d)}%
}
\caption{%
(a) Premultiplied joint PDFs at $\delta_{99}^+=1900$, with respect to the interface $\omega_0^*=0.01$:  
\dashed (blue), $\omega F_{\omega,\Delta_b}$; \solid (black) $SF_{S,\Delta_b}$.
Symbols are: \circle, $\overline \omega^*$; \solidcircle, 
$\overline S^*$. The vertical line is $\omega^*=S^*=1$.
(b) Premultiplied joint PDF of $\Delta_b$ and: \dashed (blue), vortex
stretching; \solid (black), vortex compression.  The two horizontal
lines: \solid, $\Delta_b=100\eta$; \dashed, $\Delta_b=7\eta$ are used
in (c,d).  The three contours in (a,b) contain 50\%, 90\%, and 99\% of
points, respectively.
(c) Premultiplied PDF of: $\vartriangle$, normalised vortex
stretching; $\square$, compression.
(d) Premultiplied PDFs of the absolute values of the normalised
individual eigenvalues of the rate-of-strain tensor: $\diamond$, most
positive; $\triangledown$, most negative; $\vartriangle$, positive
intermediate; $\square$, negative intermediate.
The abscissae in (c,d) are normalised with $\overline S$. 
In both cases, the PDFs are compiled at: \dashed, $\Delta_b=7\eta$;
\solid, $\Delta_b=100\eta$.  
  }
\label{fig:strain}
\end{figure}

The structure of the flow with respect to this interface is displayed
in figure \ref{fig:strain}. The joint PDFs of the vorticity and rate
of strain with $\Delta_b$ are shown in figure \ref{fig:strain}(a). The
sharp decay of the vorticity below $\overline{\omega}^*\approx 1$ is
clearly visible, spanning a thickness of about $20\eta$.  The strain
decays slowly as it gets closer to the interface and far from the
wall, but shows no especial behaviour within the buffer region. The
vorticity in the buffer layer lives in essentially the same straining
environment as in the core turbulent flow.
 
This is seen more clearly in figure \ref{fig:strain}(b) which shows
the PDF of the vortex stretching component of the strain,
$\omvec\Smat\omvec/\omega^2$. The positive (stretching) and negative
(compression) PDFs are plotted separately to allow a logarithmic
representation. Both decays slowly and apparently self-similarly as
they approach the interface, but do not change appreciably as they
enter the buffer layer. The different rate of decay of the vorticity
and the strain rate was also mentioned by
\cite{holzner2007small}. Figure \ref{fig:strain}(c) shows the same
result in the form of one-dimensional PDFs of the vortex-stretching
term at two distances from the interface, one within the buffer region
and another one in the core of the flow. The normalisation with
$\overline S$ absorbs most of the differences between the two
levels. An even more detailed comparison is figure
\ref{fig:strain}(d), which shows the PDFs of the individual
eigenvalues of the rate of strain tensor. The PDFs at the two
distances also collapse well. In both cases, the implication is that
the straining environment within the buffer layer is essentially the
same as in the core of the flow. Enstrophy is viscously diffused into
the free stream, but it keeps being stretched as it does, in agreement
with the model proposed in \citet{TownsendBook}. Note that the
magnitude of the vorticity within this inhomogeneous is of the order
of isosurfaces that were shown in \S\ref{sec:Geometry} to be within
the topological transition, and that the geometry of the flow in this
layer is therefore intermediate between the irrotational free stream
and the turbulent core, but much more complex than the former.

Other authors have explored higher order quantities close to the
interface, such as the invariants of the velocity gradient tensor
\citep{da2008invariants}, and the different terms of the vorticity
equation \citep{holzner2007small, FLM:1757608}. The study of those
quantities for the present boundary layer are unfortunately beyond the
scope of this study, but our data are openly accessible from our web
site, and interested researchers are encouraged to use them to test
their ideas.

\subsection{\label{sec:Thickness}The thickness of the interface layer.}

We have normalised our lengths up to now in terms of $\eta$, $\lambda$
or $\delta_{99}$, according to which of those scales appears to
collapse better the different Reynolds numbers in each particular
figure, or arbitrarily in figures involving a single Reynolds
number. We saw in the introduction that the thickness of the T/NT
interface layer has been the subject of much discussion, and we
mentioned that comparisons are difficult because of the variety of
definitions used by investigators. In general, there is some consensus
that the properties of the flow change across the interface over
distances of the order of the Taylor microscale, even if the narrow
range of Reynolds numbers makes the scaling ambiguous in some
cases. However, \cite{GampertRetau} showed that the thickness of the
mixing interface of a passive scalar in a jet scales with $\lambda$
over a range of $Re_\lambda$ somewhat wider than ours. This result is
surprising to us, because there are relatively few examples in which
the Taylor microscale appears in fully developed turbulence \citep[see
  however the correlation length along the strong vortices of
  isotropic turbulence in][]{JimenezWray98}.  However, the T/NT
interface is not fully developed turbulence, and \cite{HuntDurbin} and
\cite{hunt2006mechanics} have given theoretical arguments as to why
the thickness of strong vortex layers within a turbulent flow should
scale with $\lambda$. They propose that the interface is one such
layer \citep[see also][]{eisma2015interfaces}. Here we examine the
scaling of the thickness of our interface, defined as a layer in which
the enstrophy and rate of strain do not satisfy the homogeneity
constraints. We will find that the thickness scales with the Taylor
microscale, but it is unclear whether the reasons are those in
\cite{HuntDurbin}. They argue that strong shear layers are only
subject to the rate of strain of the largest turbulent scales because
the smaller ones are excluded by the shear, and that their thickness
is controlled by viscosity. The question of whether there is an
effectively high shear at the interface will be discussed in the next
section, but we have seen above that the rate of strain and the
stretching eigenvalues within the interface layer are similar to those
in the bulk of the flow, which would imply viscous lengths of the
order of $\eta$ in an equilibrium flow. We also saw that the geometry
of the vorticity within that layer is complex, and not immediately
consistent with an equilibrium viscous process. Our interface layer is
probably not the same one analysed by \cite{HuntDurbin}.  Other
dynamical models reach different conclusions about the scaling of the
thickness of the interface layer starting from different
assumptions. For example, \citet{teixeira2012turbulence} show that the
initial decay of a shear-free synthetic turbulent interface should
have thickness of order $\eta$.

\begin{figure}
\centerline{%
\includegraphics[width=0.48\textwidth]{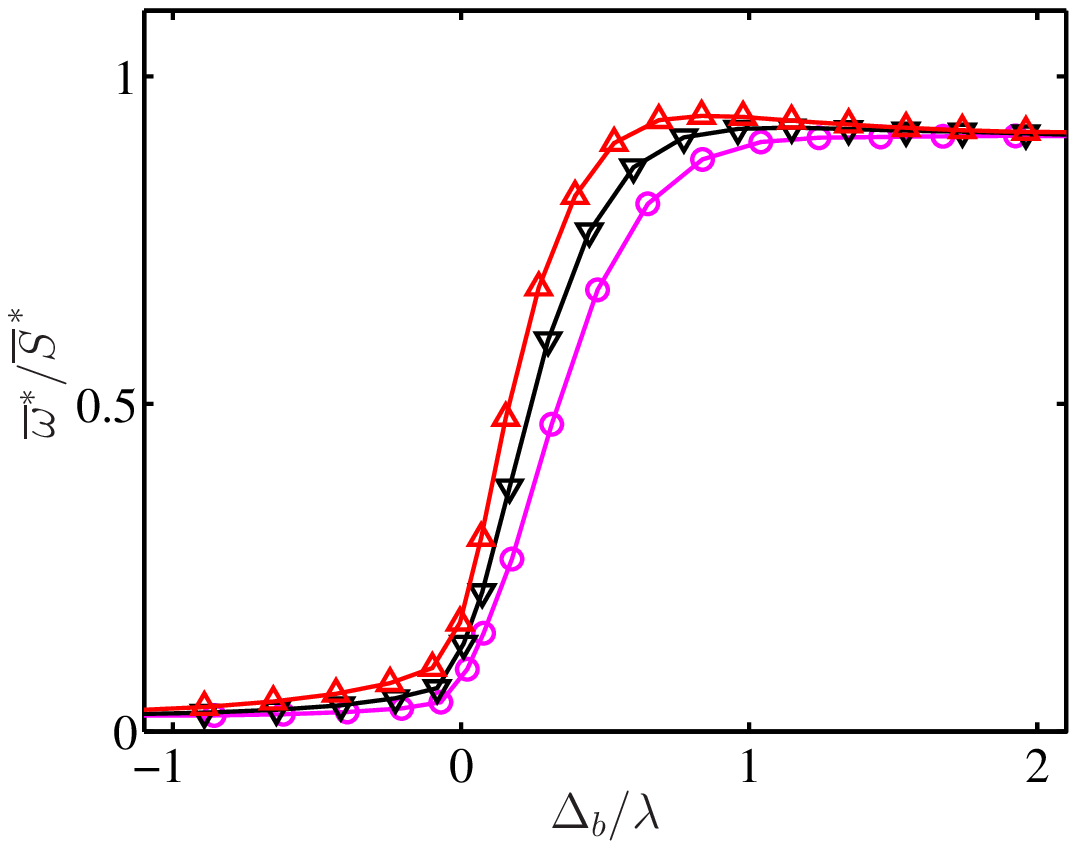}%
\mylab{-0.39\textwidth}{0.30\textwidth}{(a)}%
\hspace{4mm}%
\includegraphics[width=0.46\textwidth]{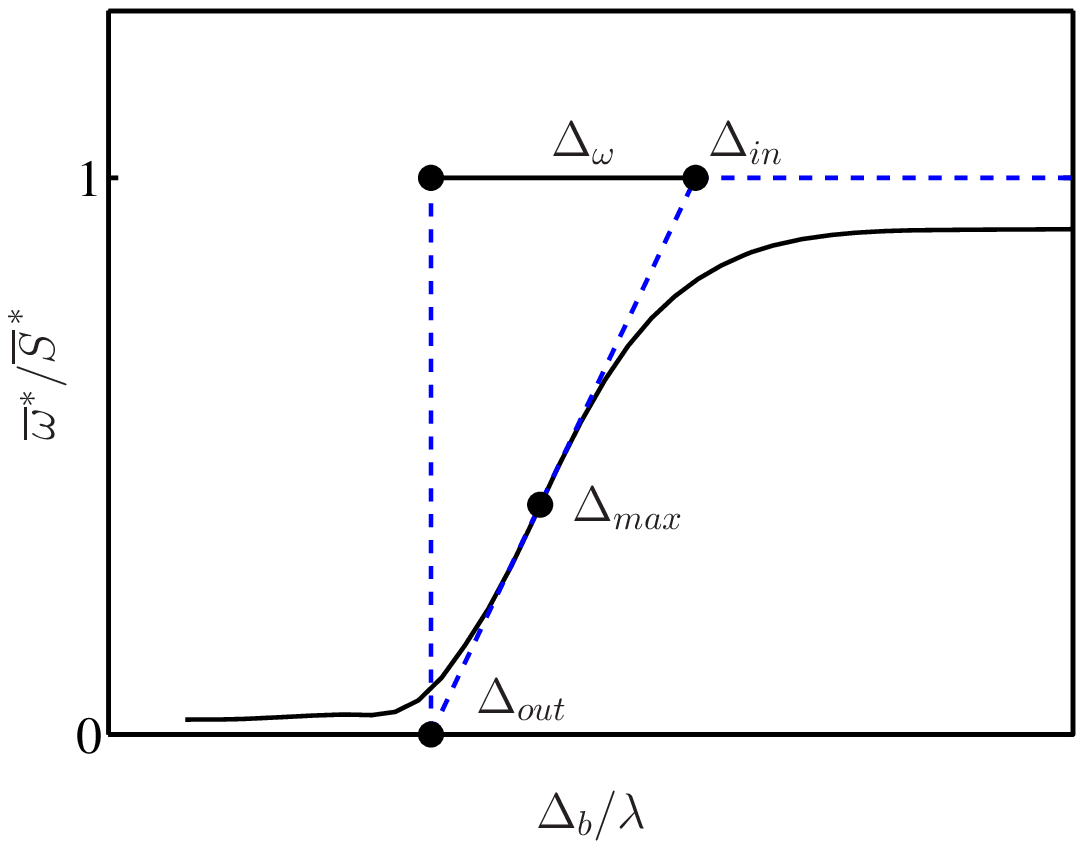}%
\mylab{-0.39\textwidth}{0.30\textwidth}{(b)}%
}
\vspace{1ex}%
\centerline{%
\includegraphics[width=0.48\textwidth]{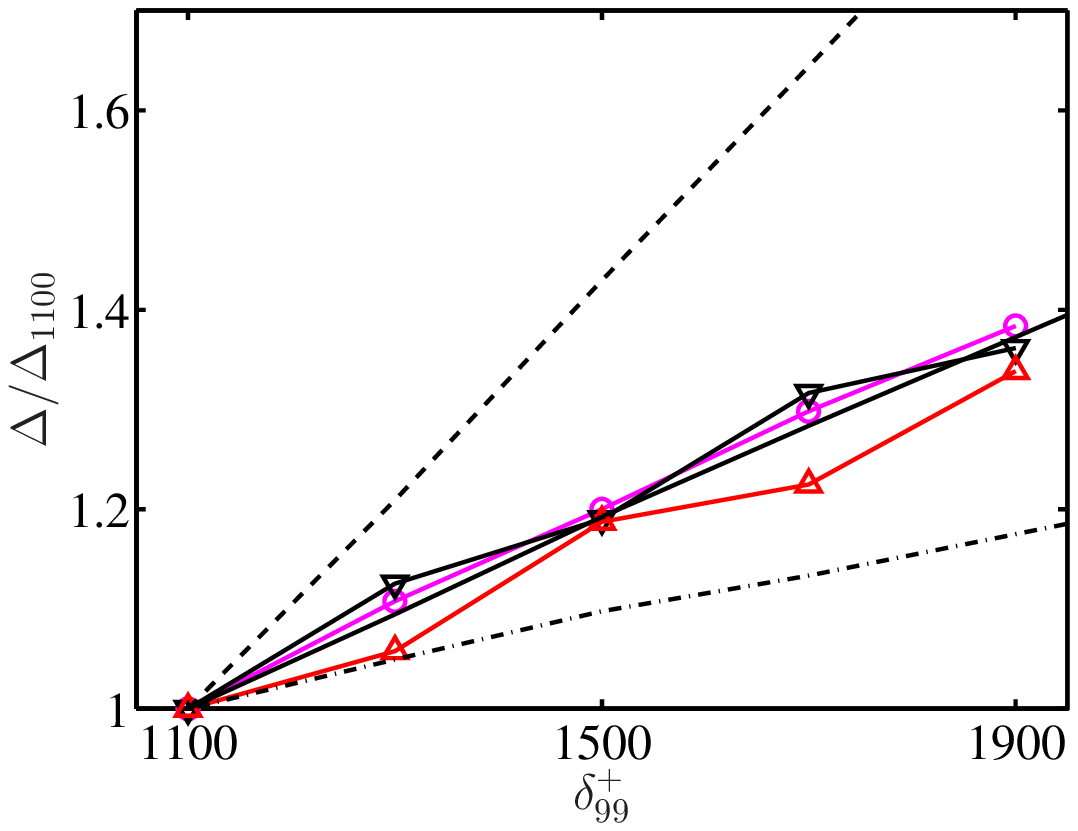}%
\mylab{-0.39\textwidth}{0.30\textwidth}{(c)}%
\hspace{2mm}%
\includegraphics[width=0.48\textwidth]{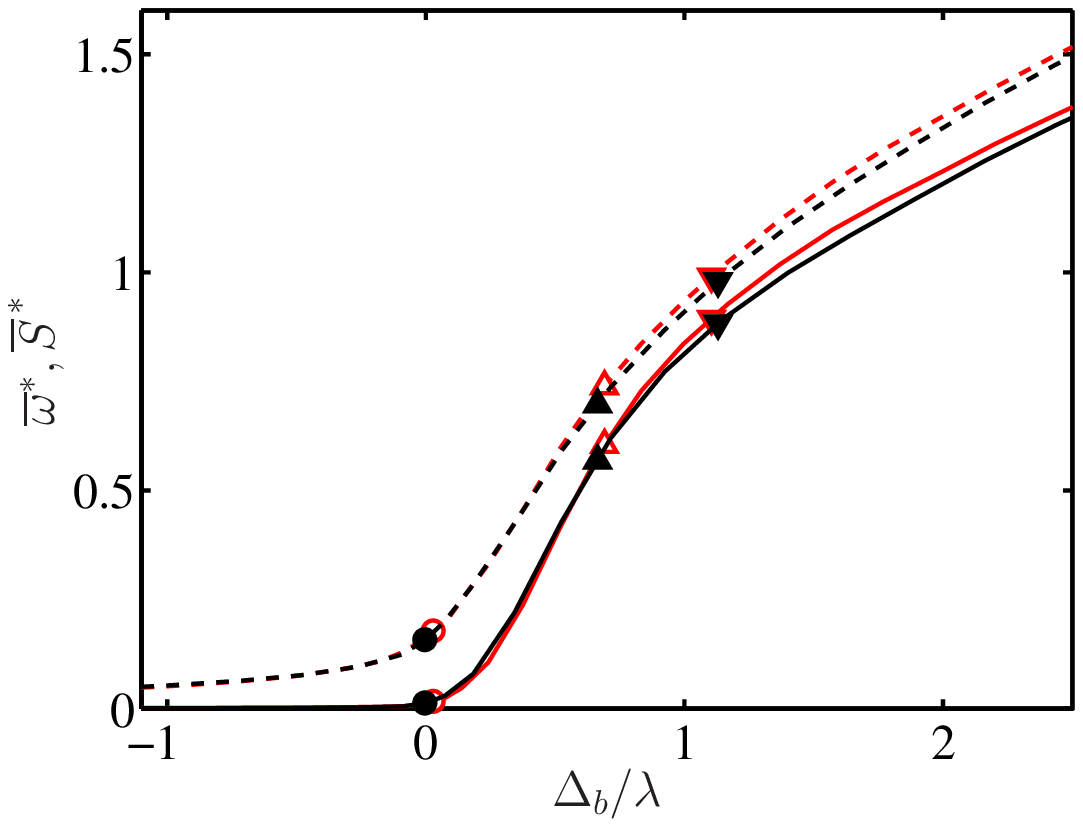}%
\mylab{-0.39\textwidth}{0.30\textwidth}{(d)}%
}
\caption{%
(a) Ratio of the conditional vorticity and rate-of-strain magnitude as a function of the
ball distance to the vorticity interface, normalised with the Taylor microscale at
$y/\delta_{99}=0.6$. \circle, $\omega_0^*=0.01$; \dtrian, 0.02; \trian, 0.04.
(b) Sketch of the definition of the interface vorticity thickness, $\Delta_\omega$. For
other symbols, see text.
(c) Lengths scales as functions of $\delta_{99}^+$, normalised with respect to
$\delta_{99}^+=1100$. Lines with symbols are $\Delta_\omega$ for interface thresholds as in
(a). Lines without symbols are flow length scales: \chndot, $\eta$; \solid, $\lambda$;
\dashed, $\delta_{99}$.
(d) Conditional profiles of: \solid, $\overline{\omega}$; \dashed, $\overline{S}$, for
$\omega_0^*=0.01$ and the two extreme Reynolds numbers. Open symbols are
$\delta_{99}^+=1100$, and closed ones are $\delta_{99}^+=1900$. \circle, $\Delta_{out}$ in
(b); \trian, $\Delta_{in}$; \dtrian, position $\Delta_u$ of the maximum gradient of the
velocity magnitude in figure \ref{fig:shear_profile}(b).
}
\label{fig:thick}
\end{figure}

Since we have defined the interface by the difference between the
conditional vorticity and rate-of-strain, the ratio
$\varpi=\overline{\omega}^*/\overline{S^*}$ is a useful indicator of
its location. It is shown in figure \ref{fig:thick}(a) for three
different thresholds. Because the geometry of the interface changes
with $\omega_0^*$, the indicator also changes, but it always undergoes
a smooth change between $\varpi\approx 1$ in the turbulent core, and
$\varpi\ll 1$ in the free stream. It is interesting that the limit in
the turbulent end is $\varpi=0.9$ rather than the homogeneous result
$\varpi=1$, but conditional and volume averages are not equivalent,
and the observed ratio is robust across thresholds and Reynolds
number. The ratio eventually climbs to about unity at distances from
the interface of the order of $\delta_{99}$, probably because far from
the interface the $\Delta_b$ isosurfaces become approximately flat,
and $\overline{\omega}\approx \bra\omega\ket$. The mean enstrophy and
dissipation agree very well at all the wall distances within the
boundary layer \citep{SilleroJimenez}.

Our definition of thickness is sketched in figure \ref{fig:thick}(b)
as the distance $\Delta_\omega$ between the intersections with
$\varpi=0$ and $\varpi=1$ of a tangent drawn through the steepest
point of the indicator. Because $\Delta_b$ is only defined with
respect to a particular isosurface and is not an additive property
(see \S\ref{sec:disfield}), any definition of thickness depends on the
detection threshold, but $\Delta_\omega$ scales well the whole
indicator profile for a given $\omega_0^*$, as a function of the
Reynolds number. The ratio of $\Delta_\omega(\delta_{99})$ to its
value at $\delta_{99}^+=1100$ is shown in \ref{fig:thick}(c). Note
that the Reynolds number dependence is the same for the three
thresholds in the figure, even if the thickness at the highest
threshold is about 1.5 times narrower than at the lowest one (not
shown). The figure also includes the Reynolds number dependence of the
three candidate length scales, and it is clear that the Taylor
microscale is the best match.

The scaling with $\lambda$ extends to the conditional profiles of
$\overline{\omega}$ and $\overline{S}$, shown in figure
\ref{fig:thick}(d) for our two extreme Reynolds numbers.  This figure
also displays the inner and outer limits of the vorticity interface
layer, defined as in figure \ref{fig:thick}(b). They span a thickness
$\Delta_\omega\approx 0.66\lambda$ for this particular
$\omega_0^*$. The peak of the velocity gradient interface discussed in
the next section is included in figure \ref{fig:thick}(d) for
reference. It is always deeper into the turbulent region than the
vorticity interface.

\section{The velocity interface}\la{sec:shear}

A model that has been extensively discussed in the literature is that
the T/NT interface layer is an active region whose dynamics is
dominated by a strong localised shear \citep{HuntDurbin}. We have
already mentioned that peaks in the vorticity magnitude have been
sought with uncertain success, but the two issues are
different. Roughly speaking, the vorticity magnitude describes `how
many' vortices there are, while a localised shear measures how are
they oriented. We have already seen that the vorticity magnitude
changes rapidly near the interface, and it follows from the
solenoidality of the vorticity field that the vortex lines at the edge
of the potential region have to be roughly parallel to the
interface. That, by itself, should lead to a reinforcement of the
tangential versus the normal vorticity component at the interface, but
whether the vortices organise themselves parallel to each other to
produce a net velocity gradient depends on the details of the
vorticity dynamics. At the moment, this can only be answered
empirically, although linearised analysis suggests that they should
\citep{hunt2006mechanics}. The experimental test is complicated by the
tendency of different groups to define the interface by thresholding
different quantities. For example, there is clear evidence of a strong
interfacial shear $(\partial u/\partial y)$ in \cite{Chauhan:14}, but
their interface is defined by thresholding $u$, and their velocity
discontinuity is probably a similar phenomenon to the vorticity
discontinuity found in the previous sections when thresholding the
vorticity. In fact, a joint PDF of the enstrophy and of the kinetic
energy (not shown) shows fairly wide distributions of each quantity
along isosurfaces of the other. For example, the vorticity magnitude
over the isosurface of the kinetic energy equivalent to that used by
\cite{Chauhan:14} ranges from $\omega^*<0.01$ to $\omega^*>1$. The two
interface definitions are probably very different in detail.

It has been known for some time that the mean streamwise velocities
within the vortical and potential regions of free shear layers
\citep{wyg:fie:70} and boundary layers \citep{jimenez2010turbulent}
are different. \citet{West:etal:09} made a detailed analysis of the
interface of a jet, and found that $\overline{\omega_z}$ is restricted
to the turbulent region (defined using the distance $\Delta_v$), with
a mild peak of the order of 20\% at the interface. They discuss this
peak as a possible surrogate for interfacial shear. They also find a
discontinuity in the streamwise velocity, but the limited resolution
of their experimental method spreads it over a fairly wide layer.

To test this matter on the present data, whose Reynolds number is
substantially higher than in \cite{jimenez2010turbulent} or
\citet{West:etal:09}, we compute the conditional velocity norm and its
gradient with respect to the distance to a vorticity interface, for a
threshold below the topological transition. We first compute the joint
PDF and the conditional profile of the velocity magnitude $|\uvec|$
with respect to $\Delta_b$, and then compute the gradient
$\dr\overline{|\uvec|}/\dr\Delta_b$.

\begin{figure}
\centering
\includegraphics[width=0.455\textwidth]{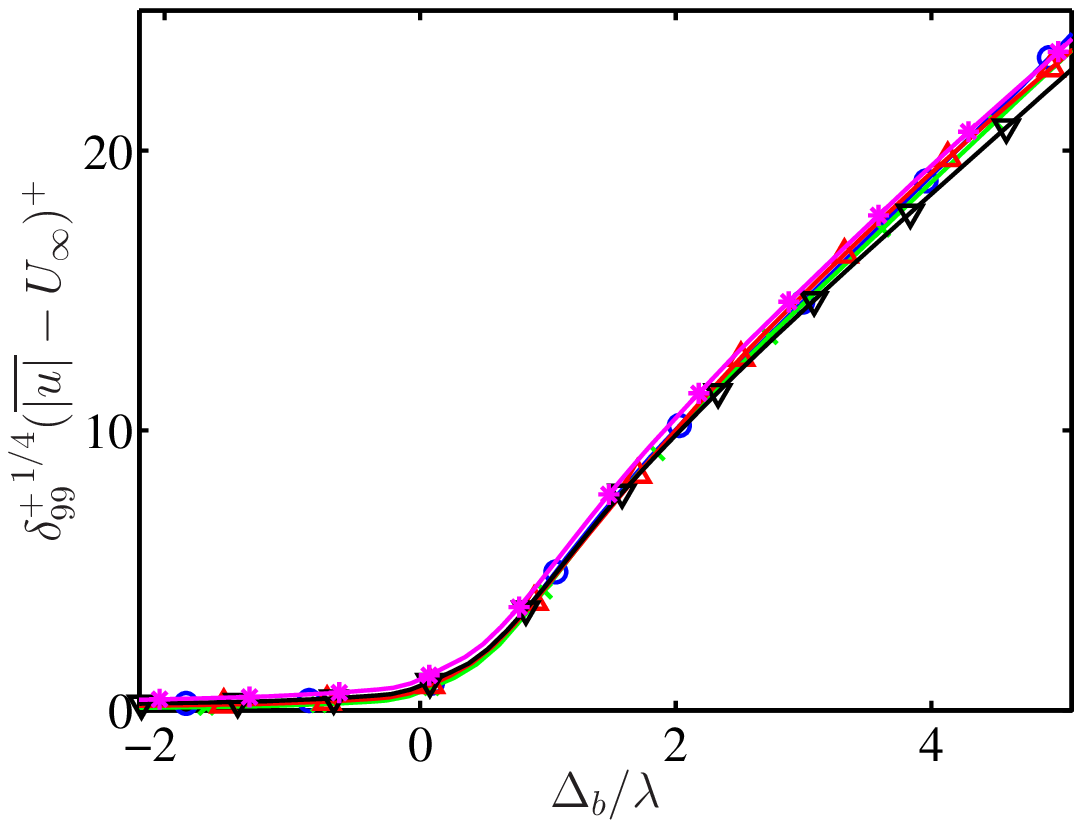}%
\mylab{-.38\textwidth}{.31\textwidth}{(a)}%
\hspace{2mm}%
\includegraphics[width=0.45\textwidth]{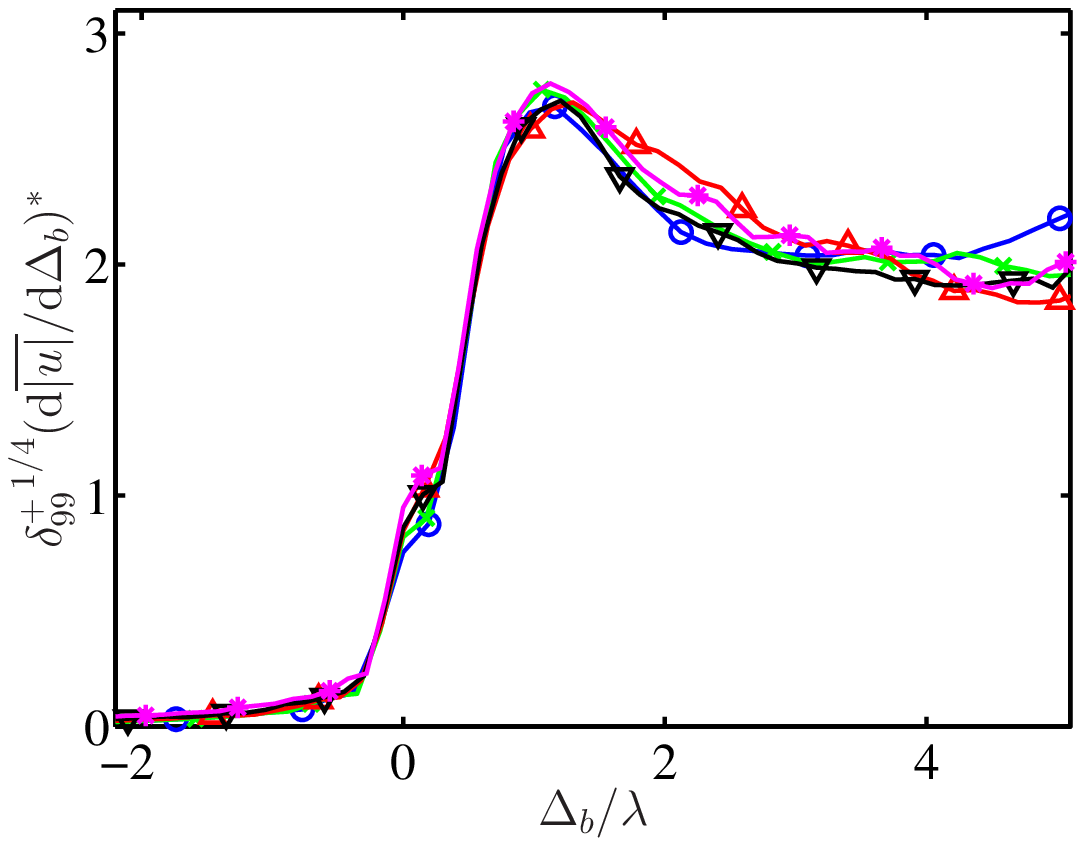}%
\mylab{-.38\textwidth}{.31\textwidth}{(b)}%
\caption{(a) Conditional profile of the norm of the velocity, $\overline{|\uvec|}$, with respect
to the ball distance $\Delta_b$. Vorticity interface $\omega_0^*=0.01$. (b)
Conditional velocity gradient, $\dr\overline{|\uvec|}/\dr\Delta_b$. 
$\circ$, $\delta_{99}^+=1100$; $\times$, 1300;
$\triangledown$, 1500; $\vartriangle$, 1700; $\star$, 1900.
}
\label{fig:shear_profile}
\end{figure}

These profiles are presented in figure \ref{fig:shear_profile} for a
range of Reynolds numbers, and agree reasonably well with the results
of \citet{West:etal:09} in a jet. The gradient of the velocity
magnitude is restricted to the turbulent side, and there is a mild
peak near the interface. The scalings used in this figure are those
found to work best for these quantities. The scaling of the distance
with $\lambda$ agrees with the results in the previous section, but
the scaling of the velocity is different from those found up to
now. Scaling the velocity gradient in star units and lengths with
$\lambda$ would correspond to a velocity scale $u_\tau$, but the
collapse of the different Reynolds numbers in figure
\ref{fig:shear_profile} requires an extra factor
${\delta_{99}^+}^{1/4}$ which implies that the velocity differences
across the interface are proportional to the Kolmogorov velocity
$(\nu\varepsilon)^{1/4}$. Since this is the velocity usually
associated with length scales of the order of $\eta$, its presence in
this context is difficult to explain, but the scaling is
clear. Omitting the ${\delta_{99}^+}^{1/4}$ factor in figure
\ref{fig:shear_profile}(b) would spread the height of the peaks over a
factor of 1.2, which is comparable to the amplitude of the peak
itself.

Note that the mean gradient in figure \ref{fig:shear_profile}(b),
$(\dr\overline{|\uvec|}/\dr\Delta_b)^*\approx 0.4$, is of the same
order of magnitude as the characteristic vorticity magnitude in this
region $(\omega^*\approx 1)$, implying a substantial alignment of the
vortices. Note also that it is unclear whether this gradient
represents a shear layer at the interface. That would imply a normal
jump of the tangential velocity, but it is difficult to define either
normals or tangents to a fractal surface. The interface used here
corresponds to the one in figure \ref{fig:distancefield}(a), and the
range of distances in the turbulent side of figure
\ref{fig:shear_profile}(b) is comparable to the first band of contours
in figure \ref{fig:distancefield}(a). The conditional shear profile in
figure \ref{fig:shear_profile}(b) is obtained numerically by
differentiating the velocity profile in figure
\ref{fig:shear_profile}(a). This amplifies the small error produced
when the conditional average is computed very near $\Delta=0$,
causing the small kink in the shear profile.

Although we have mentioned that the measured thickness in different
experiments can only be used as rough estimations, because of the
variety of definitions and flows, some comparisons may be useful. The
present results are that the thickness of the vorticity interface at
$\omega_0^*=0.01$ is $\Delta_\omega/\lambda=0.68\pm 0.01$ ($0.48\pm
0.02$ at $\omega_0^*=0.04$), where the uncertainties refer to the
variation over the range of Reynolds numbers
$Re_\lambda\in(75-108)$. The thickness defined by the position of the
maximum of the velocity gradient in figure \ref{fig:shear_profile}(b)
is $\Delta_u/\lambda = 1.16\pm 0.07$ for $\omega_0^*=0.01$, and
$0.73\pm 0.02$ at $\omega_0^*=0.04$.  \cite{GampertRetau} estimate
$\Delta/\lambda=2.9\pm 0.2$ for the interface of a passive scalar in a
round jet with $Re_\lambda\in (61-140)$, using $\Delta_v$ corrected
for the orientation of the interface in two dimensional
sections. \cite{West:etal:09} cite a thickness of the order of
$\lambda$ for a a circular jet at $Re_\lambda\approx 60$, without
scaling information, and \cite{SilvaTaveira} find $\Delta/\lambda =
0.73\pm 0.34$ for the vorticity interface of temporally growing planar
jets with $Re_\lambda\in(60-160)$, with a clear growing trend from
$\Delta/\lambda=0.54$ to 1.34 in that range of Reynolds numbers. Since
their vorticity profiles contain interface peaks such as those of the
higher thresholds in figure \ref{fig:condaverage_seepeak}(a), and
these peaks are used to determine the thickness, their results are
difficult to compare with ours. The same authors also find
substantially thinner interface layers for the shearless contact of
two different turbulent intensities. In summary all that can be said
is that the interface thickness depends on the measurement technique,
on the threshold used to define the interface, and on the flow being
investigated, but that it probably scales with $\lambda$, and actually
is of order $\lambda$, in most cases.

\section{\label{sec:Conclusions} Conclusions.}

We have studied the T/NT interface of a zero-pressure-gradient
turbulent boundary layer in the range of Reynolds numbers
$\delta^+_{99}=1000-2000$, equivalent to $Re_\lambda\approx
75-110$. The emphasis is on the statistical description of the
relatively large-scale interactions between turbulent and
non-turbulent fluid across the fractal intermittent zone, rather than
on the details of the smaller scales at which the interface can be
considered smooth. We define the interface as an approximation to an
isosurface of the vorticity magnitude, and show that its properties
depend strongly on the threshold $\omega_0$ used to define it. The
dependence on the Reynolds number can be eliminated by normalising
$\omega_0$ in `star' units, defined in terms of the root-mean-squared
magnitude of the enstrophy fluctuations at the edge of the boundary
layer, $u_\tau^2 (\delta_{99}^+)^{-1/2}/ \nu$, rather than in wall
units. In this normalisation, the geometric complexity of the
interface undergoes a transition across $\omega_0^*\approx 0.1-2$,
characterised by the increase of the fractal dimension and of the
topological genus, and that can be interpreted as the transition of
the isosurface from the free stream into the core turbulence.

Studying the behaviour of turbulence in the neighbourhood of the
interface requires the definition of the distance between a point and
a general surface. We introduce a new definition of (ball) distance,
specifically designed for complex surfaces and three-dimensional data
sets, which is compared with the more usual wall-normal (vertical)
distance to the top of the interface. While the former captures
correctly the increase of complexity across the transition, the
vertical distance misses most of it, because it hides many of the
convolutions, pockets and handles of the vorticity isosurface. In
fact, if the interface is defined as a zero-distance isosurface, the
two definitions produce different interfaces that differ even in their
average distance to the wall. While the `ball' interface follows the
vorticity isosurface as it gets closer to the wall at high thresholds,
the vertical `envelope' always stays close to the boundary layer edge.
 
We have shown that these limitations of the vertical distance are
responsible for some of the previously reported properties of the T/NT
interface. For example, the proposed layer of localised high vortex
intensity at the edge of the turbulent region disappears with the new
distance definition, and reappears with the old one.

We have used the difference between the two distance definitions to
characterise the pockets of irrotational flow as they are entrained
into the body of the flow, throwing some light on the controversy
between engulfing and nibbling. We show that the rate at which
vorticity diffuses into the irrotational pockets within the turbulent
region is independent of their position within the layer, but that
entrainment is enhanced because pockets become smaller as they are
entrained from the edge of the layer towards the wall, presumably
because they are broken down in the process. The size of the entrained
pockets scales in viscous units, but they are found at depths that
scale with the boundary layer thickness.

There is a narrow interface layer in which the enstrophy decays from
its core value, $\omega^*\approx 1$, to that of the free stream. To
ascertain whether this sharp transition is a statistical artefact of
the thresholding procedure or a true physical feature, we study
interfaces based on threshoding the norm of the rate-of-strain
tensor. We show that, whereas the enstrophy and the strain change
sharply across the vorticity interface, neither of them does so across
a strain interface. We conclude that enstrophy thresholding represents
a physical feature, while thresholding the strain does not.

We have studied in some detail the conditionally averaged properties
in the neighbourhood of an enstrophy interface, using a threshold
below the topological transition.  We find that even within the layer
in which the vorticity decays sharply, the straining structure of the
flow is essentially identical to that in the core turbulence. Because
homogeneity would imply that vorticity and strain should have
comparable magnitudes, we use this discrepancy to define a
non-equilibrium region that we identify as the interface layer. The
vorticity in this fractal `buffer' layer, even while undergoing
viscous diffusion into the free stream, retains most of the structure
of the interior of the flow.  Its enstrophy levels are in the range
previously shown to be within the complexity transition.

Finally, we also study the velocity magnitude in the neighbourhood of
the vorticity interface, completing a description of the kinematics of
the T/NT interface layer. In agreement with previous investigators, we
find that the deviations from the free-stream velocity are mostly
excluded from the irrotational zone, which includes the engulfed
pockets when using our definition of distance. The derivative of the
conditional velocity magnitude with respect to the ball distance,
which can loosely be interpreted as a shear parallel to the interface,
is restricted to the turbulent zone, with a mild maximum at the inner
edge of the interface layer.

We have explored several definitions of the thickness of the interface
layer, all of which unequivocally scale with the Taylor microscale
over our range of Reynolds numbers.

Several open questions remain. The first one is the origin of the
scaling of the interface thickness with the Taylor microscale, because
the usual argument that this layer is only subject to the strain of
the large scales is weakened here by the direct measurement of the
rate-of-strain tensor. The second one is the scaling of the
conditional velocity magnitude.  The observed scaling of the enstrophy
in star units, together of the scaling of the lengths with $\lambda$,
implies that the velocity scale should be the friction velocity, but
the collapse of the conditional velocity requires a different unit,
which differs by a factor of ${\delta_{99}^+}^{1/4}$. We can offer no
explanation for these two results but we believe that, within our
range of Reynolds numbers, the resolution of our numerical simulation
is enough to exclude most other obvious alternatives.

\subsection*{Acknowledgements}
This work of was funded by CICYT under grant TRA2009-11498, and by the
European Research Council under grant ERC-2010.AdG-20100224 and
ERC-2014.AdG-669505. Figure \ref{fig:vorticity_isocontour} was
obtained with the help of the Barcelona Supercomputing Centre. The
computational resources of the Argonne Leadership Computing Facility
at Argonne National Laboratory were supported by the U.S.  Department
of Energy under contract No. DE-AC02-06CH11357.

\bibliography{Borrell_GB}{}

\begin{thebibliography}{70}
\expandafter\ifx\csname natexlab\endcsname\relax\def\natexlab#1{#1}\fi

\bibitem[Arya {\em et~al.\/}(1998)Arya, Mount, Netanyahu, Silverman \&
  Wu]{Arya:1998}
{\sc Arya, S., Mount, D.~M., Netanyahu, N.~S., Silverman, R. \& Wu, A.~Y.} 1998
  An optimal algorithm for approximate nearest neighbor searching in fixed
  dimensions. {\em J. ACM\/} {\bf 45}~(6), 891--923.

\bibitem[Atkinson {\em et~al.\/}(2014)Atkinson, Hackl, Stegeman, Borrell \&
  Soria]{atkinson2014numerical}
{\sc Atkinson, C., Hackl, J., Stegeman, P., Borrell, G. \& Soria, J.} 2014
  Numerical issues in lagrangian tracking and topological evolution of fluid
  particles in wall-bounded turbulent flows. In {\em J. Phys.\/}, , vol. 506,
  p. 012003. IOP Publishing.

\bibitem[Batchelor(1967)]{bat67}
{\sc Batchelor, G.~K.} 1967 {\em An introduction to fluid dynamics\/}.
  Cambridge U. Press.

\bibitem[Bisset {\em et~al.\/}(2002)Bisset, Hunt \& Rogers]{FLM:95049}
{\sc Bisset, D.~K., Hunt, J. C.~R. \& Rogers, M.~M.} 2002 The
  turbulent/non-turbulent interface bounding a far wake. {\em J. Fluid Mech.\/}
  {\bf 451}, 383--410.

\bibitem[Borrell {\em et~al.\/}(2013)Borrell, Sillero \& Jim\'enez]{Borrell}
{\sc Borrell, G., Sillero, J.~A. \& Jim\'enez, J.} 2013 A code for direct
  numerical simulation of turbulent boundary layers at high reynolds numbers in
  {BG/P} supercomputers. {\em Comp. Fluids\/} {\bf 80}, 37--43.

\bibitem[Chauhan {\em et~al.\/}(2014)Chauhan, Philip, de~Silva, Hutchins \&
  Marusic]{Chauhan:14}
{\sc Chauhan, K., Philip, J., de~Silva, C.~M., Hutchins, N. \& Marusic, I.}
  2014 The turbulent/non-turbulent interface and entrainment in a boundary
  layer. {\em J. Fluid Mech.\/} {\bf 742}, 119--151.

\bibitem[Coles(1956)]{FLM:367043}
{\sc Coles, D.} 1956 The law of the wake in the turbulent boundary layer. {\em
  J. Fluid Mech.\/} {\bf 1}, 191--226.

\bibitem[Corrsin(1943)]{NACA:W-94}
{\sc Corrsin, S.} 1943 Investigation of flow in an axially symmetric heated jet
  of air. {\em NACA WR\/} {\bf W-94}.

\bibitem[Corrsin \& Kistler(1955)]{NACA:1244}
{\sc Corrsin, S. \& Kistler, A.~L.} 1955 Free-stream boundaries of turbulent
  flows. {\em NACA TR\/} {\bf 1244}.

\bibitem[Dahm \& Dimotakis(1987)]{DahmDimotakis}
{\sc Dahm, W. J.~A. \& Dimotakis, P.~E} 1987 Measurements of entrainment and
  mixing in turbulent jets. {\em AIAA J.\/} {\bf 25}, 1216--1223.

\bibitem[Dimotakis(2000)]{Dimotakis}
{\sc Dimotakis, P.~E.} 2000 The mixing transition in turbulent flows. {\em J.
  Fluid Mech.\/} {\bf 409}, 69--98.

\bibitem[Eisma {\em et~al.\/}(2015)Eisma, Westerweel, Ooms \&
  Elsinga]{eisma2015interfaces}
{\sc Eisma, J., Westerweel, J., Ooms, G. \& Elsinga, G.~E.} 2015 Interfaces and
  internal layers in a turbulent boundary layer. {\em Phys. Fluids\/} {\bf
  27}~(5), 055103.

\bibitem[Ferre {\em et~al.\/}(1990)Ferre, Mumford, Savill \& Giralt]{Ferreetal}
{\sc Ferre, J.~A., Mumford, J.~C., Savill, A.~M \& Giralt, F.} 1990
  Three-dimensional large-eddy motions and fine-scale activity in a plane
  turbulent wake. {\em J. Fluid Mech.\/} {\bf 210}, 371--414.

\bibitem[Fiedler \& Head(1966)]{Fiedler}
{\sc Fiedler, H.~E. \& Head, M.~R.} 1966 Intermittency measurements in the
  turbulent boundary layer. {\em J. Fluid Mech.\/} {\bf 25}, 719--735.

\bibitem[Gampert {\em et~al.\/}(2014)Gampert, Boschung, Hennig, Gauding \&
  Peters]{FLM:9282512}
{\sc Gampert, M., Boschung, J., Hennig, F., Gauding, M. \& Peters, N.} 2014 The
  vorticity versus the scalar criterion for the detection of the
  turbulent/non-turbulent interface. {\em J. Fluid Mech.\/} {\bf 750},
  578--596.

\bibitem[Gampert {\em et~al.\/}(2013)Gampert, Narayanaswamy, Schaefer \&
  Peters]{GampertRetau}
{\sc Gampert, M., Narayanaswamy, V., Schaefer, P. \& Peters, N.} 2013
  Conditional statistics of the turbulent/non-turbulent interface in a jet
  flow. {\em J. Fluid Mech.\/} {\bf 731}, 615--638.

\bibitem[Gartshore(1966)]{FLM:381094}
{\sc Gartshore, I.~S.} 1966 An experimental examination of the large-eddy
  equilibrium hypothesis. {\em J. Fluid Mech.\/} {\bf 24}, 89--98.

\bibitem[Holzner {\em et~al.\/}(2007)Holzner, Liberzon, Nikitin, Kinzelbach \&
  Tsinober]{holzner2007small}
{\sc Holzner, M., Liberzon, A., Nikitin, N., Kinzelbach, W. \& Tsinober, A.}
  2007 Small-scale aspects of flows in proximity of the turbulent/nonturbulent
  interface. {\em Phys. of Fluids\/} {\bf 19}~(7), 071702.

\bibitem[Holzner {\em et~al.\/}(2008)Holzner, Liberzon, Nikitin, L\"uthi,
  Kinzelbach \& Tsinober]{FLM:1757608}
{\sc Holzner, M., Liberzon, A., Nikitin, N., L\"uthi, B., Kinzelbach, W. \&
  Tsinober, A.} 2008 A lagrangian investigation of the small-scale features of
  turbulent entrainment through particle tracking and direct numerical
  simulation. {\em J. Fluid Mech.\/} {\bf 598}, 465--475.

\bibitem[Hunt \& Durbin(1999)]{HuntDurbin}
{\sc Hunt, J. C.~R \& Durbin, P.~A} 1999 Perturbed vortical layers and shear
  sheltering. {\em Fluid Dyn. Res.\/} {\bf 24}, 375--404.

\bibitem[Hunt {\em et~al.\/}(2006)Hunt, Eames \& Westerweel]{hunt2006mechanics}
{\sc Hunt, J. C.~R., Eames, I. \& Westerweel, J.} 2006 Mechanics of
  inhomogeneous turbulence and interfacial layers. {\em J. Fluid Mech.\/} {\bf
  554}, 499--519.

\bibitem[Ishihara {\em et~al.\/}(2013)Ishihara, Kaneda \& Hunt]{Ishihara}
{\sc Ishihara, Takashi, Kaneda, Yukio \& Hunt, Julian, C.R.} 2013 Thin shear
  layers in high reynolds number turbulence -- {DNS} results. {\em Flow Turb.
  and Comb.\/} {\bf 91}~(4), 895--929.

\bibitem[Ishihara {\em et~al.\/}(2015)Ishihara, Ogasawara \&
  Hunt]{Ishihara2015analysis}
{\sc Ishihara, T., Ogasawara, H. \& Hunt, J. C.~R.} 2015 Analysis of
  conditional statistics obtained near the turbulent/non-turbulent interface of
  turbulent boundary layers. {\em J. Fluid. Struct.\/} {\bf 53}, 50--57.

\bibitem[Jim{\'e}nez(2013)]{jimenez2013near}
{\sc Jim{\'e}nez, J.} 2013 Near-wall turbulence. {\em Phys. Fluids\/} {\bf
  25}~(10), 101302.

\bibitem[Jim\'enez {\em et~al.\/}(2010)Jim\'enez, Hoyas, Simens \&
  Mizuno]{jimenez2010turbulent}
{\sc Jim\'enez, J., Hoyas, S., Simens, M.~P. \& Mizuno, Y.} 2010 Turbulent
  boundary layers and channels at moderate {R}eynolds numbers. {\em J. Fluid
  Mech.\/} {\bf 657}, 335--360.

\bibitem[Jim\'enez \& Wray(1998)]{JimenezWray98}
{\sc Jim\'enez, J. \& Wray, A.~A} 1998 On the characteristics of vortex
  filaments in isotropic turbulence. {\em J. Fluid Mech.\/} {\bf 373},
  255--285.

\bibitem[Jones {\em et~al.\/}(2006)Jones, Baerentzen \& Sramek]{jones20063d}
{\sc Jones, M., Baerentzen, J.~A. \& Sramek, M.} 2006 {3D} distance fields: A
  survey of techniques and applications. {\em IEEE Trans. Vis. Comput. Gr.\/}
  {\bf 12}~(4), 581--599.

\bibitem[Klebanoff(1955)]{NACA:1247}
{\sc Klebanoff, P.~S.} 1955 Characteristics of turbulence in a boundary layer
  with zero pressure gradient. {\em NACA TR\/} {\bf 1247}.

\bibitem[Kolmogorov(1941)]{kol41}
{\sc Kolmogorov, A.~N.} 1941 The local structure of turbulence in
  incompressible viscous fluids at very large {Reynolds} numbers. {\em Dokl.
  Akad. Nauk. SSSR\/} {\bf 30}, 301--305, reprinted in {\it Proc. R. Soc.
  London.}\/ A {\bf 434}, 9--13 (1991).

\bibitem[Kovasznay {\em et~al.\/}(1970)Kovasznay, Kibens \&
  Blackwelder]{FLM:382907}
{\sc Kovasznay, L. S.~G., Kibens, V. \& Blackwelder, R.~F.} 1970 Large-scale
  motion in the intermittent region of a turbulent boundary layer. {\em J.
  Fluid Mech.\/} {\bf 41}, 283--325.

\bibitem[Lee \& Sung(2013)]{leesung2013}
{\sc Lee, J.~H. \& Sung, H.~J.} 2013 Comparison of very-large-scale motions of
  turbulent pipe and boundary layer simulations. {\em Phys. Fluids\/} {\bf
  25}~(4), 045103.

\bibitem[Leung {\em et~al.\/}(2012)Leung, Swaminathan \& Davidson]{Genus}
{\sc Leung, T, Swaminathan, N. \& Davidson, P.~A.} 2012 Geometry and
  interaction of structures in homogeneous isotropic turbulence. {\em J. Fluid
  Mech.\/} {\bf 710}, 453--481.

\bibitem[Lozano-Dur{\'a}n \& Borrell(2015)]{GenusTOMS}
{\sc Lozano-Dur{\'a}n, A. \& Borrell, G.} 2015 An efficient algorithm to
  compute the genus of discrete surfaces and applications to turbulent flows.
  {\em TOMS\/} p. (In press).

\bibitem[Mandelbrot(1975)]{FLM:386975}
{\sc Mandelbrot, B.~B.} 1975 On the geometry of homogeneous turbulence, with
  stress on the fractal dimension of the iso-surfaces of scalars. {\em J. Fluid
  Mech.\/} {\bf 72}, 401--416.

\bibitem[Mathew \& Basu(2002)]{mathew:2065}
{\sc Mathew, J. \& Basu, A.~J.} 2002 Some characteristics of entrainment at a
  cylindrical turbulence boundary. {\em Phys. Fluids\/} {\bf 14}~(7),
  2065--2072.

\bibitem[Mellado {\em et~al.\/}(2009)Mellado, Wang \& Peters]{FLM:5533652}
{\sc Mellado, J.~P., Wang, L. \& Peters, N.} 2009 Gradient trajectory analysis
  of a scalar field with external intermittency. {\em J. Fluid Mech.\/} {\bf
  626}, 333--365.

\bibitem[Moisy \& Jim\'enez(2004)]{MoisyJimenez}
{\sc Moisy, F \& Jim\'enez, J.} 2004 Geometry and clustering of intense
  structures in isotropic turbulence. {\em J. Fluid Mech.\/} {\bf 513},
  111--133.

\bibitem[Muja \& Lowe(2014)]{muja2014scalable}
{\sc Muja, M. \& Lowe, D.~G.} 2014 Scalable nearest neighbor algorithms for
  high--dimensional data. {\em IEEE T. Pattern Anal.\/} {\bf 11}, 2227--2240.

\bibitem[Mungal {\em et~al.\/}(1991)Mungal, Karasso \& Lozano]{Mungaletal}
{\sc Mungal, M.~G, Karasso, P.~S. \& Lozano, A.} 1991 The visible structure of
  turbulent jet diffusion flames: large-scale organization and flame tip
  oscillation. {\em Comb. Sci. Tech.\/} {\bf 76}, 165--185.

\bibitem[Murlis {\em et~al.\/}(1982)Murlis, Tsai \& Bradshaw]{Murlis}
{\sc Murlis, J., Tsai, J. \& Bradshaw, P} 1982 The structure of turbulent
  boundary layers at low {R}eynolds numbers. {\em J. Fluid Mech.\/} {\bf 122},
  13--56.

\bibitem[Phillips(1955)]{phillips1955irrotational}
{\sc Phillips, O.~M.} 1955 The irrotational motion outside a free turbulent
  boundary. In {\em Math. Proc. Cambridge Phil. Soc.\/}, , vol.~51, pp.
  220--229. Cambridge Univ Press.

\bibitem[Pirozzoli \& Bernardini(2013)]{pirozzoli2013probing}
{\sc Pirozzoli, S. \& Bernardini, M.} 2013 Probing high-reynolds-number effects
  in numerical boundary layers. {\em Phys. Fluids\/} {\bf 25}~(2), 021704.

\bibitem[Prasad \& Sreenivasan(1989)]{PrasadSreenivasan}
{\sc Prasad, R.R. \& Sreenivasan, K.R.} 1989 Scalar interfaces in digital
  images of turbulent flows. {\em Exp. Fluids\/} {\bf 7}~(4), 259--264.

\bibitem[van Reeuwijk \& Holzner(2014)]{Ree:Holz:14}
{\sc van Reeuwijk, M. \& Holzner, M.} 2014 The turbulence boundary of a
  temporal jet. {\em J. Fluid Mech.\/} {\bf 739}, 254--275.

\bibitem[Russ(1994)]{RussBook}
{\sc Russ, J.~C} 1994 {\em Fractal Surfaces\/}. 233 Spring Street, New York,
  N.Y. 10013: Plenum Press.

\bibitem[Sandham {\em et~al.\/}(1988)Sandham, Mungal, Broadwell \&
  Reynolds]{CTRSandham}
{\sc Sandham, N.~D., Mungal, M.~G., Broadwell, J.~E. \& Reynolds, W.~C.} 1988
  Scalar entrainment in the mixing layer. In {\em Proc. Summ. Prog.\/}, pp.
  69--76. Center for Turb. Res. Stanford U.

\bibitem[Schlatter \& {\"O}rl{\"u}(2010)]{FLM:7881038}
{\sc Schlatter, P. \& {\"O}rl{\"u}, R.} 2010 Assessment of direct numerical
  simulation data of turbulent boundary layers. {\em J. Fluid Mech.\/} {\bf
  659}, 116--126.

\bibitem[Sillero {\em et~al.\/}(2013)Sillero, Jim{\'e}nez \&
  Moser]{SilleroJimenez}
{\sc Sillero, J.~A., Jim{\'e}nez, J. \& Moser, R.~D.} 2013 One-point statistics
  for turbulent wall-bounded flows at {R}eynolds numbers up to $\delta^+\simeq
  2000$. {\em Phys. Fluids\/} {\bf 25}~(10), 105102.

\bibitem[Sillero {\em et~al.\/}(2014)Sillero, Jim\'enez \& Moser]{sillero14}
{\sc Sillero, J.~A., Jim\'enez, J. \& Moser, R.~D.} 2014 Two-point statistics
  for turbulent boundary layers and channels at {R}eynolds numbers up to
  $\delta^+ \approx 2000$. {\em Phys. Fluids\/} {\bf 26}, 105109.

\bibitem[da~Silva {\em et~al.\/}(2014{\natexlab{{\em a\/}}})da~Silva, Hunt,
  Eames \& Westerweel]{SilvaAR14}
{\sc da~Silva, C.~B., Hunt, J. C.~R., Eames, I. \& Westerweel, J.}
  2014{\natexlab{{\em a\/}}} Interfacial layers between regions of different
  turbulent intensity. {\em Ann. Rev. Fluid Mech.\/} {\bf 46}, 567--590.

\bibitem[da~Silva \& Pereira(2008)]{da2008invariants}
{\sc da~Silva, C.~B. \& Pereira, J. C.~F.} 2008 Invariants of the
  velocity-gradient, rate-of-strain, and rate-of-rotation tensors across the
  turbulent/nonturbulent interface in jets. {\em Phys. Fluids\/} {\bf 20}~(5),
  055101.

\bibitem[da~Silva {\em et~al.\/}(2011)da~Silva, dos Reis \&
  Pereira]{FLM:8400021}
{\sc da~Silva, C.~B., dos Reis, R. J.~N. \& Pereira, J. C.~F.} 2011 The intense
  vorticity structures near the turbulent/non-turbulent interface in a jet.
  {\em J. Fluid Mech.\/} {\bf 685}, 165--190.

\bibitem[da~Silva \& Taveira(2010)]{SilvaTaveira}
{\sc da~Silva, C.~B. \& Taveira, R.~R.} 2010 The thickness of the
  turbulent/nonturbulent interface is equal to the radius of the large
  vorticity structures near the edge of the shear layer. {\em Phys. Fluids\/}
  {\bf 22}, 121702.

\bibitem[da~Silva {\em et~al.\/}(2014{\natexlab{{\em b\/}}})da~Silva, Taveira
  \& Borrell]{silva2014characteristics}
{\sc da~Silva, C.~B., Taveira, R.~R. \& Borrell, G} 2014{\natexlab{{\em b\/}}}
  Characteristics of the turbulent/nonturbulent interface in boundary layers,
  jets and shear-free turbulence. {\em J. Phys.\/} {\bf 506}~(1), 012015.

\bibitem[de~Silva {\em et~al.\/}(2013)de~Silva, Philip, Chauhan, Meneveau \&
  Marusic]{de2013multiscale}
{\sc de~Silva, C.~M., Philip, J., Chauhan, K., Meneveau, C. \& Marusic, I.}
  2013 Multiscale geometry and scaling of the turbulent-nonturbulent interface
  in high reynolds number boundary layers. {\em Phys. Rev. Let\/} {\bf
  111}~(4), 044501.

\bibitem[Simens {\em et~al.\/}(2009)Simens, Jim\'enez, Hoyas \& Mizuno]{Simens}
{\sc Simens, M., Jim\'enez, J., Hoyas, S. \& Mizuno, Y.} 2009 A high-resolution
  code for turbulent boundary layers. {\em J. Comp. Phys.\/} {\bf 228},
  4218--4231.

\bibitem[Sreenivasan \& Meneveau(1986)]{FLM:392671}
{\sc Sreenivasan, K.~R. \& Meneveau, C.} 1986 The fractal facets of turbulence.
  {\em J. Fluid Mech.\/} {\bf 173}, 357--386.

\bibitem[Sreenivasan {\em et~al.\/}(1989)Sreenivasan, Ramshankar \&
  Meneveau]{Sreenivasan09011989}
{\sc Sreenivasan, K.~R., Ramshankar, R. \& Meneveau, C.} 1989 Mixing,
  entrainment and fractal dimensions of surfaces in turbulent flows. {\em Proc.
  Royal Soc. London A\/} {\bf 421}~(1860), 79--108.

\bibitem[Stewart(1956)]{stewart1956irrotational}
{\sc Stewart, R.~W.} 1956 Irrotational motion associated with free turbulent
  flows. {\em J. Fluid Mech\/} {\bf 1}~(06), 593--606.

\bibitem[Taveira {\em et~al.\/}(2013)Taveira, Diogo, Lopes \&
  da~Silva]{taveira2013lagrangian}
{\sc Taveira, R.~R., Diogo, J.~S., Lopes, D.~C. \& da~Silva, C.~B.} 2013
  Lagrangian statistics across the turbulent-nonturbulent interface in a
  turbulent plane jet. {\em Phys. Rev. E\/} {\bf 88}~(4), 043001.

\bibitem[Taveira \& da~Silva(2014)]{Tav:Sil:PF14}
{\sc Taveira, R.~R. \& da~Silva, C.~B.} 2014 Characteristics of the viscous
  superlayer in shear free turbulence and in planar turbulent jets. {\em Phys.
  Fluids\/} {\bf 26}~(2), 021702.

\bibitem[Teixeira \& da~Silva(2012)]{teixeira2012turbulence}
{\sc Teixeira, M. A.~C. \& da~Silva, C.~B.} 2012 Turbulence dynamics near a
  turbulent/non-turbulent interface. {\em Journal of Fluid Mechanics\/} {\bf
  695}, 257--287.

\bibitem[Tennekes \& Lumley(1972)]{tennekes1972first}
{\sc Tennekes, H. \& Lumley, J.~L.} 1972 {\em A first course in turbulence\/}.
  MIT press.

\bibitem[Townsend(1948)]{Townsend}
{\sc Townsend, A.~A.} 1948 Local isotropy in the turbulent wake of a cylinder.
  {\em Austral. J. Sci. Res. A\/} {\bf 1}, 161--174.

\bibitem[Townsend(1976)]{TownsendBook}
{\sc Townsend, A.~A.} 1976 {\em The structure of turbulent shear flow\/}, 2nd
  edn. Cambridge U. Press.

\bibitem[Watanabe {\em et~al.\/}(2015)Watanabe, Sakai, Nagata, Ito \&
  Hayase]{watanabe2015turbulent}
{\sc Watanabe, T., Sakai, Y., Nagata, K., Ito, Y. \& Hayase, T.} 2015 Turbulent
  mixing of passive scalar near turbulent and non-turbulent interface in mixing
  layers. {\em Phys. Fluids\/} {\bf 27}~(8), 085109.

\bibitem[Westerweel {\em et~al.\/}(2005)Westerweel, Fukushima, Pedersen \&
  Hunt]{WesterweelPRL}
{\sc Westerweel, J., Fukushima, C., Pedersen, J.~M. \& Hunt, J. C.~R.} 2005
  Mechanics of the turbulent-nonturbulent interface of a jet. {\em Phys. Rev.
  Lett.\/} {\bf 95}, 174501.

\bibitem[Westerweel {\em et~al.\/}(2009)Westerweel, Fukushima, Pedersen \&
  Hunt]{West:etal:09}
{\sc Westerweel, J., Fukushima, C., Pedersen, J.~M. \& Hunt, J. C.~R.} 2009
  Momentum and scalar transport at the turbulent/non-turbulent interface of a
  jet. {\em J. Fluid Mech.\/} {\bf 631}, 199--230.

\bibitem[Westerweel {\em et~al.\/}(2002)Westerweel, Hofmann, Fukushima \&
  Hunt]{WesterweelEiF}
{\sc Westerweel, J., Hofmann, T., Fukushima, C. \& Hunt, J.} 2002 The
  turbulent/non-turbulent interface at the outer boundary of a self-similar
  turbulent jet. {\em Exp. Fluids\/} {\bf 33}~(6), 873--878.

\bibitem[Wygnanski \& Fiedler(1970)]{wyg:fie:70}
{\sc Wygnanski, I.~J. \& Fiedler, H.~E.} 1970 The two-dimensional mixing
  region. {\em J. Fluid Mech.\/} {\bf 41}, 327--361.

\end{thebibliography}
\bibliographystyle{jfm}

\end{document}